\documentclass[aps,pra,superscriptaddress,twocolumn,longbibliography]{revtex4-2}
\usepackage{amsmath,amssymb,mathtools}
\usepackage{bm}
\usepackage{graphicx}
\usepackage{array,booktabs,multirow,siunitx}
\usepackage{subcaption}
\usepackage{algorithm,algpseudocode}
\usepackage{braket,etoolbox,enumitem,placeins}
\usepackage[dvipsnames]{xcolor}
\usepackage[colorlinks=true,allcolors=blue]{hyperref}
\usepackage[noabbrev]{cleveref}
\crefname{figure}{Fig.}{Figs.}\Crefname{figure}{Figure}{Figures}
\crefname{equation}{Eq.}{Eqs.}\Crefname{equation}{Equation}{Equations}
\crefname{table}{Table}{Tables}\Crefname{table}{Table}{Tables}
\crefname{section}{Sec.}{Secs.}\Crefname{section}{Section}{Sections}
\crefname{subsection}{Sec.}{Secs.}\Crefname{subsection}{Section}{Sections}
\crefname{subsubsection}{Sec.}{Secs.}\Crefname{subsubsection}{Section}{Sections}
\crefname{appendix}{Appendix}{Appendices}\Crefname{appendix}{Appendix}{Appendices}
\crefname{subappendix}{Appendix}{Appendices}\Crefname{subappendix}{Appendix}{Appendices}
\crefname{subsubappendix}{Appendix}{Appendices}\Crefname{subsubappendix}{Appendix}{Appendices}
\graphicspath{{figures/}}
\hypersetup{pdftitle={Mitigating large circuits with a learned small-circuit mitigator},pdfauthor={Thi Ha Kyaw, Pablo Díez-Valle, Gaurav Saxena}}

\usepackage{color}
\usepackage{comment}
\newcommand{\alldepth}{all-depth model}        

\newcommand{\REV}{\ensuremath{R_{\mathrm{EV}}}}                 
\newcommand{\TVDfour}{\ensuremath{\mathrm{TVD}_{4}}}           
\newcommand{\Psol}{\ensuremath{P_{\mathrm{sol}}}}              

\newcommand{\Psimp}{\ensuremath{\Pi_{\triangle}}}               
\newcommand{\one}{\ensuremath{\mathbf{1}}}                      

\newcommand{\etal}{\textit{et al}}                              

\newcommand{\device}{\texttt{ibm\_quebec}}
\newcommand{\cz}{\texttt{cz}}                                   
\newcommand{\rawarm}{\texttt{raw}}
\newcommand{\psarm}{\texttt{ps}}

\newcommand{\rops}{\texttt{ro\_ps}}
\newcommand{\ropsdep}{\texttt{ro\_ps\_dep}}
\newcommand{\znearm}{\texttt{zne}}

\newcommand{\shapeblind}{\texttt{shape\_blind}}

\newcommand{\gnnps}{\texttt{gnn\_ps16}}
\newcommand{\mlpps}{\texttt{mlp\_ps16}}

\newcommand{\gnncirc}{\texttt{gnn\_circonly}}

\begin{document}
\title{Mitigating large circuits with a learned small-circuit mitigator}
\author{Thi Ha Kyaw}
\email{thiha.kyaw@lge.com}
\affiliation{LG Electronics Toronto AI Lab, Toronto, Ontario M5V 1M3, Canada}

\author{Pablo Díez-Valle}
\email{pvalle@itg.es}
\affiliation{Instituto Tecnológico de Galicia, Cantón Grande 9, Planta 3, 15003 A Coruña, Spain}

\author{Gaurav Saxena}
\email{gaurav.saxena@lge.com}
\affiliation{LG Electronics Toronto AI Lab, Toronto, Ontario M5V 1M3, Canada}
\date{\today}
\begin{abstract}
Learning-based quantum error mitigation infers noise-free results from noisy measurement data.
However, exact training labels become prohibitively expensive to compute as circuits grow, and the generalization capability of a trained model to infer accurate results on circuits of depths and sizes it was not trained on poses a challenge. We show that circuit cutting allows such a model to be reused on circuits larger than those it was trained on.
We benchmark this approach against zero-noise extrapolation, Clifford data regression, and
probabilistic error cancellation on simulated circuits several times wider and deeper than the training
circuits. On random-circuit chains, the fragment-wise mitigator removes most of the reconstructed error. For a
holographic Ising-chain eigensolver under a fixed shot budget, it recovers most of the noise-induced energy
error, close to 90\% or more, and outperforms Clifford data regression and probabilistic error cancellation. Since the mitigator is trained once and needs no additional circuits or noise amplification at inference, its cost is paid up front rather than for every new circuit. The approach is also modular. Any improvement to the small-circuit mitigator, such as retraining on larger fragments or on device-specific noise, carries over directly to every circuit that can be cut into such fragments. Our proposal thus offers a practical route to bringing learned error mitigation to quantum computations far beyond the reach of exact training labels, provided they can be partitioned with a moderate number of cuts.
\end{abstract}
\maketitle

\section{Introduction}\label{sec:introduction}

Noise remains a major obstacle to extracting reliable information
from present-day quantum processors. Imperfect quantum operations,
decoherence, and measurement errors accumulate during a computation,
limiting the accuracy of the observables estimated from quantum circuits,
particularly as their size and depth increase. While quantum error
correction provides a long-term route toward fault-tolerant quantum
computation, quantum error mitigation (QEM)~\cite{Cai2023RMP} offers a near-term alternative for
improving the accuracy of noisy computations without encoding quantum
information into error-correcting codes. Established QEM
techniques, such as zero-noise extrapolation (ZNE), probabilistic
error cancellation (PEC), and Clifford data regression (CDR), exploit
different forms of information about the noise or the target circuit
to estimate ideal expectation values~\cite{Cai2023RMP,Temme2017PRL,Czarnik2021Quantum, Li2017PRX,Kandala2019Nature, Saxena/EMRE/2026,Valle/PIE/2026,GiurgicaTiron2020QCE,Kim2023Nature, vandenBerg2023NatPhys, Bravyi2021PRA, BonetMonroig2018PRA, Vovrosh2021PRE}.
However, their practical effectiveness is often constrained by
additional circuit executions, statistical fluctuations, or the need
for accurate noise characterization. More fundamentally, the resources
required for quantum error mitigation can grow rapidly with circuit
size and depth, as captured by general bounds on its sampling
overhead~\cite{Takagi2022npjQI, Quek2024NatPhys}. Developing mitigation strategies
that balance accuracy, broad applicability, and experimental cost therefore
remains an important challenge.

Machine learning has emerged as a complementary approach to QEM, in
which the relationship between noisy and ideal circuit outputs is
inferred from training data rather than specified through an explicit
analytical noise-inversion procedure. Learning-based mitigation has
been investigated using regression models, neural networks, and
graph-based architectures, with demonstrations of error reduction
across a variety of quantum circuits and noise
environments~\cite{Strikis2021PRXQ,Liao2024NMI,Placidi/DL_QEM/2026, Kim2020IEEEAccess, Bao2025ICLR,Tousi2025arXiv}.
In particular, recent deep-learning approaches have shown that neural
networks can learn corrections to noisy output probability
distributions, allowing multiple expectation values to be mitigated
simultaneously~\cite{Placidi/DL_QEM/2026}. Other studies have demonstrated the
potential of transferring learned mitigation strategies between
circuit families or system sizes~\cite{Liao2024NMI,Cantori/DL_QEM_partial_knitting/2025,Wang2026QMI}.
These results establish the promise of data-driven QEM, but they also
raise the question of whether it applies to circuits substantially
larger or deeper than those encountered during training.
Models that predict complete measurement distributions typically have
input or output dimensions that grow exponentially with the number of
qubits, while obtaining the corresponding ideal training labels
becomes increasingly difficult as the circuits grow. Consequently,
extending learned mitigation beyond the training regime generally
requires additional assumptions about circuit structure, suitable
model architectures, or new strategies for transferring information
between system sizes.

\begin{figure*}[tbp]
  \centering
  \includegraphics{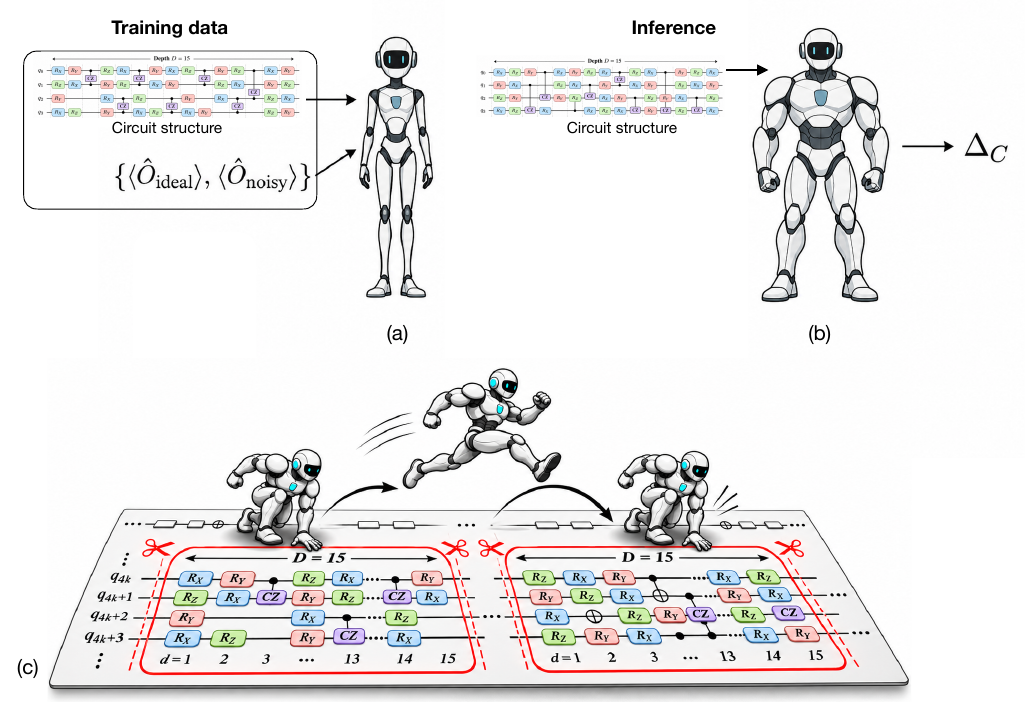}
  \caption{Overview of our proposal. (a) Training: a graph neural network
  is trained on random four-qubit circuits of depths up to $D=15$, using the
  circuit structure together with the corresponding noisy and ideal outputs. Its architecture and training settings are selected by an automated, autoresearch-style \cite{Karpathy2026autoresearch} search.
  (b) Inference: for a circuit within these limits, the trained model reads
  the circuit and its noisy output distribution and predicts the correction
  $\Delta_C$. (c) Extension
  to larger circuits: a wide and deep target circuit is cut (scissors) into
  fragments of at most four qubits and depth 15 (red boxes). The same
  pretrained mitigator is applied to every fragment configuration without
  retraining, and the mitigated fragment statistics are recombined
  classically to estimate the observables of the full circuit.}
  \label{fig:main}
\end{figure*}
Circuit cutting provides a different perspective on this problem.
Rather than treating a large quantum circuit as an indivisible
computational object, circuit-cutting techniques decompose it into
smaller fragments whose measurement statistics can be combined
through classical post-processing to reconstruct observables of the
original circuit. Wire cutting~\cite{Peng/wire_cutting/2020} achieves this decomposition by
replacing quantum communication between fragments with appropriate
state preparations and measurements, while gate
cutting~\cite{Mitarai/gate_cutting/2021} expresses interactions between fragments as combinations of
local operations. These techniques permit circuits
exceeding the width or depth supported by an individual computational
unit to be evaluated through smaller subcircuits, although the
required sampling and reconstruction overhead depends strongly on
the number and structure of the cuts. Circuit cutting has also been
combined with deep-learning-based mitigation, notably to reduce the
cost of generating training data for variational quantum
algorithms~\cite{Cantori/DL_QEM_partial_knitting/2025}.

In this work, we instead employ circuit cutting at the mitigation stage, combining
graph-neural-network-based QEM with wire and gate cutting
(\cref{fig:main}). Our central
objective is to determine whether a mitigation model trained
exclusively on small quantum circuits can be reused to estimate
noise-free observables of substantially larger and deeper circuits,
without retraining or increasing the input dimension of the model.
We first train a graph neural network on approximately $7.2$ million
random four-qubit circuits with depths ranging from $1$ to $15$,
under a fixed noise model comprising depolarizing and
amplitude-damping channels. The network encodes the circuit as a
gate-level graph and combines this representation with its noisy
measurement distribution to predict the correction that maps the noisy
output probabilities onto the ideal ones.
Its architecture and training settings are selected by an automated, autoresearch-style \cite{Karpathy2026autoresearch} search.
To extend the model beyond its
training regime, we partition larger circuits into fragments that satisfy
the four-qubit and depth-$15$ constraints of the training data, apply the
same pretrained
network independently to the noisy output distribution of each
fragment configuration, and reconstruct the global measurement
statistics through classical tensor contraction. The circuit-cutting
identities are exact in the absence of noise; the residual error in
the reconstructed observables arises from imperfect local mitigation
and finite sampling. This approach decouples the size of the target
circuit from the input size of the learned mitigator, but retains the
sampling overhead associated with circuit cutting.

We assess the proposed framework numerically on various circuits, using wire cuts, gate cuts, and combinations of the two. The largest of these, a 28-qubit, depth-103 circuit for the nitrogen molecule, is about seven times wider and seven times deeper than the four-qubit, depth-15 circuits on which the mitigator was trained.
The benchmarks
include clustered random circuits, variational circuit architectures,
molecular Hamiltonians, and quantum many-body dynamics. For the
structured random-circuit families, the learned corrections
substantially reduce the errors in reconstructed expectation values
across the system sizes and depths considered, despite the network
having encountered only four-qubit circuits during training. We
compare the resulting estimates with those obtained using ZNE, CDR,
and PEC applied to the same circuit decompositions. In addition to
comparisons at fixed shots per fragment configuration, we examine
performance under a fixed total sampling budget, accounting for the
different numbers of circuit executions required by the mitigation
methods. These comparisons reveal regimes in which the pretrained
network provides a favorable accuracy--sampling-cost trade-off, as
well as the deterioration that occurs when increasing the number of
fragment configurations leaves too few measurement shots available
for reliable local mitigation.

As a physically motivated application, we consider a holographic
variational quantum eigensolver~\cite{FossFeig2021PRR,Barratt2021npjQI} for a $24$-site transverse-field
Ising chain. A sequential matrix-product-state ansatz, optimized
on a smaller $12$-site system, is transferred to the larger chain,
and its energy is estimated using the same four-qubit mitigation
model in conjunction with circuit cutting. For the three
transverse-field strengths considered, the approach recovers
approximately $88$--$98\%$ of the difference between the noisy and
ideal ansatz energies under a fixed measurement budget, outperforming
CDR and PEC in these examples. We further
examine a $20$-qubit dimerized Ising chain and molecular Hamiltonian
circuits, illustrating both the recovery of physically relevant
observables and the occurrence of systematic biases when fragment
distributions differ substantially from those represented in the
training data.

Our results demonstrate that circuit cutting can provide a route
for extending the applicability of a fixed-size learned mitigator
to larger structured quantum circuits. The approach does not remove
the fundamental sampling limitations of QEM or the overhead
associated with circuit cutting; its effectiveness depends on the
existence of suitable circuit decompositions and on the accuracy
of the local corrections. Nevertheless, by separating the training
problem from the size of the target computation, the framework
offers a modular strategy for reusing learned noise-mitigation
information across different circuit widths, depths, and
applications. More broadly, this work identifies the interplay
between local learning accuracy, circuit structure, and
reconstruction overhead as a central consideration in extending
data-driven quantum error mitigation beyond the systems on which
it was trained.

\section{Preliminaries and Problem Formulation}
\label{sec:preliminaries}

\subsection{Quantum circuits and noisy measurement distributions}
\label{subsec:noisy_circuits}

Consider an $n$-qubit quantum circuit
$U_C = G_L G_{L-1}\cdots G_1$, where $G_m$ ($m=1,\ldots,L$) denotes
a unitary gate and $L$ is the number of gates in the circuit.
For an initial state
$\rho_0 = \ket{0}\bra{0}^{\otimes n}$, the ideal output
state is
\begin{equation}
    \rho_C^{\mathrm{id}}
    = U_C \rho_0 U_C^\dagger.
    \label{eq:ideal_state}
\end{equation}

On a noisy quantum processor, the circuit is instead
described by a quantum channel $\mathcal{E}_C$,
which incorporates the effects of imperfect gate
operations and other relevant noise processes.
The corresponding noisy output state is
\begin{equation}
    \rho_C^{\mathrm{noisy}}
    = \mathcal{E}_C(\rho_0).
    \label{eq:noisy_state}
\end{equation}
For measurements in the computational basis, the
ideal and noisy outcome distributions are
\begin{equation}
    p_C^{a}(x)
    = \bra{x}\rho_C^{a}\ket{x},
    \qquad
    a\in\{\mathrm{id},\mathrm{noisy}\},
    \label{eq:output_distributions}
\end{equation}
where $x\in\{0,1\}^{n}$ is a measurement bitstring.
Both distributions belong to the probability simplex
\begin{equation}
    \triangle_{2^n}
    =
    \left\{
    p\in\mathbb{R}_{\geq 0}^{2^n}:
    \sum_x p(x)=1
    \right\}.
\end{equation}

In an experiment, the noisy probability distribution
is not directly accessible and must be estimated from
a finite number of measurement outcomes. For $N$
measurement shots, the empirical distribution is
\begin{equation}
    \widetilde{p}_C(x)
    =
    \frac{N_x}{N},
    \label{eq:empirical_distribution}
\end{equation}
where $N_x$ denotes the number of occurrences of
bitstring $x$. Consequently, deviations from the
ideal distribution arise from both physical noise
and finite-sampling fluctuations.

\subsection{Expectation values and quantum error mitigation}
\label{subsec:qem_objective}

The objective of quantum error mitigation is to
estimate ideal expectation values using measurement
data obtained from noisy quantum circuits, without
implementing quantum error correction. For an
observable $O$, the desired quantity is
\begin{equation}
    \langle O\rangle_{\mathrm{id}}
    =
    \operatorname{Tr}
    \left[O\rho_C^{\mathrm{id}}\right].
    \label{eq:ideal_expectation}
\end{equation}

For observables diagonal in the computational basis,
$O=\sum_x o_x\ket{x}\bra{x}$, this expectation value
can be evaluated directly from the output distribution:
\begin{equation}
    \langle O\rangle_{\mathrm{id}}
    =
    \sum_x o_x p_C^{\mathrm{id}}(x).
    \label{eq:diagonal_expectation}
\end{equation}

In particular, consider a Pauli-$Z$ string
\begin{equation}
    Z_S = \prod_{j\in S} Z_j,
\end{equation}
where $S\subseteq\{1,\ldots,n\}$ denotes a subset
of qubits. Its expectation value with respect to a
distribution $p$ is
\begin{equation}
    \langle Z_S\rangle_p
    =
    \sum_{x\in\{0,1\}^n}
    (-1)^{\sum_{j\in S}x_j}p(x).
    \label{eq:z_expectation}
\end{equation}
Expectation values of other Pauli strings can
likewise be obtained by applying appropriate
single-qubit basis rotations before computational-basis
measurement. Thus, an estimate of the measurement
distribution in a given basis provides access to
all Pauli-string expectation values diagonal in
that basis.

In learning-based QEM, the goal is to construct
an estimate $\widehat{p}_C$ of the ideal distribution
$p_C^{\mathrm{id}}$ from the noisy measurement data.
The corresponding mitigated expectation value is
\begin{equation}
    \widehat{\langle O\rangle}
    =
    \sum_x o_x\widehat{p}_C(x),
    \label{eq:mitigated_expectation}
\end{equation}
for an observable diagonal in the measurement basis.

Most QEM techniques reduce the bias of the result by combining additional circuit executions with classical
post-processing~\cite{Cai2023RMP}. Commonly used methods include zero-noise extrapolation (ZNE)~\cite{Temme2017PRL,Li2017PRX,Kandala2019Nature,GiurgicaTiron2020QCE,Kim2023Nature, Valle/PIE/2026}; quasiprobabilistic methods such as probabilistic error cancellation (PEC)~\cite{Temme2017PRL,vandenBerg2023NatPhys} or error mitigation by restricted evolution (EMRE)~\cite{Saxena/EMRE/2026}; Clifford data regression (CDR)~\cite{Czarnik2021Quantum}; readout
correction~\cite{Bravyi2021PRA}; symmetry verification~\cite{BonetMonroig2018PRA}; and global depolarizing inversion~\cite{Vovrosh2021PRE}.
A complementary line of work learns the map from noisy to ideal outputs. Examples include neural networks trained on random circuits~\cite{Kim2020IEEEAccess}, learning-based QEM~\cite{Strikis2021PRXQ}, and the comparison by Liao
\etal{}~\cite{Liao2024NMI} of random forests, multilayer perceptrons (MLPs), and graph neural networks (GNNs) with ZNE on IBM hardware.

To quantify the accuracy of mitigation, we consider the mean-squared error over the Pauli-$Z$ strings,
\begin{equation}
\begin{split}
    \mathrm{MSE}_{\mathrm{EV}}
    (\widehat{p}_C)
    =
    \frac{1}{2^n-1}
    \sum_{\substack{S\subseteq[n]\\S\neq\emptyset}}
    \bigl(
    \langle Z_S\rangle_{\widehat{p}_C}
    -
    \langle Z_S\rangle_{\mathrm{id}}
    \bigr)^2 .
\end{split}
\label{eq:ev_mse}
\end{equation}
The corresponding root-mean-square error is
\begin{equation}
    \mathrm{RMSE}_{\mathrm{EV}}
    =
    \sqrt{\mathrm{MSE}_{\mathrm{EV}}}.
    \label{eq:ev_rmse}
\end{equation}
We also define the normalized error ratio
\begin{equation}
    R_{\mathrm{EV}}
    =
    \frac{
    \mathrm{MSE}_{\mathrm{EV}}
    (\widehat{p}_C^{\mathrm{mit}})
    }{
    \mathrm{MSE}_{\mathrm{EV}}
    (\widehat{p}_C^{\mathrm{raw}})
    },
    \label{eq:relative_ev_error}
\end{equation}
where $\widehat{p}_C^{\mathrm{mit}}$ and
$\widehat{p}_C^{\mathrm{raw}}$ denote the mitigated
and unmitigated estimates, respectively.
Hence, $R_{\mathrm{EV}}<1$ indicates an improvement,
while $R_{\mathrm{EV}}>1$ indicates that mitigation
increases the mean-squared error.

\subsection{Extending learned mitigation beyond training circuits}
\label{subsec:scaling_problem}

In learning-based QEM, a parameterized model is
trained to infer corrections to noisy measurement
statistics from examples for which both the noisy
and ideal outputs are available. In the present
work, we consider a mitigation map
\begin{equation}
    \mathcal{M}_{\boldsymbol{\theta}}:
    (C,\widetilde{p}_C)
    \longmapsto
    \widehat{p}_C^{\mathrm{mit}},
    \label{eq:mitigation_map}
\end{equation}
where $\boldsymbol{\theta}$ denotes the learned
parameters.
Error mitigation is thus understood here as the classical post-processing
of noisy measurement data into an estimate closer to the ideal
distribution, without additional qubits or modifications of the circuit.
The details of the model and its training are presented in Sec.~\ref{sec:ml_method}.

A central difficulty arises when the trained model
has a fixed input dimension. In our case, the model
is trained on four-qubit circuits with depths
between $1$ and $15$. Direct application to a
larger circuit is not possible without modifying
the input representation or the model itself.

We therefore consider a target circuit $C$ whose
width and depth may exceed those encountered
during training. Our objective is to construct
a collection of smaller circuit fragments
$\{C_{f,\alpha}\}$ satisfying
\begin{equation}
    n_{f,\alpha}\leq 4,
    \qquad
    D_{f,\alpha}\leq 15,
    \label{eq:fragment_constraints}
\end{equation}
where $n_{f,\alpha}$ and $D_{f,\alpha}$ denote
the width and depth of a fragment configuration,
including any operations introduced by cutting.

The same pretrained mitigation map is applied to
each fragment configuration. The resulting local
estimates are combined through a reconstruction
map $\mathcal{R}_{\mathcal{P}}$, associated with
a circuit-cutting plan $\mathcal{P}$:
\begin{equation}
    \widehat{p}_C^{\mathrm{rec}}
    =
    \mathcal{R}_{\mathcal{P}}
    \left(
    \left\{
    \mathcal{M}_{\boldsymbol{\theta}}
    (C_{f,\alpha},\widetilde{p}_{f,\alpha})
    \right\}_{f,\alpha}
    \right).
    \label{eq:global_reconstruction}
\end{equation}
Here, the reconstruction map includes the
coefficients and index contractions associated
with the chosen wire and gate decompositions.
Its explicit construction is developed in
Sec.~\ref{sec:circuit_cutting}.

The central question investigated in this work
is whether locally learned mitigation corrections
remain sufficiently accurate after reconstruction
to improve the expectation values of larger
circuits. Since circuit cutting introduces
additional measurement configurations and signed
reconstruction coefficients, the overall accuracy
depends on both the residual errors of the local
mitigator and the sampling overhead of the
cutting procedure. Our analysis therefore
considers not only mitigation accuracy but also
its dependence on circuit structure, the number
of fragments, and the available measurement
budget.

\section{Learning Quantum Error Mitigation on Small Circuits}
\label{sec:ml_method}

We first build a learned error mitigator for small circuits. Given a four-qubit circuit and its noisy output
distribution, a graph neural network (GNN) predicts how far the noisy distribution is from the ideal one, so that
this difference can be subtracted. The network sees the circuit as a graph of its gates. Once trained, it is used
unchanged as the local mitigator inside the circuit-cutting framework of Sec.~\ref{sec:circuit_cutting}. We name the trained network used throughout this work the \emph{all-depth model}: it was trained on circuits of every depth from 1
to 15, including both the dense and the sparse circuit families defined in \cref{subsec:training_data}.

\subsection{Training circuits and noise model}
\label{subsec:training_data}

The training circuits act on four qubits and use the gate set
\begin{equation}
    \mathcal{G}
    =
    \{H,S,T,R_X,R_Y,R_Z,\mathrm{CNOT}\},
    \label{eq:training_gateset}
\end{equation}
with rotations $R_P(\phi)=\exp(-i\phi P/2)$ for $P\in\{X,Y,Z\}$. The circuits have depths $D=1,\ldots,15$. To
count the depth, every gate is placed in the earliest layer allowed by the gates that precede it on its wires; the
depth is the number of layers obtained in this way, and the layer index of a gate is denoted by $\ell$.

A circuit is generated layer by layer. In each layer, every qubit is \emph{active} (receives a gate) with probability
$\nu$, except for one carrier qubit that is active in every layer; the carrier guarantees that the circuit has
exactly the requested depth. The active qubits are put in random order. Walking through this list, each qubit
forms a CNOT with the next one with probability $0.4$ (the same value is used throughout this work); otherwise it
receives a single-qubit gate drawn uniformly from $\{H,S,T,R_X,R_Y,R_Z\}$. Rotation angles are drawn uniformly from
$[-\pi,\pi]$.

We use two circuit families. In the \emph{dense} family, $\nu=1$, so every qubit receives a gate in every layer. In
the \emph{sparse} family, $\nu$ is drawn for each circuit from the uniform distribution $\mathcal{U}(0.2,0.95)$, so
some qubits stay idle for several layers. For each depth, the dataset contains $400\,000$ distinct dense circuits and
$200\,000$ distinct sparse circuits.

The noise model is depolarizing noise followed by amplitude damping. After each gate, both channels act on every
qubit that the gate touches. The depolarizing channel is
\begin{equation}
    \mathcal{D}_{p}(\rho)
    =
    (1-p)\rho
    +
    \frac{p}{3}
    \sum_{P\in\{X,Y,Z\}}
    P\rho P,
    \label{eq:ml_depolarizing}
\end{equation}
and the amplitude-damping channel is
\begin{equation}
    \mathcal{A}_{\gamma}(\rho)
    =
    \sum_{k=0}^{1}K_k\rho K_k^\dagger,
    \label{eq:ml_amplitude_damping}
\end{equation}
with Kraus operators
\begin{equation}
\begin{split}
    K_0 &=
    \ket{0}\bra{0}
    +\sqrt{1-\gamma}\ket{1}\bra{1},\\
    K_1 &=\sqrt{\gamma}\ket{0}\bra{1}.
\end{split}
\label{eq:ml_kraus}
\end{equation}
Throughout this work, $p=0.03$ and $\gamma=0.02$, so each gate is followed by
$\mathcal{A}_{\gamma}\circ\mathcal{D}_{p}$ on the qubits it acts on. Idle qubits receive no noise, and measurement
errors are not modeled.

For every training circuit $C$, we compute the ideal distribution $p_C^{\mathrm{id}}$ by statevector simulation and
the noisy distribution $p_C^{\mathrm{noisy}}$ by density-matrix simulation. Both are exact, i.e., free of sampling
noise. The network is trained to predict the difference
\begin{equation}
    \Delta_C
    =
    p_C^{\mathrm{noisy}}
    -
    p_C^{\mathrm{id}},
    \label{eq:ml_target}
\end{equation}
a 16-component vector whose entries sum to zero. Training is therefore a regression problem: from the circuit and
its noisy distribution, predict $\Delta_C$.

\subsection{Graph representation of quantum circuits}
\label{subsec:circuit_graph}

\begin{figure*}[tbp]
  \centering
  \includegraphics{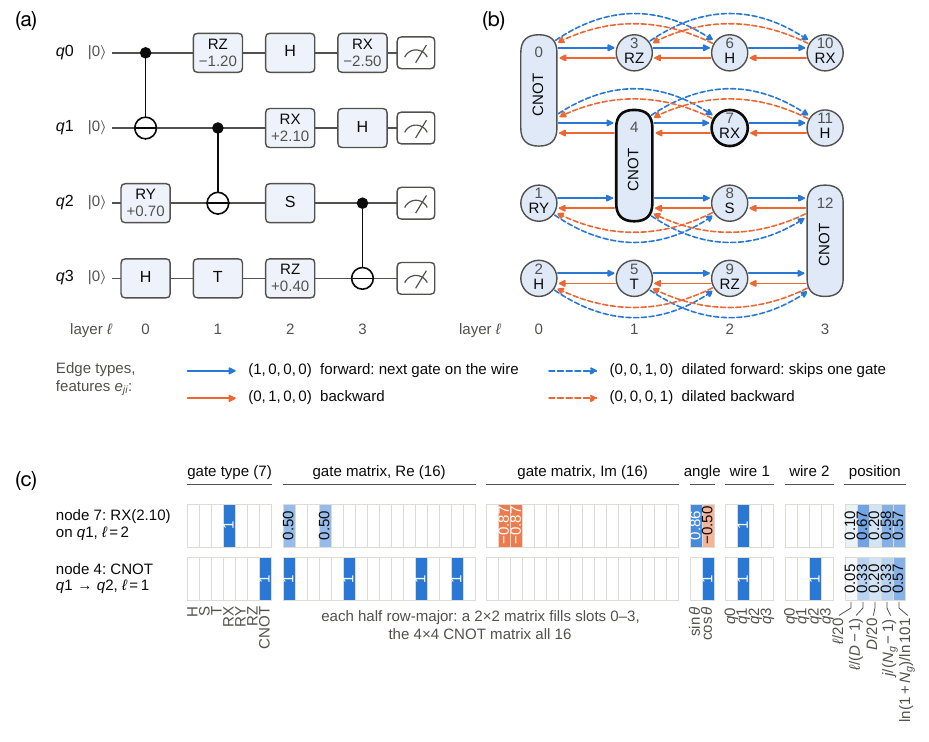}
  \caption{Encoding a circuit as a graph, shown for a four-qubit circuit of depth 4 encoded as for the all-depth model
  (edge dilation $k_{\mathrm{d}}=2$). (a) The circuit. Every qubit starts in $|0\rangle$ and is measured in the $Z$
  basis; rotation angles (in radians) are printed under the gate names, and $\ell$ is the layer index of each gate.
  (b) The graph: one node per gate, numbered $j=0,\dots,12$ layer by layer, with a CNOT as a single node on both of
  its wires. Consecutive gates on a wire are joined by a forward edge (blue, above the wire) and a backward edge
  (orange, below the wire), and each gate is joined in both directions to the gate two positions earlier on its wire
  (dashed arcs). The four edge types are one-hot edge features $\mathbf{e}_{ji}\in\{0,1\}^4$; the example has 13 nodes and 40
  directed edges. (c) The 54 node features of the outlined nodes 7 ($R_X(2.10)$ on $q_1$) and 4 (CNOT with control
  $q_1$ and target $q_2$), grouped as in \cref{subsec:circuit_graph}. Non-zero values are printed in their cells
  (blue positive, orange negative); blank cells are zero. }
  \label{fig:encoding}
\end{figure*}

We represent each circuit as a directed graph $\mathsf{G}_C=(V_C,E_C)$ (\cref{fig:encoding}). Every gate is one
node; a CNOT is a single node that sits on both of its wires. Edges connect gates that follow each other on the same
wire.

Each node $i\in V_C$ carries a 54-dimensional feature vector $\mathbf{x}_i\in\mathbb{R}^{54}$ made of five groups:
\begin{enumerate}
    \item The gate type, as a seven-dimensional one-hot vector.

    \item The gate unitary: its real and imaginary parts, each stored in 16 entries. A CNOT fills all 16 entries
    with its $4\times4$ matrix; a single-qubit gate fills the first four entries with its $2\times2$ matrix and
    leaves the rest zero.

    \item The rotation angle, as the pair $(\sin\phi,\cos\phi)$. Gates without an angle get $(0,1)$, i.e., they are
    encoded as if $\phi=0$.

    \item The wires, as two four-dimensional one-hot vectors. The first marks the qubit of a single-qubit gate or
    the control of a CNOT; the second marks the CNOT target and is zero for single-qubit gates.

    \item Five position features: $\ell/20$ and $\ell/\max(D-1,1)$ for the layer of the gate, $D/20$ for the
    circuit depth $D$, $j/\max(N_g-1,1)$ for the position of the gate in the gate list, and
    $\ln(1+N_g)/\ln 101$ for the number of gates $N_g$.
\end{enumerate}
The unitary and angle features let fixed gates and rotations by any angle share one
feature space, and the position features tell each node where it sits in the circuit and how large the circuit is,
which indicates how much noise has accumulated.

The edges follow the order of the gates along each wire. Let $v_1^{(w)},v_2^{(w)},\ldots$ be the gates acting on
wire $w$, in circuit order. Each gate $v_t^{(w)}$ is connected to the previous gate $v_{t-1}^{(w)}$ by a forward
edge ($v_{t-1}^{(w)}\to v_t^{(w)}$) and a backward edge ($v_t^{(w)}\to v_{t-1}^{(w)}$). In addition, it is connected
in both directions to the gate $k_{\mathrm{d}}$ positions earlier on the same wire, $v_{t-k_{\mathrm{d}}}^{(w)}$; we
call these \emph{dilated} edges and $k_{\mathrm{d}}$ the \emph{edge dilation}. The all-depth model uses
$k_{\mathrm{d}}=2$, so a dilated edge skips over exactly one gate, as in \cref{fig:encoding}(b). The four edge types (forward, backward, dilated forward, and
dilated backward) are stored as one-hot vectors $\mathbf{e}_{ji}\in\{0,1\}^{4}$. The dilated edges let information
travel along a wire in fewer message-passing steps.

\subsection{Graph neural network architecture}
\label{subsec:gnn_architecture}

\begin{figure*}[tbp]
  \centering
  \includegraphics{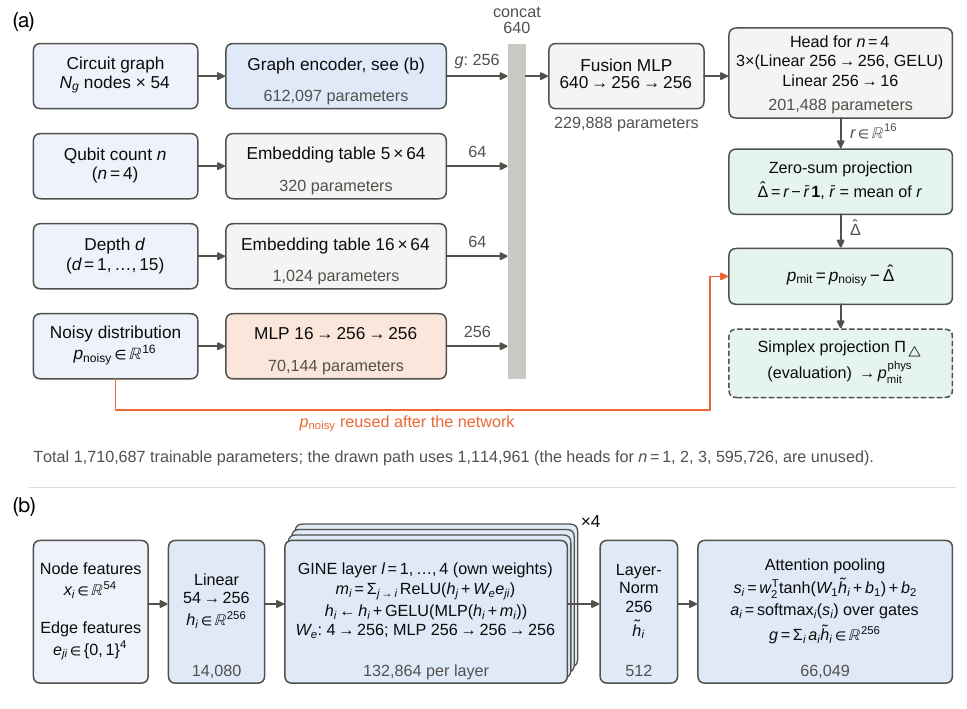}
  \caption{Architecture of the all-depth model, with dimensions and parameter counts (gray) read from its checkpoint.
  (a) Data flow: the graph encoder, learned embeddings of the qubit count $n$ and the depth $d$, an MLP on the noisy
  distribution, a fusion MLP, and the head for $n=4$, whose output $r$ the zero-sum projection turns into
  $\hat\Delta$. The simplex projection $\Pi_\triangle$ has no parameters and is applied only at evaluation. All MLPs
  use GELU, and dropout is zero. (b) Graph encoder: a linear input map, four GINE layers with separate weights and
  fixed $\epsilon=0$, LayerNorm, and attention pooling over the gates. The path drawn holds 1,114,961 of the
  1,710,687 trainable parameters (1,114,641 receive gradients; unused embedding rows hold 320); the other 595,726
  are in the heads for $n=1$, 2, and 3, which never receive a gradient.}
  \label{fig:architecture}
\end{figure*}

The network is a graph isomorphism network with edge features (GINE)~\cite{Xu2019ICLR,Hu2020ICLR}, a
message-passing architecture; its structure is shown in \cref{fig:architecture}. Each node feature vector is first
mapped to a 256-dimensional hidden vector,
\begin{equation}
    \mathbf{h}_i^{(0)}
    =
    W_{\mathrm{in}}\mathbf{x}_i
    +
    \mathbf{b}_{\mathrm{in}}.
    \label{eq:gnn_input}
\end{equation}
Four message-passing layers follow. In layer $\tau$, node $i$ collects the messages
\begin{equation}
    \mathbf{m}_i^{(\tau)}
    =
    \sum_{j:(j,i)\in E_C}
    \operatorname{ReLU}
    \left(
        \mathbf{h}_j^{(\tau)}
        +
        W_e^{(\tau)}\mathbf{e}_{ji}
        +
        \mathbf{b}_e^{(\tau)}
    \right)
    \label{eq:gnn_messages}
\end{equation}
from its neighbors, where $W_e^{(\tau)}$ and $\mathbf{b}_e^{(\tau)}$ are trainable, and updates its hidden vector as
\begin{equation}
    \mathbf{h}_i^{(\tau+1)}
    =
    \mathbf{h}_i^{(\tau)}
    +
    \operatorname{GELU}
    \Bigl[
        \operatorname{MLP}_{\tau}
        \bigl(
            \mathbf{h}_i^{(\tau)}
            +
            \mathbf{m}_i^{(\tau)}
        \bigr)
    \Bigr].
\label{eq:gnn_update}
\end{equation}
Each layer has its own weights, and each $\operatorname{MLP}_{\tau}$ consists of two 256-dimensional linear layers
with a Gaussian Error Linear Unit (GELU) activation in between.

After the last layer, the node vectors are layer-normalized, giving $\widetilde{\mathbf{h}}_i$, and combined into
one vector for the whole circuit by attention pooling. Each node receives a score
\begin{equation}
    s_i
    =
    \mathbf{w}_a^{T}
    \tanh
    \left(
        W_a\widetilde{\mathbf{h}}_i
        +
        \mathbf{b}_a
    \right)
    +b_a,
    \label{eq:gnn_attention_scores}
\end{equation}
the scores are turned into weights
\begin{equation}
    a_i
    =
    \frac{\exp(s_i)}
    {\sum_{j\in V_C}\exp(s_j)},
    \label{eq:gnn_attention}
\end{equation}
and the circuit vector is the weighted sum
\begin{equation}
    \mathbf{g}_C
    =
    \sum_{i\in V_C}
    a_i\widetilde{\mathbf{h}}_i
    \in\mathbb{R}^{256}.
    \label{eq:gnn_graph_embedding}
\end{equation}

The noisy distribution enters through a separate MLP with layer sizes $16\rightarrow256\rightarrow256$,
\begin{equation}
    \mathbf{z}_p
    =
    f_p\left(p_C^{\mathrm{noisy}}\right)
    \in\mathbb{R}^{256}.
    \label{eq:gnn_distribution_embedding}
\end{equation}
The qubit count $n$ and the depth $D$ are represented by learned 64-dimensional embeddings $\mathbf{e}_n$ and
$\mathbf{e}_D$. All four vectors are concatenated,
\begin{equation}
    \mathbf{z}_C
    =
    \left[
        \mathbf{g}_C
        \,\Vert\,
        \mathbf{e}_n
        \,\Vert\,
        \mathbf{e}_D
        \,\Vert\,
        \mathbf{z}_p
    \right]
    \in\mathbb{R}^{640},
    \label{eq:gnn_fusion_input}
\end{equation}
and passed through a fusion MLP to a 256-dimensional vector. An output head (three hidden layers with GELU and a
final linear layer) turns this vector into an unconstrained output $\mathbf{r}_C\in\mathbb{R}^{16}$. Subtracting its
mean gives the predicted correction,
\begin{equation}
    \widehat{\Delta}_C
    =
    \mathbf{r}_C
    -
    \frac{1}{16}
    \left(
        \sum_{x}\mathbf{r}_C(x)
    \right)\mathbf{1},
    \label{eq:gnn_zero_sum}
\end{equation}
which sums to zero, like the true difference $\Delta_C$.

The network has about $1.71\times10^6$ trainable parameters. This count includes output heads for one, two, and
three qubits, which exist in the code but are never trained, because all training circuits have four qubits. The
all-depth model uses no dropout.

\subsection{Training procedure}
\label{subsec:gnn_training}

\subsubsection{Training objective}

For a training circuit $C_i$, let $\boldsymbol{\xi}_i=\widehat{\Delta}_{C_i}-\Delta_{C_i}$ be the error of the
predicted correction. For a batch of $B$ circuits with $n=4$ qubits, the loss is
\begin{equation}
\begin{split}
    \mathcal{L}
    =
    \frac{1}{B}
    \sum_{i=1}^{B}
    \left[
        \frac{
            \|\boldsymbol{\xi}_i\|_2^2
        }{
            \max\{
                \|\Delta_{C_i}\|_2^2,
                10^{-8}
            \}
        }
        +
        0.1\,
        \frac{
            \|\boldsymbol{\xi}_i\|_2^2
        }{2^n}
    \right].
\end{split}
\label{eq:gnn_loss}
\end{equation}
The first term measures the error relative to the size of the true correction, so circuits with small corrections
are not ignored; the second term measures the absolute error. Since both the input and the target are exact, the
training data contain no sampling noise.

\subsubsection{Data split and symmetry augmentation}
\label{subsubsec:symmetry}

The circuits are split into training, validation, and test sets by hashing each circuit, so the assignment is
deterministic and reproducible (details in \cref{app:dataset}). The split is done separately for each depth, so
that at every depth about 80\% of the circuits go to training and 10\% each to validation and test.
This gives $7\,198\,448$ training, $900\,913$ validation, and $900\,639$ test circuits.

We also use a symmetry of the problem. Let $\pi\in S_4$ be a relabeling (permutation) of the four qubits, $\pi C$
the circuit with its qubits relabeled, and $P_\pi$ the corresponding permutation of the 16 measurement outcomes.
Because the noise acts in the same way on every qubit, relabeling the circuit simply permutes its output
distributions,
\begin{equation}
\begin{split}
    p_{\pi C}^{\mathrm{id}}
    &=
    P_\pi p_C^{\mathrm{id}},\\
    p_{\pi C}^{\mathrm{noisy}}
    &=
    P_\pi p_C^{\mathrm{noisy}},
\end{split}
\label{eq:gnn_permutation_symmetry}
\end{equation}
and therefore
\begin{equation}
    \Delta_{\pi C}
    =
    P_\pi\Delta_C.
    \label{eq:gnn_permutation_target}
\end{equation}
Every time a training circuit is drawn, we apply a random relabeling to its graph, its noisy distribution, and its
target. This costs no extra simulation. The augmentation is done on the fly and does not change the stored dataset
of $7\,198\,448$ training circuits. Since there are $4!=24$ relabelings, these circuits provide up to
$24\times7\,198\,448\approx1.7\times10^8$ distinct labeled examples (fewer for circuits that look the same under
some relabelings). In the final 20-hour training run, the network processed about $7.1\times10^8$ samples, so each
training circuit was seen about 99 times on average, each time with a freshly drawn relabeling.

Relabeling creates one risk: for circuits without rotation gates, which are common at small depths, a validation
or test circuit could be a relabeled copy of a training circuit. To prevent this leakage, all relabeled versions of
such a circuit are assigned to the same set (\cref{app:dataset}).

\subsubsection{Optimization and model selection}

The network is trained with the AdamW optimizer (batch size 256), a learning rate that warms up and then decays
along a cosine schedule over a fixed 20-hour budget, and an exponential moving average of the weights, which is
used for evaluation. Deeper circuits are drawn more often: depth $D$ is sampled with probability $D/120$. The
hyperparameters were chosen by a 64-trial search, and the selected configuration was then trained with four random
seeds for 20 hours each on an NVIDIA T4 GPU. The seed with the best validation score became the all-depth model; the
test set was used only after this choice. The full optimization settings, the search, and the ablations are given
in \cref{app:ml_training}.

\subsection{Mitigation with the trained model}
\label{subsec:gnn_inference}

For a four-qubit circuit $C$ with an estimated noisy distribution $\widetilde{p}_C$, the trained network predicts
$\widehat{\Delta}_C$, and the mitigated distribution is
\begin{equation}
    p_C^{\mathrm{mit}}
    =
    \widetilde{p}_C
    -
    \widehat{\Delta}_C.
    \label{eq:gnn_mitigated_distribution}
\end{equation}
No additional circuits, noise amplification, or noise model are needed at this stage. Because
$\widehat{\Delta}_C$ sums to zero, $p_C^{\mathrm{mit}}$ is normalized, but some of its entries can be slightly
negative. We therefore project it onto the probability simplex,
\begin{equation}
    \widehat{p}_C^{\mathrm{mit}}
    =
    \Pi_{\triangle_{16}}
    \left(
        p_C^{\mathrm{mit}}
    \right),
    \qquad
    \Pi_{\triangle_{16}}(p)
    =
    \arg\min_{q\in\triangle_{16}}
    \|q-p\|_2^2,
    \label{eq:gnn_simplex_projection}
\end{equation}
which gives a valid probability distribution from which computational-basis expectation values are computed.

The network only accepts circuits with at most four qubits and depth at most 15, because its input and output have
a fixed size. Circuits on fewer than four qubits fit by adding idle wires. To mitigate larger circuits, we therefore
do not enlarge or retrain the network; instead, the next section cuts a large circuit into fragments that the
network can handle, mitigates each fragment, and recombines the results. The noise model stays fixed throughout, so
the study tests generalization to larger and deeper circuits, not to other noise.

\section{Extending Learned Quantum Error Mitigation
via Circuit Cutting}
\label{sec:circuit_cutting}

The learning-based mitigator developed in Sec.~\ref{sec:ml_method} operates on circuits of at most four qubits and depth 15.
As discussed above, directly applying the
model to larger circuits is not possible because its
input and output distributions have a fixed dimension.
To overcome this restriction, we combine the pretrained
model with circuit cutting, which decomposes a target
circuit into smaller fragments whose measurement
statistics can be reconstructed through classical
post-processing~\cite{Peng/wire_cutting/2020,Mitarai/gate_cutting/2021}.

The central idea is to perform error mitigation locally,
at the level of the circuit fragments, before reconstructing
the global observables. In contrast to approaches that
use circuit cutting to generate training data, our
method employs cutting at the inference stage. The
same pretrained model is applied to every fragment
configuration without modifying its architecture or
updating its parameters.

We employ two complementary decompositions:
\emph{wire cutting}, which replaces quantum
communication between consecutive circuit fragments
with measurements and state preparations, and
\emph{gate cutting}, which decomposes entangling
operations between fragments into combinations of
local operations. Both constructions are exact in
the absence of noise. Their combination allows
larger circuits to be represented by collections
of small, independently executable circuits,
subject to the sampling and reconstruction overhead
associated with the chosen cuts.

\subsection{Overview of the framework}
\label{subsec:cutting_overview}

Consider a target $n$-qubit circuit $C$ whose
width or depth exceeds the limits of the
trained mitigator. We introduce a cut plan
$\mathcal{P}$ that partitions the circuit into
fragments $\{F_j\}_{j=1}^{N_F}$, where $N_F$
denotes the number of fragments.

Each cut introduces a collection of local
operations that replace the original quantum
connection. Consequently, each fragment must
be executed under several different configurations,
which we refer to as \emph{fragment variants}.
We denote the corresponding circuit configurations
by $\{C_{j,\alpha}\}$, where $\alpha$ labels the
preparation, measurement, and gate-decomposition
settings associated with the cuts incident on
fragment $j$.

The fragment variants are constructed so that
\begin{equation}
    n_{j,\alpha}\leq 4,\qquad
    D_{j,\alpha}\leq 15,
    \label{eq:cut_fragment_constraints}
\end{equation}
including any additional operations required
to implement the cuts. Each variant therefore
lies within the input domain of the pretrained
four-qubit mitigator.

Let $p_{j,\alpha}^{\mathrm{id}}$ denote the
ideal output distribution of variant
$C_{j,\alpha}$. For a valid cut plan,
the ideal distribution of the original circuit
can be expressed as
\begin{equation}
    p_C^{\mathrm{id}}
    =
    \mathcal{R}_{\mathcal{P}}
    \left(
        \left\{
        p_{j,\alpha}^{\mathrm{id}}
        \right\}_{j,\alpha}
    \right),
    \label{eq:cut_ideal_reconstruction}
\end{equation}
where $\mathcal{R}_{\mathcal{P}}$ denotes
the classical reconstruction procedure.

The cutting identities are linear in the
underlying quantum operations, whereas
$\mathcal{R}_{\mathcal{P}}$ is generally a
multilinear function of the fragment
statistics. As described below, it can
be evaluated as a contraction of tensors
constructed from the fragment distributions.

In our approach, the ideal fragment
distributions are replaced by their
learned mitigation estimates,
\begin{equation}
    \widehat{p}_{j,\alpha}^{\mathrm{mit}}
    =
    \mathcal{M}_{\boldsymbol{\theta}}
    \left(
        C_{j,\alpha},
        \widetilde{p}_{j,\alpha}
    \right),
    \label{eq:cut_local_mitigation}
\end{equation}
where $\mathcal{M}_{\boldsymbol{\theta}}$
is the fixed pretrained model and
$\widetilde{p}_{j,\alpha}$ is the
empirical noisy distribution obtained
from the corresponding fragment variant.

The reconstructed estimate is then
\begin{equation}
    \widehat{p}_C^{\mathrm{mit}}
    =
    \mathcal{R}_{\mathcal{P}}
    \left(
        \left\{
        \widehat{p}_{j,\alpha}^{\mathrm{mit}}
        \right\}_{j,\alpha}
    \right).
    \label{eq:cut_mitigated_reconstruction}
\end{equation}

This construction decouples the size of the
training circuits from that of the target circuit:
the neural network processes only small fragment
configurations, while the global circuit information
is recovered through classical reconstruction.

\subsection{Wire cutting}
\label{subsec:wire_cutting}

Wire cutting partitions a circuit by
replacing the identity channel connecting
two consecutive operations on a qubit
with a linear combination of measurement
and state-preparation operations. We
follow the Pauli-basis decomposition of
Peng \emph{et al.}~\cite{Peng/wire_cutting/2020}.

For an arbitrary single-qubit state
$\rho$, the identity channel admits
the expansion
\begin{equation}
    \rho
    =
    \frac{1}{2}
    \sum_{P\in\{I,X,Y,Z\}}
    \operatorname{Tr}(P\rho)\,P.
    \label{eq:wire_cut_identity}
\end{equation}

For $P\in\{X,Y,Z\}$, let
$\Pi_{P,\pm}$ denote the projectors
onto the eigenstates with eigenvalues
$\pm1$. Then
\begin{equation}
    P=\Pi_{P,+}-\Pi_{P,-}.
    \label{eq:wire_pauli_projectors}
\end{equation}
The identity operator can similarly
be expressed as
$I=\Pi_{Z,+}+\Pi_{Z,-}$.

Equation~\eqref{eq:wire_cut_identity} combined with Eq.~\eqref{eq:wire_pauli_projectors} therefore expresses quantum information
flow across a cut as a signed combination
of local measurement statistics and
prepared eigenstates.
The fragment preceding the cut measures the outgoing wire in an appropriate Pauli basis,
while the fragment following the cut
is initialized in a corresponding
eigenstate.
A detailed discussion can be found in Ref.~\cite{Peng/wire_cutting/2020}.

For the six-state decomposition, the
downstream fragment requires the
preparation states
\begin{equation}
    \mathcal{S}_6
    =
    \left\{
    \ket{0},\ket{1},
    \ket{+},\ket{-},
    \ket{+i},\ket{-i}
    \right\},
    \label{eq:wire_six_states}
\end{equation}
where
\begin{equation}
\begin{aligned}
    \ket{\pm}
        &=\frac{\ket{0}\pm\ket{1}}{\sqrt{2}},\\
    \ket{\pm i}
        &=\frac{\ket{0}\pm i\ket{1}}{\sqrt{2}}.
\end{aligned}
\end{equation}

The upstream fragment is measured in
one of the $X$, $Y$, or $Z$ bases.
The trace contribution associated
with $I$ can be obtained by summing
the outcomes of a $Z$-basis
measurement, so no additional
measurement basis is required.

All preparation and measurement
operations are implemented using
the gate set of the learned mitigator.
An $X$-basis measurement is realized
by a Hadamard gate followed by a
terminal $Z$ measurement. A
$Y$-basis measurement is implemented
by applying $R_Z(-\pi/2)$ and then
$H$ before measurement.
Similarly, the required initial states
can be prepared from $\ket{0}$
using combinations of $H$, $S$,
and single-qubit rotations. Each
preparation or basis transformation
requires at most two elementary
gates in our implementation.
These additional gates are included
in the fragment circuits and experience
the same noise channels as the
original circuit gates.

A fragment with $a$ incoming
wire-cut connections and $b$
outgoing wire-cut connections
requires, in the six-state scheme, up to
\begin{equation}
    N_{\mathrm{wire}}
    =
    6^a 3^b
    \label{eq:wire_variant_count}
\end{equation}
local preparation and measurement
configurations.

\subsection{Gate cutting}
\label{subsec:gate_cutting}

Wire cutting is particularly useful
when a qubit can be transferred
between successive circuit blocks.
For circuits consisting of spatially
separated blocks connected by
entangling operations, it is often
more convenient to cut the gates
connecting the blocks.

We employ the gate-decomposition
method of Mitarai and
Fujii~\cite{Mitarai/gate_cutting/2021}, which
expresses a two-qubit operation
as a signed combination of
single-qubit operations.

Consider the unitary
\begin{equation}
    U_{ZZ}(\theta)
    =
    \exp(-i\theta Z\otimes Z).
    \label{eq:zz_unitary}
\end{equation}
Its action on an arbitrary
two-qubit state $\rho$ can be
decomposed as
\begin{equation}
\begin{split}
    U_{ZZ}(\theta)\rho
    U_{ZZ}^{\dagger}(\theta)
    ={}&
    \cos^2\theta\,\rho +
    \sin^2\theta\,
    (Z\otimes Z)\rho(Z\otimes Z)\\
    &+
    \frac{\sin(2\theta)}{2}
    \Bigl[
        \mathcal{M}\otimes\mathcal{R}_{+}
        -
        \mathcal{M}\otimes\mathcal{R}_{-}\\
    &\hspace{12mm}
        +
        \mathcal{R}_{+}\otimes\mathcal{M}
        -
        \mathcal{R}_{-}\otimes\mathcal{M}
    \Bigr](\rho),
\end{split}
\label{eq:gate_cut_decomposition}
\end{equation}
where
\begin{equation}
    \mathcal{R}_{\pm}(\sigma)
    =
    R_Z(\pm\pi/2)\,
    \sigma\,
    R_Z^{\dagger}(\pm\pi/2)
    \label{eq:gate_rotation_maps}
\end{equation}
and
\begin{equation}
    \mathcal{M}(\sigma)
    =
    \sum_{m=0}^{1}
    (-1)^m
    \Pi_m\sigma\Pi_m,
    \qquad
    \Pi_m=\ket{m}\bra{m}.
    \label{eq:gate_signed_measurement}
\end{equation}

The map $\mathcal{M}$ represents
a $Z$-basis projective measurement
whose outcomes are weighted by
their eigenvalue signs. It is
therefore a signed linear map,
rather than a trace-preserving
quantum channel.

Equation~\eqref{eq:gate_cut_decomposition}
contains six product terms, each
acting locally on the individual qubits.
A gate connecting two fragments
can consequently be replaced by
local operations, followed by
a classical signed recombination. The coefficient one-norm of this
decomposition is
\begin{equation}
    \gamma_{\mathrm{g}}(\theta)
    =
    1+2|\sin(2\theta)|.
    \label{eq:gate_gamma}
\end{equation}
It approaches unity for weak
interactions and reaches three
when $|\sin(2\theta)|=1$.

The CNOT gate can be treated
within the same framework through
the Clifford equivalence
\begin{equation}
    \mathrm{CNOT}
    =
    (I\otimes H)\,
    \mathrm{CZ}\,
    (I\otimes H),
\end{equation}
where, up to an irrelevant
global phase,
\begin{equation}
    \mathrm{CZ}
    =
    (S\otimes S)
    \exp\left(
        i\frac{\pi}{4}Z\otimes Z
    \right).
    \label{eq:cz_zz_decomposition}
\end{equation}
Thus, CNOT cutting corresponds
to $\theta=-\pi/4$ in
Eq.~\eqref{eq:zz_unitary}, with
\begin{equation}
    \gamma_{\mathrm{g}}^{\mathrm{CNOT}}
    =3.
\end{equation}

The additional local Clifford
operations are absorbed into
the single-qubit operations
implementing each decomposition
term, thereby reducing the
required circuit depth.

An important implementation
constraint arises from the
signed measurement map
$\mathcal{M}$. The pretrained
neural network accepts circuits
with terminal measurements,
but does not support
mid-circuit measurement and reset. We therefore implement the
measurement component by
terminating the measured wire
and recording its $Z$-basis
outcome. The subsequent
post-measurement evolution of
the corresponding logical qubit
is represented using a fresh
wire prepared in either
$\ket{0}$ or $\ket{1}$.
The two branches are evaluated
separately, and their outcome
signs are incorporated during
classical post-processing.

This construction avoids
mid-circuit measurement and
reset operations, but uses
an additional wire within
the fragment. The additional
wire must be counted when
enforcing the four-qubit
constraint. For each incident gate cut,
a fragment requires up to
six local configurations:
$I$, $Z$, $\mathcal{R}_{+}$,
$\mathcal{R}_{-}$, and the
two measurement branches
associated with $\mathcal{M}$.
We verified the gate-cutting
identity numerically against
the exact two-qubit channel,
with discrepancies at the
level of $10^{-16}$.

\subsection{Fragment construction, circuit constraints, and local mitigation}
\label{subsec:fragment_construction}

A cut plan specifies how the
wires and operations of the
original circuit are assigned
to fragments. The construction
must satisfy the input
restrictions of the pretrained
mitigator after accounting
for the additional operations
introduced by cutting.

We impose three constraints:

\emph{(R1) Width constraint.}
Every fragment contains at
most four wires, including
the additional wires required
to implement the measurement
branches of gate cuts.

\emph{(R2) Depth constraint.}
Every executable fragment
variant has at most 15
layers, including all state
preparations and measurement
basis transformations introduced
by wire or gate cutting.

\emph{(R3) Gate-cut continuity.}
For the gate-cut construction
used here, a wire participating
in a cut must have an entangling
operation with another wire of
the same fragment both before
and after the cut. This ensures
that the pre- and post-cut
segments can be represented
within the chosen fragment
layout.

Fragments containing fewer
than four wires are embedded
into the four-qubit input
space by adding idle wires.
This is consistent with the
training dataset, which
includes circuits with idle
wires. The corresponding
idle-qubit outcomes are
marginalized when constructing
the fragment statistics.

The cut locations are chosen
according to the structure
of each circuit family.
We employ three principal
patterns.

First, \emph{wire hand-off}
cuts are used for sequential
or hierarchical circuits,
where a qubit interacts with
one block and subsequently
participates in another.
The connecting wire is cut
between the two blocks.
This arrangement is used in
the clustered random circuits,
staircase ans\"atze,
hierarchical architectures,
and holographic circuits.

Second, \emph{crossing-gate}
cuts are used when two
otherwise local blocks are
connected by a limited number
of entangling operations.
The connecting gates are
decomposed according to
Eq.~\eqref{eq:gate_cut_decomposition}.
This construction is used
for ladder circuits and
weak intercluster interactions.

Third, \emph{temporal cuts}
are introduced for circuits
in which the same qubits
participate throughout a long
evolution. By cutting the
wires at selected time-slice
boundaries, the circuit is
partitioned into fragments
whose depths remain within
the model's training range.
This approach is used for
the molecular and spin-chain
dynamics benchmarks.

The fragment layouts are
constructed so that all
three constraints are
satisfied. The same cut
plan is used for the
learned mitigator and
the comparison methods,
ensuring that differences
between mitigation strategies
are not caused by changes
in the circuit partition.

Once a valid cut plan is
specified, each fragment
is executed under the
preparation, measurement,
and gate-decomposition
configurations required
by its incident cuts. The resulting probabilities
are combined into a \emph{fragment tensor},
written for fragment $j$ as
\begin{equation}
    R_j
    \left(
        \mathbf{P}_{j}^{\mathrm{in}},
        \mathbf{P}_{j}^{\mathrm{out}},
        \mathbf{k}_j;
        \mathbf{y}_j
    \right),
    \label{eq:fragment_tensor}
\end{equation}
where $\mathbf{P}_{j}^{\mathrm{in}}$
and $\mathbf{P}_{j}^{\mathrm{out}}$
label incoming and outgoing
wire-cut Pauli indices,
$\mathbf{k}_j$ specifies
the incident gate-cut terms,
and $\mathbf{y}_j$ contains
the output bits of the
original circuit associated
with that fragment. In practice, every variant
is executed under the specified noise model
and sampled using a finite
number of shots.
The pretrained model then
predicts the corrected
probability distribution
of each variant, and these distributions are
combined to obtain the mitigated global statistics.
The explicit contraction is given in
\cref{app:reconstruction}.

\subsection{Sampling overhead and limitations}
\label{subsec:cutting_overhead}

Circuit cutting introduces
additional circuit
configurations and
signed reconstruction
coefficients. Although
the pretrained model
does not require
additional calibration
circuits or noise
amplification at
inference, its total
execution cost depends
on the number of fragment
configurations.

For a fragment with
$a_j$ incoming wire
cuts, $b_j$ outgoing
wire cuts, and $g_j$
incident gate cuts,
the number of local
configurations in the
six-state scheme is
bounded by
\begin{equation}
    N_j
    \lesssim
    6^{a_j}
    3^{b_j}
    6^{g_j},
    \label{eq:cut_variant_cost}
\end{equation}
before accounting for
possible reuse of
identical configurations.

The corresponding
four-state scheme
replaces the factor
$6^{a_j}$ by $4^{a_j}$.
Thus, the total
execution cost is
controlled not only
by the number of
qubits and layers
in the original circuit,
but also by the
connectivity and
number of cuts
incident on each
fragment.

For the decompositions
used here, a useful
measure of the
quasiprobability
overhead is
\begin{equation}
    \Gamma_{\mathcal{P}}
    =
    \prod_{e\in\mathcal{W}}
    \gamma_{\mathrm{w},e}
    \prod_{g\in\mathcal{G}_{\mathrm{cut}}}
    \gamma_{\mathrm{g},g},
    \label{eq:cut_total_gamma}
\end{equation}
where $\mathcal{W}$
and $\mathcal{G}_{\mathrm{cut}}$
denote the sets of
wire and gate cuts, and
$\gamma_{\mathrm{w},e}$ and $\gamma_{\mathrm{g},g}$
are the one-norms of the individual wire-
and gate-cut decompositions
[Eqs.~\eqref{eq:gate_gamma} and~\eqref{eq:wire_gamma}].

The cutting identities
are evaluated by
explicitly summing
the required terms,
rather than by
Monte Carlo sampling
of the cut indices.
Nevertheless, this
does not eliminate
the statistical cost
of signed reconstruction.
Finite-shot fluctuations
in the fragment estimates
can be amplified by
the reconstruction
coefficients, with
$\Gamma_{\mathcal{P}}$
setting conservative
worst-case bounds on
the error and variance.

Consequently, accurate
local mitigation does
not automatically imply
equally accurate
global reconstruction.
The resulting error
depends on both the
residual bias and
statistical fluctuations
of the local estimates,
as well as the structure
and coefficients of
the cut contraction.
The propagation of residual bias is discussed
briefly in \cref{app:reconstruction}.

To examine the trade-off
between mitigation accuracy
and sampling cost, we consider two measurement
budget conventions in our numerical experiments.
In the first, every executed fragment
configuration receives the same number of
shots; in the second, a total measurement budget
is assigned to the entire target circuit and
divided among all circuit executions required by
the corresponding method (\cref{sec:results}).

The proposed framework
is consequently most
useful when a target
circuit admits a
decomposition into
small fragments with
a manageable number
of configurations and
sufficient measurement
statistics per variant.
Its effectiveness is
not determined by
target-circuit size
alone.

In the following
section, we investigate
these trade-offs
numerically, including
generalization beyond
the training width
and depth, comparisons
with established
mitigation methods,
and the effects of
sampling constraints
on physically motivated
quantum circuits.

\section{Numerical Results and Benchmarks}
\label{sec:results}

We investigate whether the four-qubit mitigator of
Sec.~\ref{sec:ml_method} can be reused on circuits with larger widths,
greater depths, and different physical structures.
Throughout, the pretrained network is kept fixed: no parameters are updated using the target circuits or their fragments.
All results are numerical simulations under the same gate-local noise model used in training, namely depolarizing noise with probability $p=0.03$ followed by amplitude damping with $\gamma=0.02$ on each wire acted on by a gate. Fragment circuits, including the state preparations and basis changes introduced by cutting, are simulated with a density-matrix simulator; finite-shot data are generated by multinomial sampling.
Ideal references are obtained by noiseless simulation of the uncut circuit, except for the 28-qubit chemistry calculations, for which the ideal observables are evaluated within the fixed-electron-pair subspace (\cref{subsec:results_nitrogen}).

We compare the proposed GINE mitigator with ZNE, CDR, and PEC.
In ZNE, selected gates are folded to obtain effective noise-scale factors of one, two, and three.
We use quadratic Richardson extrapolation, and also examine linear extrapolation in the initial benchmarks.
CDR uses $10$--$30$ near-Clifford variants per
fragment configuration, generated by snapping all but two non-Clifford
rotations to multiples of $\pi/2$; a linear map from noisy to ideal
expectation values is then fitted for each Pauli string~\cite{Czarnik2021Quantum}.
PEC uses a quasiprobability decomposition of the inverse of the known
local noise channel, with a one-norm of
$\gamma\simeq1.114$ per noise location. The PEC results are obtained by emulating its signed sampling estimator at the stated shot budget, rather than explicitly executing the sampled inverse-channel circuits~\cite{Temme2017PRL}.
The present comparisons therefore concern the stated numerical
implementations of these protocols, not their globally optimized forms.

In every comparison, the cut plan is held fixed between methods. The
\emph{cut-noisy} reference is obtained by contracting the noisy fragment
statistics without mitigation, and is distinct from noisy execution of
the original uncut circuit because cutting introduces additional noisy
operations. We report the expectation-value root-mean-square error
(EV-RMSE), the normalized mean-squared-error ratio $\REV$ defined in
Sec.~\ref{subsec:qem_objective}, the total-variation (TV) distance for
normalized distributions, and errors in selected physical observables.
The quantity $1-\REV$ measures the fraction of the \emph{mean-squared}
expectation-value error removed, not the fraction of RMSE removed.
For sets of circuits, we distinguish the mean of per-circuit ratios
from the ratio of pooled errors and state which one is reported. For
reconstructed signed vectors, the quantity reported as TV is understood
as the half-$\ell_1$ discrepancy from the ideal distribution; it need
not satisfy all properties of the TV distance between two probability
distributions.

We use two sampling conventions. In the first, each executed fragment
configuration receives a fixed number $S$ of shots, so methods using
additional configurations incur greater total sampling costs. In the
second, a fixed total budget $B$ is divided over \emph{all} configurations
required by a given method, including CDR training copies and ZNE
noise-scaled variants.
For a method requiring
$N_{\mathrm{exec}}$
configurations, uniform
allocation gives
\begin{equation}
    N_{\mathrm{shots}}
    =
    \frac{B}
    {N_{\mathrm{exec}}}
    \label{eq:cut_fixed_budget}
\end{equation}
shots per configuration,
up to integer rounding.
The latter convention is our primary test of
sampling efficiency. The model was trained on exact probabilities, so we
also examine explicitly how its performance changes when the fragment
inputs are estimated from finite shots.

\subsection{Performance of the pretrained four-qubit mitigator}
\label{subsec:local_model_results}

Before investigating mitigation beyond the training circuit
size, we first examine the performance of the pretrained
four-qubit model on an independent test set. This establishes
the accuracy of the local mitigation primitive that is
subsequently employed in the circuit-cutting framework.

\begin{figure}[t]
\centering
\includegraphics[width=\linewidth]{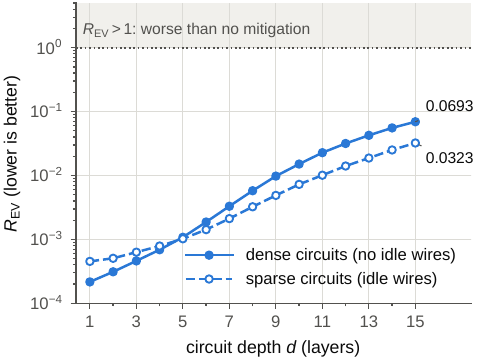}
\caption{$R_{\mathrm{EV}}$ [Eq.~\eqref{eq:relative_ev_error}] of the all-depth model versus circuit depth on the locked test set of four-qubit circuits, with depolarizing ($p=0.03$) and amplitude-damping ($\gamma=0.02$) noise after every gate and exact output probabilities. The solid line with filled circles shows dense circuits
(39,813--40,349 per depth), and the dashed line with open circles shows sparse circuits with idle wires
(19,881--20,140 per depth); the test set contains 900,639 circuits in total. Lower is better; the dotted line marks
$R_{\mathrm{EV}}=1$ (no mitigation), and the shaded region above it is worse than no mitigation.}
\label{fig:local_model_results}
\end{figure}

The model is evaluated on a locked test set containing
$900{,}639$ four-qubit circuits with depths ranging from
$1$ to $15$, including both dense and sparse circuits. The evaluation uses exact noisy output
probabilities under the same depolarizing and
amplitude-damping noise model employed during training,
thereby isolating the accuracy of the learned correction
from finite-shot fluctuations.

Figure~\ref{fig:local_model_results} shows $R_{\mathrm{EV}}$ as a function of circuit depth.
For dense circuits, the error ratio increases from $2.16\times10^{-4}$ at depth $1$ to
$0.0693$ at depth $15$. Thus, even at the maximum
training depth, the model removes approximately $93.1\%$
of the unmitigated expectation-value mean-squared error.
For sparse circuits, the error ratio remains below
$0.0323$ across all depths. Overall, the model improves
the expectation-value accuracy of all but $187$ circuits
in the entire test set.

These results establish the accuracy of the pretrained
local mitigator within its training domain. In the
following subsections, we investigate whether the
same model can be applied without retraining to
fragments of larger circuits, and whether the resulting
local corrections remain effective after global
reconstruction.

\subsection{Generalization beyond training circuit size}
\label{subsec:results_transfer}

We first isolate the effect of fragment-wise mitigation and classical
reconstruction from the effects of finite sampling. The test circuits
are chains of randomly generated, densely entangling blocks, each of
at most four wires, with one wire handed from one block to the next.
Gate types and rotation angles within the blocks follow the same distribution
used for training. The fragments are evaluated using exact noisy
probabilities, which are passed through the pretrained network before
reconstruction. The results for 4--8 qubits and for the extended width-and-depth series are summarized in Table~\ref{tab:results_transfer}.

\begin{table*}[t]
\caption{Generalization using wire cuts and exact noisy fragment probabilities.
$N_{\mathrm{var}}$ is the number of distinct fragment configurations per target
circuit. The entries for $\REV$ are means of per-circuit MSE ratios;
TV denotes the mean distance from the ideal full-circuit distribution.
In this study, the reconstructed full-circuit distribution is projected
onto the probability simplex before $\REV$ and TV are evaluated
(\cref{app:reconstruction}). The two four-qubit control rows use the same
ten circuits, mitigated without and with a wire cut.
Each of the first six rows averages 10 circuits and the remaining rows
average five circuits. No circuit has $\REV>1$ in these sets.
Both cluster and staggered chains are chains of small random blocks (at most four qubits each) linked by one shared, cut wire, so every fragment fits the four-qubit, depth-15 mitigator.
In cluster chains the blocks run at the same time, so the circuit grows in width at depth 15; in staggered chains the blocks are offset in time, so the circuit reaches depths 30--60 while each fragment stays at depth 15.}
\label{tab:results_transfer}
\centering
\footnotesize
\begin{tabular}{lrrrrrrr}
\toprule
Set & Qubits & Depth & Wire cuts & $N_{\mathrm{var}}$ & $\langle\REV\rangle$ & TV (noisy) & TV (GNN) \\
\midrule
Four-qubit control, uncut & 4 & 15 & 0 & 1 & 0.054 & 0.340 & 0.066 \\
Four-qubit control, cut & 4 & 15 & 1 & 9 & 0.048 & 0.342 & 0.061 \\
Cluster chain & 5 & 15 & 1 & 9 & 0.058 & 0.388 & 0.086 \\
Cluster chain & 6 & 15 & 1 & 9 & 0.061 & 0.387 & 0.091 \\
Cluster chain & 7 & 15 & 1 & 9 & 0.089 & 0.416 & 0.115 \\
Cluster chain & 8 & 15 & 2 & 27 & 0.072 & 0.455 & 0.108 \\
Cluster chain & 12 & 15 & 3 & 45 & 0.075 & 0.537 & 0.139 \\
Cluster chain & 16 & 15 & 4 & 63 & 0.108 & 0.607 & 0.196 \\
Staggered chain & 8 & 30 & 2 & 27 & 0.102 & 0.411 & 0.125 \\
Staggered chain & 12 & 45 & 3 & 45 & 0.093 & 0.546 & 0.160 \\
Staggered chain & 12 & 60 & 5 & 81 & 0.146 & 0.560 & 0.175 \\
\bottomrule
\end{tabular}
\end{table*}

For the 4--8-qubit circuits the model leaves approximately $5$--$9\%$
of the original expectation-value MSE, and the larger circuits retain
similar accuracy: $\langle\REV\rangle$ ranges from $0.075$ to $0.146$
for the extended series of 8--16 qubits and depths of 15--60. The four-qubit control is
particularly useful: cutting and mitigating the circuit gives a mean
$\REV=0.048$, compared with $0.054$ when the \emph{same} circuits are
mitigated without cutting. This near agreement demonstrates that the
cut--mitigate--reconstruct sequence need not substantially degrade
accuracy in these examples. It does not imply that cutting is
cost-free: the cut circuit requires nine fragment configurations
instead of one, although on these circuits cutting barely changes the
unmitigated TV distance ($0.340$ uncut versus $0.342$ cut).

The mean reconstructed error ratio is often comparable to the mean
fragment-level error ratio, so the signed reconstruction does not produce
large empirical amplification for these block-structured random circuits.
It can nevertheless propagate an unusually poor local correction: the
largest ratio in the 12-qubit, depth-60 set is $0.348$ ($0.393$ without the
final projection of the reconstructed distribution), which we traced
to a single inaccurately mitigated fragment. This provides
a first indication that the performance of a fixed learned model depends
on the statistics of the fragments produced by the cut plan, not only on
the width and depth of the original circuit.

\subsection{Comparison with established QEM methods}
\label{subsec:results_comparison}

We next examine whether this wire-cut generalization extends to variational quantum circuits \cite{bharti2022noisy}, and compare it with established mitigation techniques.
We consider three variational families: a matrix-product-state (MPS)
staircase, a hierarchical quantum convolutional neural network (QCNN),
and a particle-conserving Givens-rotation staircase (see Fig.~\ref{fig:app_ansatz_circuits}).
These contain 6--12 qubits at depths from 11 to 67. The comparison of our model with ZNE, CDR, and PEC is detailed in \cref{app:subsec:wire_cut_variational}.
The GNN substantially reduces the EV-RMSE for all nine targets.
PEC is nevertheless the most accurate method for eight of the nine circuits (CDR for the remaining one), because the fragments contain few noise locations and its sampling overhead is therefore modest.

\begin{figure*}[t]
\centering
\includegraphics{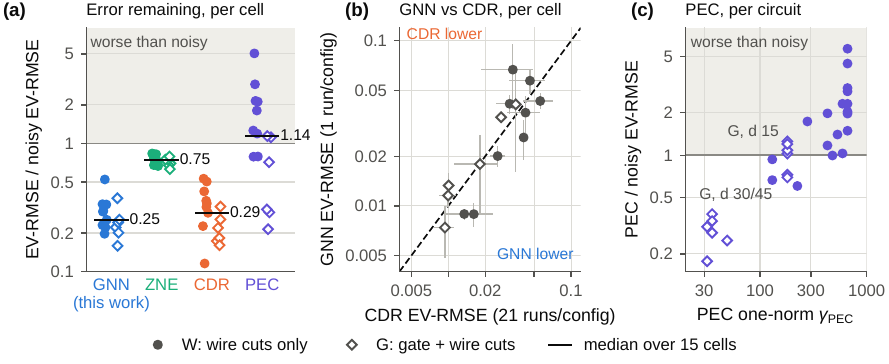}
\caption{Equal-per-configuration-shot comparison for wire-cut random
chains (W, filled circles) and mixed wire/gate-cut ladders (G, open
diamonds) at $10^5$ shots per executed fragment configuration. ZNE
(Richardson) executes three and CDR 21 circuits per fragment
configuration, and the GNN and PEC one. Each of the 15 cells contains
two random target circuits. (a)~EV-RMSE of each method divided by the
cut-noisy EV-RMSE of the same cell (ratio of the cell means); black
bars mark the median over the 15 cells, and values in the shaded region
above 1 are worse than no mitigation. (b)~Cell-mean EV-RMSE of the GNN
versus that of CDR; whiskers give the range over the two circuits on
each axis, and the dashed diagonal marks equal error. (c)~EV-RMSE of
PEC divided by the cut-noisy EV-RMSE for each of the 30 circuits,
against the one-norm $\gamma_{\mathrm{PEC}}$ of the quasiprobability
decomposition of the circuit's largest fragment. Absolute values are
listed in \cref{tab:app_grid_full}.}
\label{fig:results_methods}
\end{figure*}

To isolate gate-cutting effects, we also compare wire-only chains
(scheme W) with ladders having crossing CNOTs that are decomposed by
gate cuts and, at larger depths, additional wire cuts (scheme G). The
comparison comprises 30 circuits in 15 cells, with two circuits per
cell and $10^5$ shots per executed fragment configuration. CDR uses 20
near-Clifford training copies, or 21 executions per target fragment
configuration. The complete comparison is shown in
Fig.~\ref{fig:results_methods} and listed in \cref{tab:app_grid_full}.

Across all 15 comparison cells, the GNN gives the lowest EV-RMSE in
eight and CDR in seven [Fig.~\ref{fig:results_methods}(b)]; in one of
the GNN cells (G, eight qubits, depth 30), the two differ by only about
$1\%$. Relative to the cut-noisy estimate, the median EV-RMSE ratio over
the cells is $0.25$ for the GNN and $0.29$ for CDR
[Fig.~\ref{fig:results_methods}(a)]. The two methods are therefore
generally competitive in accuracy under this \emph{per-configuration}
comparison, but CDR requires 21 times as many circuit configurations as
the GNN. ZNE is less accurate for the investigated gate-local noise and
removes only about a quarter of the cut-noisy error (median ratio
$0.75$). The accuracy of PEC follows its sampling overhead
[Fig.~\ref{fig:results_methods}(c)]. It deteriorates strongly for the
densely populated wire-cut fragments, whose inverse-noise one-norms
range from roughly 130 to 660, and is worse than no mitigation for 14
of the 18 wire-cut circuits. The gate-cut ladders of depth 30 and 45
have one-norms of only 32--49, and there PEC reduces the EV-RMSE to
$0.18$--$0.38$ of the cut-noisy value. At depth 15, however, the ladder
fragments have a one-norm of about 180, and PEC is worse than no
mitigation for four of the six circuits.

Gate cutting therefore broadens the class of circuits accessible to
the trained model, but the additional configurations may be substantial:
a 12-qubit ladder of depth 30 requires 1296 variants, compared with 45
for the 12-qubit wire-only chain of the same depth. Their final errors
are not directly interchangeable because the circuits have different
structures. Rather, the comparison illustrates the tension between
fragment feasibility and the experimental cost of evaluating the
cutting decomposition.

\subsection{Finite-shot behavior and fixed-total-budget scaling}
\label{subsec:results_budget}

\begin{figure*}[t]
\centering
\includegraphics{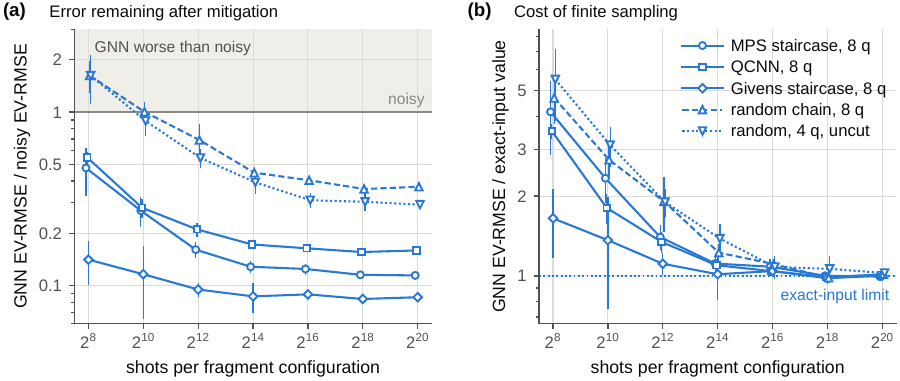}
\caption{Finite-shot sensitivity of fragment-wise GNN mitigation for
five test circuits (marker and line style), averaged over five
multinomial sampling seeds. (a)~GNN EV-RMSE divided by the unmitigated
EV-RMSE obtained from the same sampled inputs at each shot count; values in the shaded
region above 1 mean that the GNN is worse than no mitigation. (b)~GNN
EV-RMSE divided by the GNN EV-RMSE obtained with exact fragment
probabilities; the dotted line marks this infinite-shot limit.
Whiskers give the seed standard deviations, propagated to first order
in (a) with the numerator and denominator treated as independent.
Absolute values are listed in \cref{tab:app_shot_scan}. The trained
network did not encounter shot-noisy inputs.}
\label{fig:results_shots}
\end{figure*}

A model trained on exact output probabilities need not remain effective
when its input is a noisy estimate obtained from a limited number of
shots. We therefore first vary the shots per fragment variant from
$256$ to $1{,}048{,}576$ on an eight-qubit MPS staircase, a QCNN, a
Givens staircase, a random chain, and an uncut four-qubit reference.
Each point averages five independent multinomial sampling seeds.
Figure~\ref{fig:results_shots} compares the mitigated errors with the
exact-probability limits (numerical values in \cref{tab:app_shot_scan}). On the MPS staircase, for example, EV-RMSE
falls from $0.0561\pm0.0174$ at 256 shots to $0.0189\pm0.0020$ at
4096 shots and $0.0146\pm0.0011$ at 65\,536 shots, approaching the
exact-probability value $0.0135$. Across this range, the unmitigated
error is dominated by its systematic noise contribution and changes
little with the number of shots, whereas the learned correction is
noticeably sensitive to input fluctuations. In particular, the model may amplify sampling noise
at very low shot counts: at 256 shots, the GNN error exceeds the
unmitigated error by a factor of $1.6$ for the random eight-qubit chain
and the uncut four-qubit reference [Fig.~\ref{fig:results_shots}(a)].
In the tested circuits, its performance becomes consistently favorable
by roughly 4096 shots per variant and approaches the infinite-shot limit
between approximately $6.5\times10^4$ and $2.6\times10^5$ shots, where
the excess over the exact-input error falls to a few percent
[Fig.~\ref{fig:results_shots}(b)].

The practical importance of this sensitivity is best assessed at a
fixed total execution budget, rather than at a fixed number of
shots for every fragment circuit. We assign $B=10^5$ shots to each
target circuit and distribute them uniformly over all executed
configurations for each method. CDR uses ten near-Clifford training
copies plus the target execution, so it receives a factor of eleven
fewer shots per circuit than the GNN. We consider (i) a width series
from 4 to 24 qubits, (ii) an eight-qubit depth series up to depth 60,
and (iii) ladder circuits containing gate cuts, with as many as 1296
configurations. Each of the 15 cells contains three target circuits;
their errors are listed in \cref{tab:app_budget_full}.

\begin{figure}[t]
\centering
\includegraphics{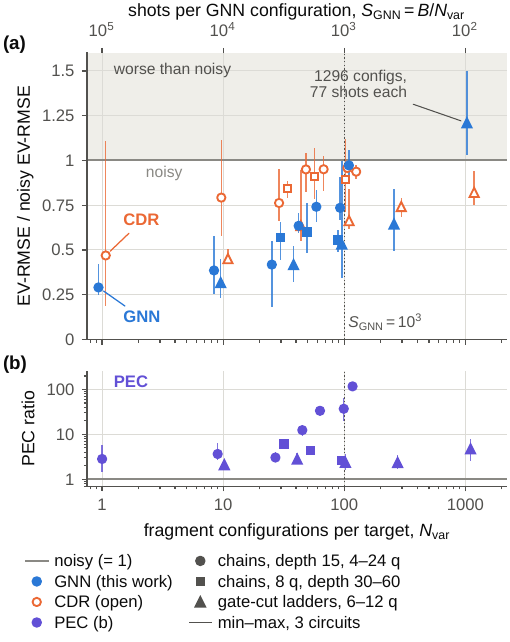}
\caption{Fixed-total-shot-budget comparison with $B=10^5$ shots per
target circuit, distributed uniformly over each method's executed
configurations, against the number of fragment configurations per
target, $N_{\mathrm{var}}$. The top axis gives the resulting shots per
GNN configuration, $S_{\mathrm{GNN}}=B/N_{\mathrm{var}}$, and the dotted
line marks $S_{\mathrm{GNN}}=10^3$. (a)~EV-RMSE of the GNN (filled)
and CDR (open) divided by the cut-noisy EV-RMSE of the same cell (ratio
of the means over its three target circuits); whiskers give the minimum
and maximum of the three per-circuit ratios, and values in the shaded
region above 1 are worse than no mitigation. (b)~The same ratio for
PEC on a logarithmic scale. Marker shape identifies the series:
depth-15 wire-cut chains of 4--24 qubits (circles), eight-qubit
wire-cut chains of depth 30--60 (squares), and gate-cut ladders of
6--12 qubits (triangles). Points are offset slightly along
$N_{\mathrm{var}}$ for legibility. Absolute values are listed in
\cref{tab:app_budget_full}.}
\label{fig:results_fixed_budget}
\end{figure}

Figure~\ref{fig:results_fixed_budget} and \cref{tab:app_budget_full} show
that the GNN gives the lowest EV-RMSE for most cells until the number of
fragment configurations becomes too large. Note that the RMSE contains both bias and variance contributions, so a large variance raises the RMSE even when the bias is small.
The biases of the different QEM methods for these circuits are compared in \cref{app:subsec:fixed_budget_errors}.
As the number of configurations grows and the shots per configuration
fall, the GNN error relative to no mitigation rises
[Fig.~\ref{fig:results_fixed_budget}(a)]. Most cells have more than
$10^3$ shots per GNN configuration, and there the GNN removes
$26$--$71\%$ of the cut-noisy error. Below this level, its advantage
narrows: it still removes about a third of the error at 308 shots per
configuration (eight-qubit, depth-30 ladder), but at 77 shots (1296
configurations) it is worse than no mitigation. Under this protocol,
PEC is worse than no mitigation in every cell
[Fig.~\ref{fig:results_fixed_budget}(b)].
This comparison shows that the learned procedure has a clear
finite-budget failure regime: its applicability cannot be inferred from
the target width or depth alone.
In the investigated fixed-budget examples, the GNN remains useful down
to a few hundred shots per fragment configuration, with better
performance at higher shot counts. The corresponding limit of a few
hundred configurations at $B=10^5$ is an
\emph{empirical budget-dependent threshold}, not a universal scaling
law. It depends on the number of cuts, the structure and noise level of
the fragments, the model's sensitivity to multinomial fluctuations, and
the reconstruction coefficients. More generally, the training cost of
the GNN is paid once, whereas the per-target execution cost of the
method remains governed by circuit-cutting overhead.

\subsection{Application to quantum many-body systems}
\label{subsec:results_many_body}

We turn to a physically motivated example in which the target observable
is an energy rather than a distribution-level metric.
We consider two cases. The first is a holographic variational quantum eigensolver (VQE)~\cite{FossFeig2021PRR,Barratt2021npjQI} for an Ising chain, in which the trial ground state is prepared by a four-qubit block applied sequentially along the chain, with each block handing one qubit to the next. This example is a genuine many-body VQE whose cut decomposition fits within a budget of $2\times10^5$ shots per measurement basis.
The second case, the real-time dynamics of a 20-qubit dimerized Ising chain, is presented in \cref{app:subsec:results_quench}.

For the holographic VQE, we consider the
24-site transverse-field Ising Hamiltonian with open boundaries,
\begin{equation}
H=-J\sum_{i=1}^{N-1}Z_iZ_{i+1}-h\sum_{i=1}^{N}X_i,
\qquad N=24,\label{eq:results_tfim}
\end{equation}
for $h/J=0.5,1,1.5$. We use a sequential matrix-product-state (MPS)
ansatz of bond dimension two, implemented as eight overlapping blocks
of at most four qubits. Each block uses a 14-layer circuit with shared
parameters; a single qubit is passed to the subsequent block through
a wire cut. The full circuit has depth 77 (78 for the $X$-basis
measurement configuration), seven wire cuts, eight fragments, and 117
fragment configurations per measurement basis.

The 32 shared circuit parameters are optimized using ideal simulation
on a 12-site Ising chain and then transferred without reoptimization
to 24 sites. We evaluate energy terms from separate $Z$- and $X$-basis
fragment runs, each allotted a total of $2\times10^5$ shots. This gives
approximately 1709 shots per variant for the GNN, compared with
approximately 155 shots for CDR with ten near-Clifford copies. The
exact ground-state energy of the 24-site system is obtained from the
free-fermion solution; the appropriate mitigation reference is the
\emph{noise-free energy of the transferred ansatz}, so that ansatz
error and mitigation error are kept distinct.

\begin{figure}[t]
\centering
\includegraphics{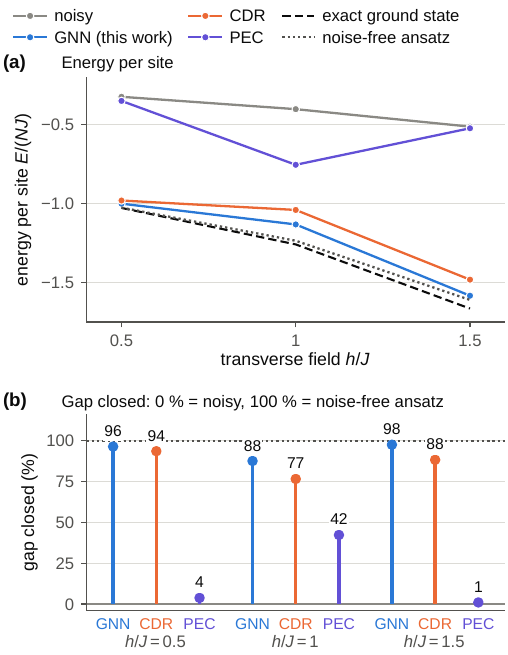}
\caption{24-site holographic Ising VQE at $h/J=0.5$, $1$, and $1.5$,
with a fixed total budget of $2\times10^5$ shots per measurement basis.
(a)~Energy per site, $E/(NJ)$, of the cut-noisy execution and of the
GNN, CDR, and PEC estimates, together with the exact ground-state
energy (dashed) and the noise-free energy of the transferred ansatz
(dotted). (b)~Share of the gap between the cut-noisy and the
noise-free ansatz energy closed by each method,
$100\%\times(E_{\mathrm{noisy}}-E_{\mathrm{method}})/(E_{\mathrm{noisy}}-E_{\mathrm{ansatz}})$;
computed from the energies in \cref{tab:results_holographic}. ZNE was
not run for this experiment.}
\label{fig:results_holo}
\end{figure}

As shown in Fig.~\ref{fig:results_holo} (energies in \cref{tab:results_holographic}), the cut-noisy energy lies
far above the noise-free ansatz energy for all three fields. The GNN
recovers $96\%$, $88\%$, and $98\%$ of this gap, compared with $94\%$,
$77\%$, and $88\%$ for CDR [Fig.~\ref{fig:results_holo}(b)]. PEC
performs less favorably at the available per-configuration shot count,
closing only $4\%$, $42\%$, and $1\%$ of the gap. In the ordered regime $h/J=0.5$, the
noise-free ansatz energy lies only $0.002J$ per site above the exact
ground-state energy. At the critical and paramagnetic points, variational gaps of $0.023J$ and $0.055J$ per site, respectively, remain even for
ideal execution, as expected for a bond-dimension-two ansatz. The results therefore support a concrete
application of the fixed small-circuit mitigator to a 24-qubit
energy-estimation task.
The corresponding one- and two-body expectation-value
biases are shown in \cref{fig:app_holo_bias}.

\subsection{Application to quantum chemistry examples}

We also evaluate our method on quantum-chemistry examples: molecular observables evaluated on clustered hardware-efficient circuits, and number-conserving circuits.
To probe the limits of the mitigator, we further estimate the energy of the nitrogen molecule, which requires a 28-qubit circuit of depth 103, under a fixed shot budget.
The first two cases are presented below, and the nitrogen calculation in \cref{subsec:results_nitrogen}.

\subsubsection{Molecular observables and state-dependent bias}
\label{subsec:results_molecular_diagnostic}

\begin{figure*}[t]
\centering
\includegraphics{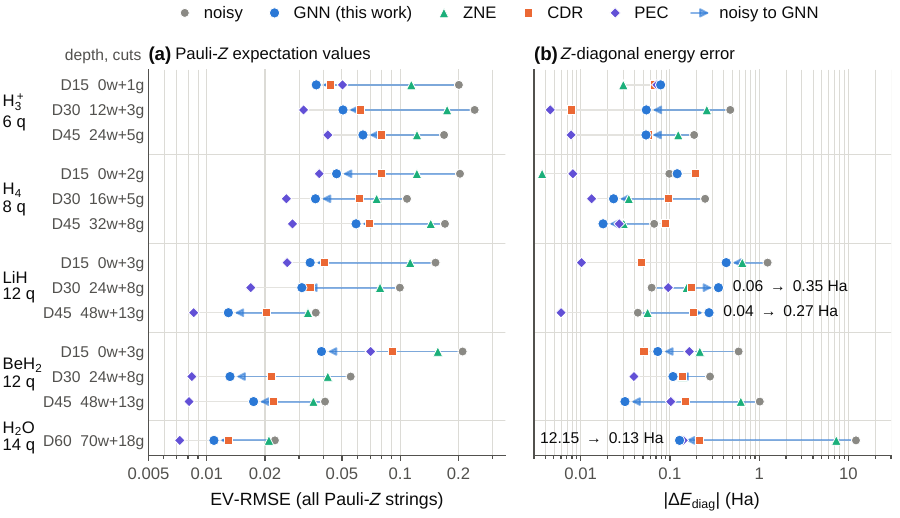}
\caption{Hardware-efficient molecular circuits. (a)~EV-RMSE over all
$2^n-1$ Pauli-$Z$ strings and (b)~absolute error of the $Z$-diagonal
Hamiltonian component, $|\Delta E_{\mathrm{diag}}|$ (Ha); both axes are
logarithmic. Rows give the molecule, the number of qubits (q), the
circuit depth (D), and the numbers of wire (w) and gate (g) cuts. All
methods act on the same fragments with $10^5$ shots per fragment
circuit; ZNE uses Richardson extrapolation. Arrows run from the
cut-noisy value to the GNN value and are omitted where the two differ
by less than $0.12$ decades. Each value is a single realization, so no
error bars are shown. $E_{\mathrm{diag}}$ is not a ground-state energy.}
\label{fig:results_hea}
\end{figure*}

Before turning to number-conserving chemistry ans\"atze, we examine molecular observables evaluated on clustered hardware-efficient circuits with random parameters.
The circuits act on 6--14 qubits at depths from 15 to 60, and contain spatial gate cuts between orbital clusters and temporal wire cuts.
The molecules are $\mathrm{H}_3^+$, $\mathrm{H}_4$, LiH, $\mathrm{BeH}_2$, and $\mathrm{H}_2\mathrm{O}$.
The ansatz includes single-qubit $R_Y$ and $R_Z$ rotations and does
\emph{not} conserve particle number. Only the computational-basis
($Z$-diagonal) component $E_{\mathrm{diag}}$ of the molecular Hamiltonian
is evaluated. These are therefore tests of mitigating molecularly
weighted observables on specified states, not estimates of the
physical molecules' ground-state energies. The results are shown in
\cref{fig:results_hea}.

Across the investigated circuits, the GNN consistently reduces the
unweighted EV-RMSE [Fig.~\ref{fig:results_hea}(a)], although PEC is
often more accurate on the sparser fragment configurations. The weighted diagonal energy exposes
a limitation that the distribution-level score alone does not reveal.
For LiH at depth 30, the absolute error in $E_{\mathrm{diag}}$ increases
from $0.06$ Ha before mitigation to $0.35$ Ha after GNN mitigation
[Fig.~\ref{fig:results_hea}(b)];
at depth 45 it increases from $0.04$ to $0.27$ Ha. In contrast, for
the 14-qubit, depth-60 water circuit, containing 70 wire cuts, 18 gate
cuts, and 49,617 fragment configurations, the diagonal-energy error decreases from $12.15$ Ha to $0.13$ Ha.
These observations show that a lower mean-squared error over Pauli strings does not guarantee a lower error for a Hamiltonian with highly nonuniform coefficients.
This motivates the more controlled pair-conserving chemistry benchmarks below, where both the full projected Hamiltonian and the ansatz reference energy are available.

\subsubsection{Number-conserving chemistry benchmarks}
\label{subsec:results_pair_chemistry}

We examine quantum-chemical energy estimation in the seniority-zero,
or electron-pair, subspace. Each spatial orbital is represented by one
qubit, with $\ket{1}$ corresponding to a doubly occupied orbital.
The \emph{complete Hamiltonian projected into this subspace} contains
occupation, density-density, and pair-hopping terms; its expectation
is reconstructed from $Z$-, $X$-, and $Y$-basis measurements. The
minimum energy in the pair subspace is the doubly occupied
configuration-interaction (DOCI) energy. Because the ans\"atze are
restricted to this subspace, their noise-free energies and the DOCI
reference must be distinguished from the full-configuration-interaction
(FCI) ground-state energy.

\begin{figure}[t]
\centering
\includegraphics{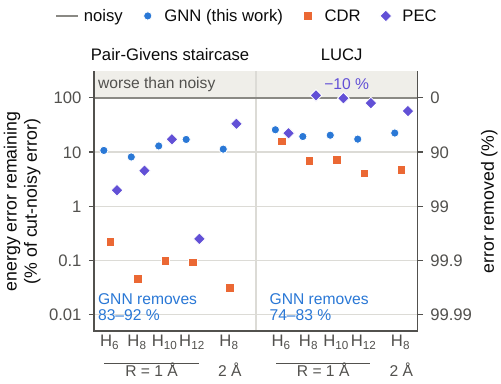}
\caption{Energy error remaining after mitigation for the pair-Givens
staircase (left) and the LUCJ ansatz (right) on
$\mathrm{H}_n$ chains ($n$ qubits) at $R=1$\,\AA{} and on
$\mathrm{H}_8$ at $R=2$\,\AA, as a percentage of the cut-noisy energy
error,
$100\,|E_{\mathrm{m}}-E_{\mathrm{ansatz}}^{\mathrm{id}}|/|E_{\mathrm{noisy}}-E_{\mathrm{ansatz}}^{\mathrm{id}}|$,
on a logarithmic scale; the right axis gives the same positions as the
percentage of the error removed. The gray line at $100\%$ is the
cut-noisy estimate, and the shaded band above it is worse than no
mitigation (PEC, LUCJ $\mathrm{H}_8$ at $1$\,\AA: $-10\%$
removed). The mitigated energies are compared with the noise-free ansatz
energies, not the DOCI or FCI ground-state energies. Single runs with
$2\times10^5$ shots per measurement basis; the signed errors are listed
in \cref{tab:app_pair_energies}.}
\label{fig:results_energy_removed}
\end{figure}

We consider linear hydrogen chains $\mathrm{H}_6$, $\mathrm{H}_8$,
$\mathrm{H}_{10}$, and $\mathrm{H}_{12}$ at interatomic spacing
$R=1\,\text{\AA}$, together with a stretched $\mathrm{H}_8$ chain at
$R=2\,\text{\AA}$, in the STO-3G basis. Each measurement basis has a
fixed total budget of $2\times10^5$ shots, divided across the
fragment variants. CDR uses ten near-Clifford training copies per
variant. We compare two number-conserving circuit constructions with
different local statistics and degrees of non-Cliffordness.
The results are summarized in \cref{fig:results_energy_removed}; the
signed energy errors and the reference energies are listed in
\cref{tab:app_pair_energies}. For completeness, the Pauli-expectation errors in all three
measurement bases for both ans\"atze are given in \cref{app:subsec:bias_plots}.

\paragraph{Pair-Givens staircase.}

The first ansatz applies one pair-conserving Givens rotation to each
adjacent orbital pair, starting from the Hartree--Fock configuration.
The rotations use four CNOT gates per bond, and wire hand-offs produce
fragments of two or three qubits. The five chain calculations require
45--153 fragment variants per measurement basis. The resulting states
are shallow within each fragment and close to computational-basis
states, which makes them favorable for near-Clifford regression.

\paragraph{LUCJ ansatz.}

The second ansatz is a windowed version of the local unitary cluster
Jastrow (LUCJ) construction~\cite{Motta/LUCJ/2023}. Its local sequence has
the schematic form $e^{-K}e^{iJ}e^{K}$ in the electron-pair subspace,
with four-orbital windows sharing one carried orbital. Each complete
window contains twelve non-Clifford rotations and occupies 13--15
layers, producing denser fragments than the pair-Givens staircase.
The five chain calculations require 9--45 fragment variants per
measurement basis. This comparison tests whether a change in the
Clifford content of the fragments is sufficient to change the
relative performance of the mitigators.

\Cref{fig:results_energy_removed} shows that the GNN substantially
reduces the energy error in both ansatz families, but CDR is consistently more accurate, for reasons discussed below.
On the pair-Givens staircase, the GNN removes $83$--$92\%$
of the initial error, while CDR leaves only approximately
$0.3$--$5$ mHa, less than $0.3\%$ of the cut-noisy error and about two
decades below the GNN residual (\cref{fig:results_energy_removed}; the
dependence on chain length is shown in \cref{fig:app_givens_energy}). At $R=1\,\text{\AA}$, the GNN residual grows with chain length, from $0.24$ Ha for $\mathrm{H}_6$ to $0.80$ Ha for $\mathrm{H}_{12}$, whereas the CDR residual stays between $1.4$ and $4.9$ mHa. This unusually strong performance of CDR is explained in
part by the near-Clifford construction: several of its training copies
coincide with or closely resemble the small target fragments. The
comparison therefore does not establish a universal advantage of CDR,
but it does identify a circuit family on which the chosen GNN is
clearly suboptimal.

The LUCJ circuits contain denser, more non-Clifford fragments,
so CDR no longer obtains nearly exact local corrections. Nevertheless,
it retains the lowest absolute energy error in every chain. The GNN
removes $74$--$83\%$ of the cut-noisy energy error but leaves
$0.22$--$0.89$ Ha, while CDR leaves $0.045$--$0.407$ Ha ($4$--$16\%$ of
the cut-noisy error), so the gap between the two methods narrows. PEC suffers
large finite-shot errors for some LUCJ fragments, whose inverse-noise
one-norms are approximately $180$--$385$. No method reaches chemical
accuracy in this LUCJ benchmark.

The residual learned-mitigation error is concentrated in the $Z$-basis
occupation statistics. In these states, many ideal one-body expectation
values are close to $\pm1$; with strong gate noise the required
corrections are large and differ from those typical of random
four-qubit training circuits. Exact-probability calculations and an
increase of the measurement budget from $2\times10^5$ to
$5\times10^5$ shots per basis (\cref{app:subsec:bias_plots}) confirm that the remaining GNN energy
error is dominated by systematic bias rather than sampling
fluctuations. In contrast, the $X$- and $Y$-basis contributions are
comparatively small. These results identify output-state statistics,
and not only the number of non-Clifford gates, as an important
determinant of model transferability.

\subsection{Summary of numerical findings}
\label{subsec:results_summary}

The simulations establish two complementary conclusions. First, a
fixed four-qubit, depth-15 mitigator can be applied without retraining
to structured circuits reaching 28 qubits and depth 103 by combining local predictions with wire/gate-cut reconstruction.
Second, the
advantage of a pretrained model depends critically on the total sampling budget. The GNN avoids per-target regression training and
noise-scaled executions, but it still pays the fragment-execution and
signed-reconstruction costs of circuit cutting, and its accuracy
degrades when the number of shots per fragment configuration becomes too small.
Conversely, with a larger shot budget, the same small mitigator can be applied to even larger circuits.
These competing considerations motivate the discussion
of applicability and limitations in the following section.

\section{Discussion and Conclusion}
\label{sec:discussion}

In this work, we have investigated whether a learned quantum error mitigator trained exclusively on small circuits can be reused for substantially larger quantum computations.
Our results show that circuit cutting provides such a route: by decomposing
larger circuits into smaller fragments, a single GNN trained
only on four-qubit circuits of depths $1$--$15$ can be
reused, without retraining, to estimate observables of
structured circuits with up to 28 qubits and depth 103, about $7$ times larger than those used to arrive at the trained model.
The central advantage of this approach is that it decouples
the training problem from the size of the target
computation: the learned mitigator acts locally,
while global observables are recovered through classical
reconstruction.
Comparisons with established mitigation techniques
show that the learned model can offer favorable
accuracy--sampling-cost trade-offs, particularly when
its one-time training cost is amortized over many target
circuits. The successful mitigation of clustered
random circuits, variational architectures, and the
24-site holographic VQE illustrates the potential of
this modular approach.
At the same time, the results identify limitations
arising from circuit-cutting overhead, finite-shot
fluctuations, and systematic errors in physically
important observables. More broadly, they show that the
success of learning-based QEM beyond its training system size depends
jointly on the transferability of local corrections, the structure
and cost of the circuit decomposition, and the observables being
estimated.

An important distinction must be drawn, however,
between extending the applicability of a fixed-size model
and achieving resource-efficient scalability. Although
circuit cutting removes the input-size restriction of the
GNN, it introduces additional fragment configurations
and a sampling overhead that depends on the number and
structure of the cuts. Our fixed-budget experiments
highlight this trade-off: the GNN remains effective for
many larger circuits, but its advantage diminishes when
the measurement budget is distributed over too many
fragment configurations. Since the model was trained
using exact noisy probabilities, it is particularly
sensitive to finite-shot fluctuations at low shot counts.
Thus, the practical applicability of the framework
depends not only on the accuracy of local mitigation
but also on finding circuit decompositions with
manageable reconstruction costs.

The results further reveal that a small average error
over many expectation values does not guarantee
accurate estimates of every physical observable.
While the GNN performs competitively with established
methods on several benchmark families, CDR is more
accurate for the molecular-energy calculations.
In these examples, the GNN lowers the EV-RMSE, an
unweighted average over many Pauli strings, but leaves
systematic errors in the occupations of the occupied orbitals.
Because the molecular Hamiltonians weight these
occupations with large coefficients, these residual
errors dominate the energy error. They persist when
statistical fluctuations are reduced and arise for
fragments whose output distributions, with many one-body
expectation values close to $\pm1$, differ from those of
the random training circuits. The limitation is
therefore that the local mitigation is accurate on
average but not uniformly across observables, and a
distribution-level score does not reveal where it falls
short. This suggests that incorporating physically motivated
training circuits, relevant symmetries, and
observable-dependent objectives could improve the
transferability of learned mitigation.

Finally, our numerical demonstrations assume a fixed,
gate-local depolarizing and amplitude-damping noise
model. Generalization across circuit sizes should
therefore be distinguished from generalization across
noise environments. The separate hardware test in
\cref{sec:grover}, which uses the proposed network trained on 
device data.
Natural next steps include device-specific training,
training with finite-shot inputs to improve robustness
to sampling noise, and improved fragment selection and
shot allocation. Together with validation on quantum
hardware, these developments would help establish the
practical reach of the approach.

\FloatBarrier

\section*{Acknowledgments}
The authors thank Jack S. Baker for fruitful discussions. The authors thank Sean Kim, Jacob Song, Jay Kim, and Paria Nejat of LG Electronics for their continuous support. P.D.-V. acknowledges the support provided by project ARQADE (CER-20251019), funded by the CERVERA Research Programme of CDTI (Centre for Technological Development and Innovation).

\textbf{AI disclosure statement.} The ideation was done by the authors. Claude Code (Fable 5.1, Opus 5.5) and ChatGPT-6.1 Sol were used to polish the manuscript and write codes with which we train our deep learning models on the AWS EC2 instances. The authors take full responsibility for the contents of this article.

\section*{Competing interests}

The authors declare no competing interests.

\section*{Data availability statement}
The data that support the findings of this study are available from the corresponding
author upon reasonable request.

\bibliographystyle{apsrev4-2-titles}
\bibliography{refs}

\begin{thebibliography}{46}%
\makeatletter
\providecommand \@ifxundefined [1]{%
 \@ifx{#1\undefined}
}%
\providecommand \@ifnum [1]{%
 \ifnum #1\expandafter \@firstoftwo
 \else \expandafter \@secondoftwo
 \fi
}%
\providecommand \@ifx [1]{%
 \ifx #1\expandafter \@firstoftwo
 \else \expandafter \@secondoftwo
 \fi
}%
\providecommand \natexlab [1]{#1}%
\providecommand \enquote  [1]{``#1''}%
\providecommand \bibnamefont  [1]{#1}%
\providecommand \bibfnamefont [1]{#1}%
\providecommand \citenamefont [1]{#1}%
\providecommand \href@noop [0]{\@secondoftwo}%
\providecommand \href [0]{\begingroup \@sanitize@url \@href}%
\providecommand \@href[1]{\@@startlink{#1}\@@href}%
\providecommand \@@href[1]{\endgroup#1\@@endlink}%
\providecommand \@sanitize@url [0]{\catcode `\\12\catcode `\$12\catcode `\&12\catcode `\#12\catcode `\^12\catcode `\_12\catcode `\%12\relax}%
\providecommand \@@startlink[1]{}%
\providecommand \@@endlink[0]{}%
\providecommand \url  [0]{\begingroup\@sanitize@url \@url }%
\providecommand \@url [1]{\endgroup\@href {#1}{\urlprefix }}%
\providecommand \urlprefix  [0]{URL }%
\providecommand \Eprint [0]{\href }%
\providecommand \doibase [0]{https://doi.org/}%
\providecommand \selectlanguage [0]{\@gobble}%
\providecommand \bibinfo  [0]{\@secondoftwo}%
\providecommand \bibfield  [0]{\@secondoftwo}%
\providecommand \translation [1]{[#1]}%
\providecommand \BibitemOpen [0]{}%
\providecommand \bibitemStop [0]{}%
\providecommand \bibitemNoStop [0]{.\EOS\space}%
\providecommand \EOS [0]{\spacefactor3000\relax}%
\providecommand \BibitemShut  [1]{\csname bibitem#1\endcsname}%
\let\auto@bib@innerbib\@empty
\bibitem [{\citenamefont {Cai}\ \emph {et~al.}(2023)\citenamefont {Cai}, \citenamefont {Babbush}, \citenamefont {Benjamin}, \citenamefont {Endo}, \citenamefont {Huggins}, \citenamefont {Li}, \citenamefont {McClean},\ and\ \citenamefont {O'Brien}}]{Cai2023RMP}%
  \BibitemOpen
  \bibfield  {author} {\bibinfo {author} {\bibfnamefont {Z.}~\bibnamefont {Cai}}, \bibinfo {author} {\bibfnamefont {R.}~\bibnamefont {Babbush}}, \bibinfo {author} {\bibfnamefont {S.~C.}\ \bibnamefont {Benjamin}}, \bibinfo {author} {\bibfnamefont {S.}~\bibnamefont {Endo}}, \bibinfo {author} {\bibfnamefont {W.~J.}\ \bibnamefont {Huggins}}, \bibinfo {author} {\bibfnamefont {Y.}~\bibnamefont {Li}}, \bibinfo {author} {\bibfnamefont {J.~R.}\ \bibnamefont {McClean}},\ and\ \bibinfo {author} {\bibfnamefont {T.~E.}\ \bibnamefont {O'Brien}},\ }\bibfield  {title} {\bibinfo {title} {Quantum error mitigation},\ }\href {https://doi.org/10.1103/RevModPhys.95.045005} {\bibfield  {journal} {\bibinfo  {journal} {Rev. Mod. Phys.}\ }\textbf {\bibinfo {volume} {95}},\ \bibinfo {pages} {045005} (\bibinfo {year} {2023})}\BibitemShut {NoStop}%
\bibitem [{\citenamefont {Temme}\ \emph {et~al.}(2017)\citenamefont {Temme}, \citenamefont {Bravyi},\ and\ \citenamefont {Gambetta}}]{Temme2017PRL}%
  \BibitemOpen
  \bibfield  {author} {\bibinfo {author} {\bibfnamefont {K.}~\bibnamefont {Temme}}, \bibinfo {author} {\bibfnamefont {S.}~\bibnamefont {Bravyi}},\ and\ \bibinfo {author} {\bibfnamefont {J.~M.}\ \bibnamefont {Gambetta}},\ }\bibfield  {title} {\bibinfo {title} {Error mitigation for short-depth quantum circuits},\ }\href {https://doi.org/10.1103/PhysRevLett.119.180509} {\bibfield  {journal} {\bibinfo  {journal} {Phys. Rev. Lett.}\ }\textbf {\bibinfo {volume} {119}},\ \bibinfo {pages} {180509} (\bibinfo {year} {2017})}\BibitemShut {NoStop}%
\bibitem [{\citenamefont {Czarnik}\ \emph {et~al.}(2021)\citenamefont {Czarnik}, \citenamefont {Arrasmith}, \citenamefont {Coles},\ and\ \citenamefont {Cincio}}]{Czarnik2021Quantum}%
  \BibitemOpen
  \bibfield  {author} {\bibinfo {author} {\bibfnamefont {P.}~\bibnamefont {Czarnik}}, \bibinfo {author} {\bibfnamefont {A.}~\bibnamefont {Arrasmith}}, \bibinfo {author} {\bibfnamefont {P.~J.}\ \bibnamefont {Coles}},\ and\ \bibinfo {author} {\bibfnamefont {L.}~\bibnamefont {Cincio}},\ }\bibfield  {title} {\bibinfo {title} {Error mitigation with {Clifford} quantum-circuit data},\ }\href {https://doi.org/10.22331/q-2021-11-26-592} {\bibfield  {journal} {\bibinfo  {journal} {Quantum}\ }\textbf {\bibinfo {volume} {5}},\ \bibinfo {pages} {592} (\bibinfo {year} {2021})}\BibitemShut {NoStop}%
\bibitem [{\citenamefont {Li}\ and\ \citenamefont {Benjamin}(2017)}]{Li2017PRX}%
  \BibitemOpen
  \bibfield  {author} {\bibinfo {author} {\bibfnamefont {Y.}~\bibnamefont {Li}}\ and\ \bibinfo {author} {\bibfnamefont {S.~C.}\ \bibnamefont {Benjamin}},\ }\bibfield  {title} {\bibinfo {title} {Efficient variational quantum simulator incorporating active error minimization},\ }\href {https://doi.org/10.1103/PhysRevX.7.021050} {\bibfield  {journal} {\bibinfo  {journal} {Phys. Rev. X}\ }\textbf {\bibinfo {volume} {7}},\ \bibinfo {pages} {021050} (\bibinfo {year} {2017})}\BibitemShut {NoStop}%
\bibitem [{\citenamefont {Kandala}\ \emph {et~al.}(2019)\citenamefont {Kandala}, \citenamefont {Temme}, \citenamefont {C{\'o}rcoles}, \citenamefont {Mezzacapo}, \citenamefont {Chow},\ and\ \citenamefont {Gambetta}}]{Kandala2019Nature}%
  \BibitemOpen
  \bibfield  {author} {\bibinfo {author} {\bibfnamefont {A.}~\bibnamefont {Kandala}}, \bibinfo {author} {\bibfnamefont {K.}~\bibnamefont {Temme}}, \bibinfo {author} {\bibfnamefont {A.~D.}\ \bibnamefont {C{\'o}rcoles}}, \bibinfo {author} {\bibfnamefont {A.}~\bibnamefont {Mezzacapo}}, \bibinfo {author} {\bibfnamefont {J.~M.}\ \bibnamefont {Chow}},\ and\ \bibinfo {author} {\bibfnamefont {J.~M.}\ \bibnamefont {Gambetta}},\ }\bibfield  {title} {\bibinfo {title} {Error mitigation extends the computational reach of a noisy quantum processor},\ }\href {https://doi.org/10.1038/s41586-019-1040-7} {\bibfield  {journal} {\bibinfo  {journal} {Nature}\ }\textbf {\bibinfo {volume} {567}},\ \bibinfo {pages} {491} (\bibinfo {year} {2019})}\BibitemShut {NoStop}%
\bibitem [{\citenamefont {Saxena}\ and\ \citenamefont {Kyaw}(2026)}]{Saxena/EMRE/2026}%
  \BibitemOpen
  \bibfield  {author} {\bibinfo {author} {\bibfnamefont {G.}~\bibnamefont {Saxena}}\ and\ \bibinfo {author} {\bibfnamefont {T.~H.}\ \bibnamefont {Kyaw}},\ }\bibfield  {title} {\bibinfo {title} {{Constant runtime error mitigation via restricted evolution}},\ }\href {https://doi.org/10.1038/s41534-026-01284-1} {\bibfield  {journal} {\bibinfo  {journal} {npj Quantum Inf.}\ }\textbf {\bibinfo {volume} {12}},\ \bibinfo {pages} {148} (\bibinfo {year} {2026})}\BibitemShut {NoStop}%
\bibitem [{\citenamefont {D{\ifmmode\acute{\imath}\else\'{\i}\fi}ez-Valle}\ \emph {et~al.}(2026)\citenamefont {D{\ifmmode\acute{\imath}\else\'{\i}\fi}ez-Valle}, \citenamefont {Saxena}, \citenamefont {Baker}, \citenamefont {Lee},\ and\ \citenamefont {Kyaw}}]{Valle/PIE/2026}%
  \BibitemOpen
  \bibfield  {author} {\bibinfo {author} {\bibfnamefont {P.}~\bibnamefont {D{\ifmmode\acute{\imath}\else\'{\i}\fi}ez-Valle}}, \bibinfo {author} {\bibfnamefont {G.}~\bibnamefont {Saxena}}, \bibinfo {author} {\bibfnamefont {J.~S.}\ \bibnamefont {Baker}}, \bibinfo {author} {\bibfnamefont {J.-H.}\ \bibnamefont {Lee}},\ and\ \bibinfo {author} {\bibfnamefont {T.~H.}\ \bibnamefont {Kyaw}},\ }\bibfield  {title} {\bibinfo {title} {{Physics-inspired extrapolation for efficient error mitigation and hardware certification}},\ }\href {https://doi.org/10.1088/2058-9565/ae5555} {\bibfield  {journal} {\bibinfo  {journal} {Quantum Sci. Technol.}\ }\textbf {\bibinfo {volume} {11}},\ \bibinfo {pages} {025032} (\bibinfo {year} {2026})}\BibitemShut {NoStop}%
\bibitem [{\citenamefont {Giurgica-Tiron}\ \emph {et~al.}(2020)\citenamefont {Giurgica-Tiron}, \citenamefont {Hindy}, \citenamefont {LaRose}, \citenamefont {Mari},\ and\ \citenamefont {Zeng}}]{GiurgicaTiron2020QCE}%
  \BibitemOpen
  \bibfield  {author} {\bibinfo {author} {\bibfnamefont {T.}~\bibnamefont {Giurgica-Tiron}}, \bibinfo {author} {\bibfnamefont {Y.}~\bibnamefont {Hindy}}, \bibinfo {author} {\bibfnamefont {R.}~\bibnamefont {LaRose}}, \bibinfo {author} {\bibfnamefont {A.}~\bibnamefont {Mari}},\ and\ \bibinfo {author} {\bibfnamefont {W.~J.}\ \bibnamefont {Zeng}},\ }\bibfield  {title} {\bibinfo {title} {Digital zero noise extrapolation for quantum error mitigation},\ }in\ \href {https://doi.org/10.1109/QCE49297.2020.00045} {\emph {\bibinfo {booktitle} {2020 IEEE International Conference on Quantum Computing and Engineering (QCE)}}}\ (\bibinfo  {publisher} {IEEE},\ \bibinfo {year} {2020})\ pp.\ \bibinfo {pages} {306--316}\BibitemShut {NoStop}%
\bibitem [{\citenamefont {Kim}\ \emph {et~al.}(2023)\citenamefont {Kim}, \citenamefont {Eddins}, \citenamefont {Anand}, \citenamefont {Wei}, \citenamefont {van~den Berg}, \citenamefont {Rosenblatt}, \citenamefont {Nayfeh}, \citenamefont {Wu}, \citenamefont {Zaletel}, \citenamefont {Temme} \emph {et~al.}}]{Kim2023Nature}%
  \BibitemOpen
  \bibfield  {author} {\bibinfo {author} {\bibfnamefont {Y.}~\bibnamefont {Kim}}, \bibinfo {author} {\bibfnamefont {A.}~\bibnamefont {Eddins}}, \bibinfo {author} {\bibfnamefont {S.}~\bibnamefont {Anand}}, \bibinfo {author} {\bibfnamefont {K.~X.}\ \bibnamefont {Wei}}, \bibinfo {author} {\bibfnamefont {E.}~\bibnamefont {van~den Berg}}, \bibinfo {author} {\bibfnamefont {S.}~\bibnamefont {Rosenblatt}}, \bibinfo {author} {\bibfnamefont {H.}~\bibnamefont {Nayfeh}}, \bibinfo {author} {\bibfnamefont {Y.}~\bibnamefont {Wu}}, \bibinfo {author} {\bibfnamefont {M.}~\bibnamefont {Zaletel}}, \bibinfo {author} {\bibfnamefont {K.}~\bibnamefont {Temme}}, \emph {et~al.},\ }\bibfield  {title} {\bibinfo {title} {Evidence for the utility of quantum computing before fault tolerance},\ }\href {https://doi.org/10.1038/s41586-023-06096-3} {\bibfield  {journal} {\bibinfo  {journal} {Nature}\ }\textbf {\bibinfo {volume} {618}},\ \bibinfo {pages} {500} (\bibinfo {year} {2023})}\BibitemShut {NoStop}%
\bibitem [{\citenamefont {van~den Berg}\ \emph {et~al.}(2023)\citenamefont {van~den Berg}, \citenamefont {Minev}, \citenamefont {Kandala},\ and\ \citenamefont {Temme}}]{vandenBerg2023NatPhys}%
  \BibitemOpen
  \bibfield  {author} {\bibinfo {author} {\bibfnamefont {E.}~\bibnamefont {van~den Berg}}, \bibinfo {author} {\bibfnamefont {Z.~K.}\ \bibnamefont {Minev}}, \bibinfo {author} {\bibfnamefont {A.}~\bibnamefont {Kandala}},\ and\ \bibinfo {author} {\bibfnamefont {K.}~\bibnamefont {Temme}},\ }\bibfield  {title} {\bibinfo {title} {Probabilistic error cancellation with sparse {Pauli}--{Lindblad} models on noisy quantum processors},\ }\href {https://doi.org/10.1038/s41567-023-02042-2} {\bibfield  {journal} {\bibinfo  {journal} {Nat. Phys.}\ }\textbf {\bibinfo {volume} {19}},\ \bibinfo {pages} {1116} (\bibinfo {year} {2023})}\BibitemShut {NoStop}%
\bibitem [{\citenamefont {Bravyi}\ \emph {et~al.}(2021)\citenamefont {Bravyi}, \citenamefont {Sheldon}, \citenamefont {Kandala}, \citenamefont {Mckay},\ and\ \citenamefont {Gambetta}}]{Bravyi2021PRA}%
  \BibitemOpen
  \bibfield  {author} {\bibinfo {author} {\bibfnamefont {S.}~\bibnamefont {Bravyi}}, \bibinfo {author} {\bibfnamefont {S.}~\bibnamefont {Sheldon}}, \bibinfo {author} {\bibfnamefont {A.}~\bibnamefont {Kandala}}, \bibinfo {author} {\bibfnamefont {D.~C.}\ \bibnamefont {Mckay}},\ and\ \bibinfo {author} {\bibfnamefont {J.~M.}\ \bibnamefont {Gambetta}},\ }\bibfield  {title} {\bibinfo {title} {Mitigating measurement errors in multiqubit experiments},\ }\href {https://doi.org/10.1103/PhysRevA.103.042605} {\bibfield  {journal} {\bibinfo  {journal} {Phys. Rev. A}\ }\textbf {\bibinfo {volume} {103}},\ \bibinfo {pages} {042605} (\bibinfo {year} {2021})}\BibitemShut {NoStop}%
\bibitem [{\citenamefont {Bonet-Monroig}\ \emph {et~al.}(2018)\citenamefont {Bonet-Monroig}, \citenamefont {Sagastizabal}, \citenamefont {Singh},\ and\ \citenamefont {O'Brien}}]{BonetMonroig2018PRA}%
  \BibitemOpen
  \bibfield  {author} {\bibinfo {author} {\bibfnamefont {X.}~\bibnamefont {Bonet-Monroig}}, \bibinfo {author} {\bibfnamefont {R.}~\bibnamefont {Sagastizabal}}, \bibinfo {author} {\bibfnamefont {M.}~\bibnamefont {Singh}},\ and\ \bibinfo {author} {\bibfnamefont {T.~E.}\ \bibnamefont {O'Brien}},\ }\bibfield  {title} {\bibinfo {title} {Low-cost error mitigation by symmetry verification},\ }\href {https://doi.org/10.1103/PhysRevA.98.062339} {\bibfield  {journal} {\bibinfo  {journal} {Phys. Rev. A}\ }\textbf {\bibinfo {volume} {98}},\ \bibinfo {pages} {062339} (\bibinfo {year} {2018})}\BibitemShut {NoStop}%
\bibitem [{\citenamefont {Vovrosh}\ \emph {et~al.}(2021)\citenamefont {Vovrosh}, \citenamefont {Khosla}, \citenamefont {Greenaway}, \citenamefont {Self}, \citenamefont {Kim},\ and\ \citenamefont {Knolle}}]{Vovrosh2021PRE}%
  \BibitemOpen
  \bibfield  {author} {\bibinfo {author} {\bibfnamefont {J.}~\bibnamefont {Vovrosh}}, \bibinfo {author} {\bibfnamefont {K.~E.}\ \bibnamefont {Khosla}}, \bibinfo {author} {\bibfnamefont {S.}~\bibnamefont {Greenaway}}, \bibinfo {author} {\bibfnamefont {C.}~\bibnamefont {Self}}, \bibinfo {author} {\bibfnamefont {M.~S.}\ \bibnamefont {Kim}},\ and\ \bibinfo {author} {\bibfnamefont {J.}~\bibnamefont {Knolle}},\ }\bibfield  {title} {\bibinfo {title} {Simple mitigation of global depolarizing errors in quantum simulations},\ }\href {https://doi.org/10.1103/PhysRevE.104.035309} {\bibfield  {journal} {\bibinfo  {journal} {Phys. Rev. E}\ }\textbf {\bibinfo {volume} {104}},\ \bibinfo {pages} {035309} (\bibinfo {year} {2021})}\BibitemShut {NoStop}%
\bibitem [{\citenamefont {Takagi}\ \emph {et~al.}(2022)\citenamefont {Takagi}, \citenamefont {Endo}, \citenamefont {Minagawa},\ and\ \citenamefont {Gu}}]{Takagi2022npjQI}%
  \BibitemOpen
  \bibfield  {author} {\bibinfo {author} {\bibfnamefont {R.}~\bibnamefont {Takagi}}, \bibinfo {author} {\bibfnamefont {S.}~\bibnamefont {Endo}}, \bibinfo {author} {\bibfnamefont {S.}~\bibnamefont {Minagawa}},\ and\ \bibinfo {author} {\bibfnamefont {M.}~\bibnamefont {Gu}},\ }\bibfield  {title} {\bibinfo {title} {Fundamental limits of quantum error mitigation},\ }\href {https://doi.org/10.1038/s41534-022-00618-z} {\bibfield  {journal} {\bibinfo  {journal} {npj Quantum Inf.}\ }\textbf {\bibinfo {volume} {8}},\ \bibinfo {pages} {114} (\bibinfo {year} {2022})}\BibitemShut {NoStop}%
\bibitem [{\citenamefont {Quek}\ \emph {et~al.}(2024)\citenamefont {Quek}, \citenamefont {Stilck~Fran{\c{c}}a}, \citenamefont {Khatri}, \citenamefont {Meyer},\ and\ \citenamefont {Eisert}}]{Quek2024NatPhys}%
  \BibitemOpen
  \bibfield  {author} {\bibinfo {author} {\bibfnamefont {Y.}~\bibnamefont {Quek}}, \bibinfo {author} {\bibfnamefont {D.}~\bibnamefont {Stilck~Fran{\c{c}}a}}, \bibinfo {author} {\bibfnamefont {S.}~\bibnamefont {Khatri}}, \bibinfo {author} {\bibfnamefont {J.~J.}\ \bibnamefont {Meyer}},\ and\ \bibinfo {author} {\bibfnamefont {J.}~\bibnamefont {Eisert}},\ }\bibfield  {title} {\bibinfo {title} {Exponentially tighter bounds on limitations of quantum error mitigation},\ }\href {https://doi.org/10.1038/s41567-024-02536-7} {\bibfield  {journal} {\bibinfo  {journal} {Nat. Phys.}\ }\textbf {\bibinfo {volume} {20}},\ \bibinfo {pages} {1648} (\bibinfo {year} {2024})}\BibitemShut {NoStop}%
\bibitem [{\citenamefont {Strikis}\ \emph {et~al.}(2021)\citenamefont {Strikis}, \citenamefont {Qin}, \citenamefont {Chen}, \citenamefont {Benjamin},\ and\ \citenamefont {Li}}]{Strikis2021PRXQ}%
  \BibitemOpen
  \bibfield  {author} {\bibinfo {author} {\bibfnamefont {A.}~\bibnamefont {Strikis}}, \bibinfo {author} {\bibfnamefont {D.}~\bibnamefont {Qin}}, \bibinfo {author} {\bibfnamefont {Y.}~\bibnamefont {Chen}}, \bibinfo {author} {\bibfnamefont {S.~C.}\ \bibnamefont {Benjamin}},\ and\ \bibinfo {author} {\bibfnamefont {Y.}~\bibnamefont {Li}},\ }\bibfield  {title} {\bibinfo {title} {Learning-based quantum error mitigation},\ }\href {https://doi.org/10.1103/PRXQuantum.2.040330} {\bibfield  {journal} {\bibinfo  {journal} {PRX Quantum}\ }\textbf {\bibinfo {volume} {2}},\ \bibinfo {pages} {040330} (\bibinfo {year} {2021})}\BibitemShut {NoStop}%
\bibitem [{\citenamefont {Liao}\ \emph {et~al.}(2024)\citenamefont {Liao}, \citenamefont {Wang}, \citenamefont {Sitdikov}, \citenamefont {Salcedo}, \citenamefont {Seif},\ and\ \citenamefont {Minev}}]{Liao2024NMI}%
  \BibitemOpen
  \bibfield  {author} {\bibinfo {author} {\bibfnamefont {H.}~\bibnamefont {Liao}}, \bibinfo {author} {\bibfnamefont {D.~S.}\ \bibnamefont {Wang}}, \bibinfo {author} {\bibfnamefont {I.}~\bibnamefont {Sitdikov}}, \bibinfo {author} {\bibfnamefont {C.}~\bibnamefont {Salcedo}}, \bibinfo {author} {\bibfnamefont {A.}~\bibnamefont {Seif}},\ and\ \bibinfo {author} {\bibfnamefont {Z.~K.}\ \bibnamefont {Minev}},\ }\bibfield  {title} {\bibinfo {title} {Machine learning for practical quantum error mitigation},\ }\href {https://doi.org/10.1038/s42256-024-00927-2} {\bibfield  {journal} {\bibinfo  {journal} {Nat. Mach. Intell.}\ }\textbf {\bibinfo {volume} {6}},\ \bibinfo {pages} {1478} (\bibinfo {year} {2024})}\BibitemShut {NoStop}%
\bibitem [{\citenamefont {Placidi}\ \emph {et~al.}(2026)\citenamefont {Placidi}, \citenamefont {Williams}, \citenamefont {Rinaldi}, \citenamefont {Mills}, \citenamefont {C{\ifmmode\hat{\imath}\else\^{\i}\fi}rstoiu}, \citenamefont {Eccles},\ and\ \citenamefont {Duncan}}]{Placidi/DL_QEM/2026}%
  \BibitemOpen
  \bibfield  {author} {\bibinfo {author} {\bibfnamefont {L.}~\bibnamefont {Placidi}}, \bibinfo {author} {\bibfnamefont {I.}~\bibnamefont {Williams}}, \bibinfo {author} {\bibfnamefont {E.}~\bibnamefont {Rinaldi}}, \bibinfo {author} {\bibfnamefont {D.}~\bibnamefont {Mills}}, \bibinfo {author} {\bibfnamefont {C.}~\bibnamefont {C{\ifmmode\hat{\imath}\else\^{\i}\fi}rstoiu}}, \bibinfo {author} {\bibfnamefont {V.}~\bibnamefont {Eccles}},\ and\ \bibinfo {author} {\bibfnamefont {R.}~\bibnamefont {Duncan}},\ }\bibfield  {title} {\bibinfo {title} {{Deep Learning Approaches to Quantum Error Mitigation}},\ }\bibfield  {journal} {\bibinfo  {journal} {arXiv}\ }\href {https://doi.org/10.48550/arXiv.2601.14226} {10.48550/arXiv.2601.14226} (\bibinfo {year} {2026}),\ \Eprint {https://arxiv.org/abs/2601.14226} {2601.14226} \BibitemShut {NoStop}%
\bibitem [{\citenamefont {Kim}\ \emph {et~al.}(2020)\citenamefont {Kim}, \citenamefont {Park},\ and\ \citenamefont {Rhee}}]{Kim2020IEEEAccess}%
  \BibitemOpen
  \bibfield  {author} {\bibinfo {author} {\bibfnamefont {C.}~\bibnamefont {Kim}}, \bibinfo {author} {\bibfnamefont {K.~D.}\ \bibnamefont {Park}},\ and\ \bibinfo {author} {\bibfnamefont {J.-K.}\ \bibnamefont {Rhee}},\ }\bibfield  {title} {\bibinfo {title} {Quantum error mitigation with artificial neural network},\ }\href {https://doi.org/10.1109/ACCESS.2020.3031607} {\bibfield  {journal} {\bibinfo  {journal} {IEEE Access}\ }\textbf {\bibinfo {volume} {8}},\ \bibinfo {pages} {188853} (\bibinfo {year} {2020})}\BibitemShut {NoStop}%
\bibitem [{\citenamefont {Bao}\ \emph {et~al.}(2025)\citenamefont {Bao}, \citenamefont {Ye}, \citenamefont {Ruan}, \citenamefont {Liu}, \citenamefont {Wu},\ and\ \citenamefont {Yan}}]{Bao2025ICLR}%
  \BibitemOpen
  \bibfield  {author} {\bibinfo {author} {\bibfnamefont {T.}~\bibnamefont {Bao}}, \bibinfo {author} {\bibfnamefont {X.}~\bibnamefont {Ye}}, \bibinfo {author} {\bibfnamefont {H.}~\bibnamefont {Ruan}}, \bibinfo {author} {\bibfnamefont {C.}~\bibnamefont {Liu}}, \bibinfo {author} {\bibfnamefont {W.}~\bibnamefont {Wu}},\ and\ \bibinfo {author} {\bibfnamefont {J.}~\bibnamefont {Yan}},\ }\bibfield  {title} {\bibinfo {title} {Beyond circuit connections: a non-message passing graph transformer approach for quantum error mitigation},\ }in\ \href {https://openreview.net/forum?id=XnVttczoAV} {\emph {\bibinfo {booktitle} {International Conference on Learning Representations (ICLR)}}}\ (\bibinfo {year} {2025})\BibitemShut {NoStop}%
\bibitem [{\citenamefont {Tousi}\ and\ \citenamefont {DeSouza}(2025)}]{Tousi2025arXiv}%
  \BibitemOpen
  \bibfield  {author} {\bibinfo {author} {\bibfnamefont {S.~M.~A.}\ \bibnamefont {Tousi}}\ and\ \bibinfo {author} {\bibfnamefont {G.~N.}\ \bibnamefont {DeSouza}},\ }\href@noop {} {\bibinfo {title} {{QAGT-MLP}: An attention-based graph transformer for small and large-scale quantum error mitigation}} (\bibinfo {year} {2025}),\ \Eprint {https://arxiv.org/abs/arXiv:2511.03119} {arXiv:2511.03119} \BibitemShut {NoStop}%
\bibitem [{\citenamefont {Cantori}\ \emph {et~al.}(2025)\citenamefont {Cantori}, \citenamefont {Mari}, \citenamefont {Vitali},\ and\ \citenamefont {Pilati}}]{Cantori/DL_QEM_partial_knitting/2025}%
  \BibitemOpen
  \bibfield  {author} {\bibinfo {author} {\bibfnamefont {S.}~\bibnamefont {Cantori}}, \bibinfo {author} {\bibfnamefont {A.}~\bibnamefont {Mari}}, \bibinfo {author} {\bibfnamefont {D.}~\bibnamefont {Vitali}},\ and\ \bibinfo {author} {\bibfnamefont {S.}~\bibnamefont {Pilati}},\ }\bibfield  {title} {\bibinfo {title} {{Deep-learned error mitigation via partially knitted circuits for the variational quantum eigensolver}},\ }\bibfield  {journal} {\bibinfo  {journal} {arXiv}\ }\href {https://doi.org/10.48550/arXiv.2506.04146} {10.48550/arXiv.2506.04146} (\bibinfo {year} {2025}),\ \Eprint {https://arxiv.org/abs/2506.04146} {2506.04146} \BibitemShut {NoStop}%
\bibitem [{\citenamefont {Wang}\ \emph {et~al.}(2026)\citenamefont {Wang}, \citenamefont {Wu}, \citenamefont {Liu}, \citenamefont {He}, \citenamefont {Shang}, \citenamefont {Guo},\ and\ \citenamefont {Chen}}]{Wang2026QMI}%
  \BibitemOpen
  \bibfield  {author} {\bibinfo {author} {\bibfnamefont {H.}~\bibnamefont {Wang}}, \bibinfo {author} {\bibfnamefont {X.}~\bibnamefont {Wu}}, \bibinfo {author} {\bibfnamefont {J.}~\bibnamefont {Liu}}, \bibinfo {author} {\bibfnamefont {R.}~\bibnamefont {He}}, \bibinfo {author} {\bibfnamefont {J.}~\bibnamefont {Shang}}, \bibinfo {author} {\bibfnamefont {H.}~\bibnamefont {Guo}},\ and\ \bibinfo {author} {\bibfnamefont {Q.}~\bibnamefont {Chen}},\ }\bibfield  {title} {\bibinfo {title} {Scalable quantum error mitigation with physically informed graph neural networks},\ }\href {https://doi.org/10.1007/s42484-026-00462-w} {\bibfield  {journal} {\bibinfo  {journal} {Quantum Mach. Intell.}\ }\textbf {\bibinfo {volume} {8}},\ \bibinfo {pages} {114} (\bibinfo {year} {2026})}\BibitemShut {NoStop}%
\bibitem [{\citenamefont {Karpathy}(2026)}]{Karpathy2026autoresearch}%
  \BibitemOpen
  \bibfield  {author} {\bibinfo {author} {\bibfnamefont {A.}~\bibnamefont {Karpathy}},\ }\href@noop {} {\bibinfo {title} {autoresearch}},\ \bibinfo {howpublished} {GitHub repository} (\bibinfo {year} {2026}),\ \bibinfo {note} {\url{https://github.com/karpathy/autoresearch}}\BibitemShut {NoStop}%
\bibitem [{\citenamefont {Peng}\ \emph {et~al.}(2020)\citenamefont {Peng}, \citenamefont {Harrow}, \citenamefont {Ozols},\ and\ \citenamefont {Wu}}]{Peng/wire_cutting/2020}%
  \BibitemOpen
  \bibfield  {author} {\bibinfo {author} {\bibfnamefont {T.}~\bibnamefont {Peng}}, \bibinfo {author} {\bibfnamefont {A.~W.}\ \bibnamefont {Harrow}}, \bibinfo {author} {\bibfnamefont {M.}~\bibnamefont {Ozols}},\ and\ \bibinfo {author} {\bibfnamefont {X.}~\bibnamefont {Wu}},\ }\bibfield  {title} {\bibinfo {title} {Simulating large quantum circuits on a small quantum computer},\ }\href {https://doi.org/10.1103/PhysRevLett.125.150504} {\bibfield  {journal} {\bibinfo  {journal} {Phys. Rev. Lett.}\ }\textbf {\bibinfo {volume} {125}},\ \bibinfo {pages} {150504} (\bibinfo {year} {2020})}\BibitemShut {NoStop}%
\bibitem [{\citenamefont {Mitarai}\ and\ \citenamefont {Fujii}(2021)}]{Mitarai/gate_cutting/2021}%
  \BibitemOpen
  \bibfield  {author} {\bibinfo {author} {\bibfnamefont {K.}~\bibnamefont {Mitarai}}\ and\ \bibinfo {author} {\bibfnamefont {K.}~\bibnamefont {Fujii}},\ }\bibfield  {title} {\bibinfo {title} {{Constructing a virtual two-qubit gate by sampling single-qubit operations}},\ }\href {https://doi.org/10.1088/1367-2630/abd7bc} {\bibfield  {journal} {\bibinfo  {journal} {New J. Phys.}\ }\textbf {\bibinfo {volume} {23}},\ \bibinfo {pages} {023021} (\bibinfo {year} {2021})}\BibitemShut {NoStop}%
\bibitem [{\citenamefont {Foss-Feig}\ \emph {et~al.}(2021)\citenamefont {Foss-Feig}, \citenamefont {Hayes}, \citenamefont {Dreiling}, \citenamefont {Figgatt}, \citenamefont {Gaebler}, \citenamefont {Moses}, \citenamefont {Pino},\ and\ \citenamefont {Potter}}]{FossFeig2021PRR}%
  \BibitemOpen
  \bibfield  {author} {\bibinfo {author} {\bibfnamefont {M.}~\bibnamefont {Foss-Feig}}, \bibinfo {author} {\bibfnamefont {D.}~\bibnamefont {Hayes}}, \bibinfo {author} {\bibfnamefont {J.~M.}\ \bibnamefont {Dreiling}}, \bibinfo {author} {\bibfnamefont {C.}~\bibnamefont {Figgatt}}, \bibinfo {author} {\bibfnamefont {J.~P.}\ \bibnamefont {Gaebler}}, \bibinfo {author} {\bibfnamefont {S.~A.}\ \bibnamefont {Moses}}, \bibinfo {author} {\bibfnamefont {J.~M.}\ \bibnamefont {Pino}},\ and\ \bibinfo {author} {\bibfnamefont {A.~C.}\ \bibnamefont {Potter}},\ }\bibfield  {title} {\bibinfo {title} {Holographic quantum algorithms for simulating correlated spin systems},\ }\href {https://doi.org/10.1103/PhysRevResearch.3.033002} {\bibfield  {journal} {\bibinfo  {journal} {Phys. Rev. Research}\ }\textbf {\bibinfo {volume} {3}},\ \bibinfo {pages} {033002} (\bibinfo {year} {2021})}\BibitemShut {NoStop}%
\bibitem [{\citenamefont {Barratt}\ \emph {et~al.}(2021)\citenamefont {Barratt}, \citenamefont {Dborin}, \citenamefont {Bal}, \citenamefont {Stojevic}, \citenamefont {Pollmann},\ and\ \citenamefont {Green}}]{Barratt2021npjQI}%
  \BibitemOpen
  \bibfield  {author} {\bibinfo {author} {\bibfnamefont {F.}~\bibnamefont {Barratt}}, \bibinfo {author} {\bibfnamefont {J.}~\bibnamefont {Dborin}}, \bibinfo {author} {\bibfnamefont {M.}~\bibnamefont {Bal}}, \bibinfo {author} {\bibfnamefont {V.}~\bibnamefont {Stojevic}}, \bibinfo {author} {\bibfnamefont {F.}~\bibnamefont {Pollmann}},\ and\ \bibinfo {author} {\bibfnamefont {A.~G.}\ \bibnamefont {Green}},\ }\bibfield  {title} {\bibinfo {title} {Parallel quantum simulation of large systems on small {NISQ} computers},\ }\href {https://doi.org/10.1038/s41534-021-00420-3} {\bibfield  {journal} {\bibinfo  {journal} {npj Quantum Inf.}\ }\textbf {\bibinfo {volume} {7}},\ \bibinfo {pages} {79} (\bibinfo {year} {2021})}\BibitemShut {NoStop}%
\bibitem [{\citenamefont {Xu}\ \emph {et~al.}(2019)\citenamefont {Xu}, \citenamefont {Hu}, \citenamefont {Leskovec},\ and\ \citenamefont {Jegelka}}]{Xu2019ICLR}%
  \BibitemOpen
  \bibfield  {author} {\bibinfo {author} {\bibfnamefont {K.}~\bibnamefont {Xu}}, \bibinfo {author} {\bibfnamefont {W.}~\bibnamefont {Hu}}, \bibinfo {author} {\bibfnamefont {J.}~\bibnamefont {Leskovec}},\ and\ \bibinfo {author} {\bibfnamefont {S.}~\bibnamefont {Jegelka}},\ }\bibfield  {title} {\bibinfo {title} {How powerful are graph neural networks?},\ }in\ \href {https://openreview.net/forum?id=ryGs6iA5Km} {\emph {\bibinfo {booktitle} {International Conference on Learning Representations (ICLR)}}}\ (\bibinfo {year} {2019})\BibitemShut {NoStop}%
\bibitem [{\citenamefont {Hu}\ \emph {et~al.}(2020)\citenamefont {Hu}, \citenamefont {Liu}, \citenamefont {Gomes}, \citenamefont {Zitnik}, \citenamefont {Liang}, \citenamefont {Pande},\ and\ \citenamefont {Leskovec}}]{Hu2020ICLR}%
  \BibitemOpen
  \bibfield  {author} {\bibinfo {author} {\bibfnamefont {W.}~\bibnamefont {Hu}}, \bibinfo {author} {\bibfnamefont {B.}~\bibnamefont {Liu}}, \bibinfo {author} {\bibfnamefont {J.}~\bibnamefont {Gomes}}, \bibinfo {author} {\bibfnamefont {M.}~\bibnamefont {Zitnik}}, \bibinfo {author} {\bibfnamefont {P.}~\bibnamefont {Liang}}, \bibinfo {author} {\bibfnamefont {V.}~\bibnamefont {Pande}},\ and\ \bibinfo {author} {\bibfnamefont {J.}~\bibnamefont {Leskovec}},\ }\bibfield  {title} {\bibinfo {title} {Strategies for pre-training graph neural networks},\ }in\ \href {https://openreview.net/forum?id=HJlWWJSFDH} {\emph {\bibinfo {booktitle} {International Conference on Learning Representations (ICLR)}}}\ (\bibinfo {year} {2020})\BibitemShut {NoStop}%
\bibitem [{\citenamefont {Bharti}\ \emph {et~al.}(2022)\citenamefont {Bharti}, \citenamefont {Cervera-Lierta}, \citenamefont {Kyaw}, \citenamefont {Haug}, \citenamefont {Alperin-Lea}, \citenamefont {Anand}, \citenamefont {Degroote}, \citenamefont {Heimonen}, \citenamefont {Kottmann}, \citenamefont {Menke}, \citenamefont {Mok}, \citenamefont {Sim}, \citenamefont {Kwek},\ and\ \citenamefont {Aspuru-Guzik}}]{bharti2022noisy}%
  \BibitemOpen
  \bibfield  {author} {\bibinfo {author} {\bibfnamefont {K.}~\bibnamefont {Bharti}}, \bibinfo {author} {\bibfnamefont {A.}~\bibnamefont {Cervera-Lierta}}, \bibinfo {author} {\bibfnamefont {T.~H.}\ \bibnamefont {Kyaw}}, \bibinfo {author} {\bibfnamefont {T.}~\bibnamefont {Haug}}, \bibinfo {author} {\bibfnamefont {S.}~\bibnamefont {Alperin-Lea}}, \bibinfo {author} {\bibfnamefont {A.}~\bibnamefont {Anand}}, \bibinfo {author} {\bibfnamefont {M.}~\bibnamefont {Degroote}}, \bibinfo {author} {\bibfnamefont {H.}~\bibnamefont {Heimonen}}, \bibinfo {author} {\bibfnamefont {J.~S.}\ \bibnamefont {Kottmann}}, \bibinfo {author} {\bibfnamefont {T.}~\bibnamefont {Menke}}, \bibinfo {author} {\bibfnamefont {W.-K.}\ \bibnamefont {Mok}}, \bibinfo {author} {\bibfnamefont {S.}~\bibnamefont {Sim}}, \bibinfo {author} {\bibfnamefont {L.-C.}\ \bibnamefont {Kwek}},\ and\ \bibinfo {author} {\bibfnamefont {A.}~\bibnamefont {Aspuru-Guzik}},\ }\bibfield  {title} {\bibinfo {title} {Noisy intermediate-scale quantum algorithms},\ }\href
  {https://doi.org/10.1103/RevModPhys.94.015004} {\bibfield  {journal} {\bibinfo  {journal} {Rev. Mod. Phys.}\ }\textbf {\bibinfo {volume} {94}},\ \bibinfo {pages} {015004} (\bibinfo {year} {2022})}\BibitemShut {NoStop}%
\bibitem [{\citenamefont {Motta}\ \emph {et~al.}(2023)\citenamefont {Motta}, \citenamefont {Sung}, \citenamefont {Whaley}, \citenamefont {Head-Gordon},\ and\ \citenamefont {Shee}}]{Motta/LUCJ/2023}%
  \BibitemOpen
  \bibfield  {author} {\bibinfo {author} {\bibfnamefont {M.}~\bibnamefont {Motta}}, \bibinfo {author} {\bibfnamefont {K.~J.}\ \bibnamefont {Sung}}, \bibinfo {author} {\bibfnamefont {K.~B.}\ \bibnamefont {Whaley}}, \bibinfo {author} {\bibfnamefont {M.}~\bibnamefont {Head-Gordon}},\ and\ \bibinfo {author} {\bibfnamefont {J.}~\bibnamefont {Shee}},\ }\bibfield  {title} {\bibinfo {title} {{Bridging physical intuition and hardware efficiency for correlated electronic states: the local unitary cluster Jastrow ansatz for electronic structure}},\ }\href {https://doi.org/10.1039/d3sc02516k} {\bibfield  {journal} {\bibinfo  {journal} {Chem. Sci.}\ }\textbf {\bibinfo {volume} {14}},\ \bibinfo {pages} {11213} (\bibinfo {year} {2023})}\BibitemShut {NoStop}%
\bibitem [{\citenamefont {Bergstra}\ and\ \citenamefont {Bengio}(2012)}]{Bergstra2012JMLR}%
  \BibitemOpen
  \bibfield  {author} {\bibinfo {author} {\bibfnamefont {J.}~\bibnamefont {Bergstra}}\ and\ \bibinfo {author} {\bibfnamefont {Y.}~\bibnamefont {Bengio}},\ }\bibfield  {title} {\bibinfo {title} {Random search for hyper-parameter optimization},\ }\href@noop {} {\bibfield  {journal} {\bibinfo  {journal} {J. Mach. Learn. Res.}\ }\textbf {\bibinfo {volume} {13}},\ \bibinfo {pages} {281} (\bibinfo {year} {2012})}\BibitemShut {NoStop}%
\bibitem [{\citenamefont {{Qiskit Textbook contributors}}(2023)}]{QiskitTextbook2023}%
  \BibitemOpen
  \bibfield  {author} {\bibinfo {author} {\bibnamefont {{Qiskit Textbook contributors}}},\ }\href@noop {} {\bibinfo {title} {Qiskit {Textbook}, chapter ``{Grover's} algorithm'', section 4.1 ``{Solving} {Sudoku} using {Grover's} algorithm''}},\ \bibinfo {howpublished} {GitHub repository Qiskit/textbook (archived 18 January 2024)} (\bibinfo {year} {2023}),\ \bibinfo {note} {\url{https://github.com/Qiskit/textbook}, file notebooks/ch-algorithms/grover.ipynb; accessed 14 October 2026}\BibitemShut {NoStop}%
\bibitem [{\citenamefont {Nielsen}\ and\ \citenamefont {Chuang}(2010)}]{Nielsen2010}%
  \BibitemOpen
  \bibfield  {author} {\bibinfo {author} {\bibfnamefont {M.~A.}\ \bibnamefont {Nielsen}}\ and\ \bibinfo {author} {\bibfnamefont {I.~L.}\ \bibnamefont {Chuang}},\ }\href {https://doi.org/10.1017/CBO9780511976667} {\emph {\bibinfo {title} {Quantum Computation and Quantum Information: 10th Anniversary Edition}}}\ (\bibinfo  {publisher} {Cambridge University Press},\ \bibinfo {address} {Cambridge},\ \bibinfo {year} {2010})\BibitemShut {NoStop}%
\bibitem [{\citenamefont {Barenco}\ \emph {et~al.}(1995)\citenamefont {Barenco}, \citenamefont {Bennett}, \citenamefont {Cleve}, \citenamefont {DiVincenzo}, \citenamefont {Margolus}, \citenamefont {Shor}, \citenamefont {Sleator}, \citenamefont {Smolin},\ and\ \citenamefont {Weinfurter}}]{Barenco1995PRA}%
  \BibitemOpen
  \bibfield  {author} {\bibinfo {author} {\bibfnamefont {A.}~\bibnamefont {Barenco}}, \bibinfo {author} {\bibfnamefont {C.~H.}\ \bibnamefont {Bennett}}, \bibinfo {author} {\bibfnamefont {R.}~\bibnamefont {Cleve}}, \bibinfo {author} {\bibfnamefont {D.~P.}\ \bibnamefont {DiVincenzo}}, \bibinfo {author} {\bibfnamefont {N.}~\bibnamefont {Margolus}}, \bibinfo {author} {\bibfnamefont {P.}~\bibnamefont {Shor}}, \bibinfo {author} {\bibfnamefont {T.}~\bibnamefont {Sleator}}, \bibinfo {author} {\bibfnamefont {J.~A.}\ \bibnamefont {Smolin}},\ and\ \bibinfo {author} {\bibfnamefont {H.}~\bibnamefont {Weinfurter}},\ }\bibfield  {title} {\bibinfo {title} {Elementary gates for quantum computation},\ }\href {https://doi.org/10.1103/PhysRevA.52.3457} {\bibfield  {journal} {\bibinfo  {journal} {Phys. Rev. A}\ }\textbf {\bibinfo {volume} {52}},\ \bibinfo {pages} {3457} (\bibinfo {year} {1995})}\BibitemShut {NoStop}%
\bibitem [{\citenamefont {Grover}(1996)}]{Grover1996STOC}%
  \BibitemOpen
  \bibfield  {author} {\bibinfo {author} {\bibfnamefont {L.~K.}\ \bibnamefont {Grover}},\ }\bibfield  {title} {\bibinfo {title} {A fast quantum mechanical algorithm for database search},\ }in\ \href {https://doi.org/10.1145/237814.237866} {\emph {\bibinfo {booktitle} {Proceedings of the 28th Annual ACM Symposium on Theory of Computing (STOC '96)}}}\ (\bibinfo  {publisher} {ACM},\ \bibinfo {year} {1996})\ pp.\ \bibinfo {pages} {212--219}\BibitemShut {NoStop}%
\bibitem [{\citenamefont {Boyer}\ \emph {et~al.}(1998)\citenamefont {Boyer}, \citenamefont {Brassard}, \citenamefont {H{\o}yer},\ and\ \citenamefont {Tapp}}]{Boyer1998FdP}%
  \BibitemOpen
  \bibfield  {author} {\bibinfo {author} {\bibfnamefont {M.}~\bibnamefont {Boyer}}, \bibinfo {author} {\bibfnamefont {G.}~\bibnamefont {Brassard}}, \bibinfo {author} {\bibfnamefont {P.}~\bibnamefont {H{\o}yer}},\ and\ \bibinfo {author} {\bibfnamefont {A.}~\bibnamefont {Tapp}},\ }\bibfield  {title} {\bibinfo {title} {Tight bounds on quantum searching},\ }\href {https://doi.org/10.1002/(SICI)1521-3978(199806)46:4/5<493::AID-PROP493>3.0.CO;2-P} {\bibfield  {journal} {\bibinfo  {journal} {Fortschr. Phys.}\ }\textbf {\bibinfo {volume} {46}},\ \bibinfo {pages} {493} (\bibinfo {year} {1998})}\BibitemShut {NoStop}%
\bibitem [{\citenamefont {Nishio}\ \emph {et~al.}(2020)\citenamefont {Nishio}, \citenamefont {Pan}, \citenamefont {Satoh}, \citenamefont {Amano},\ and\ \citenamefont {Van~Meter}}]{Nishio2020JETC}%
  \BibitemOpen
  \bibfield  {author} {\bibinfo {author} {\bibfnamefont {S.}~\bibnamefont {Nishio}}, \bibinfo {author} {\bibfnamefont {Y.}~\bibnamefont {Pan}}, \bibinfo {author} {\bibfnamefont {T.}~\bibnamefont {Satoh}}, \bibinfo {author} {\bibfnamefont {H.}~\bibnamefont {Amano}},\ and\ \bibinfo {author} {\bibfnamefont {R.}~\bibnamefont {Van~Meter}},\ }\bibfield  {title} {\bibinfo {title} {Extracting success from {IBM}'s 20-qubit machines using error-aware compilation},\ }\href {https://doi.org/10.1145/3386162} {\bibfield  {journal} {\bibinfo  {journal} {ACM J. Emerg. Technol. Comput. Syst.}\ }\textbf {\bibinfo {volume} {16}},\ \bibinfo {pages} {1} (\bibinfo {year} {2020})}\BibitemShut {NoStop}%
\bibitem [{\citenamefont {Javadi-Abhari}\ \emph {et~al.}(2024)\citenamefont {Javadi-Abhari}, \citenamefont {Treinish}, \citenamefont {Krsulich}, \citenamefont {Wood}, \citenamefont {Lishman}, \citenamefont {Gacon}, \citenamefont {Martiel}, \citenamefont {Nation}, \citenamefont {Bishop}, \citenamefont {Cross} \emph {et~al.}}]{JavadiAbhari2024arXiv}%
  \BibitemOpen
  \bibfield  {author} {\bibinfo {author} {\bibfnamefont {A.}~\bibnamefont {Javadi-Abhari}}, \bibinfo {author} {\bibfnamefont {M.}~\bibnamefont {Treinish}}, \bibinfo {author} {\bibfnamefont {K.}~\bibnamefont {Krsulich}}, \bibinfo {author} {\bibfnamefont {C.~J.}\ \bibnamefont {Wood}}, \bibinfo {author} {\bibfnamefont {J.}~\bibnamefont {Lishman}}, \bibinfo {author} {\bibfnamefont {J.}~\bibnamefont {Gacon}}, \bibinfo {author} {\bibfnamefont {S.}~\bibnamefont {Martiel}}, \bibinfo {author} {\bibfnamefont {P.~D.}\ \bibnamefont {Nation}}, \bibinfo {author} {\bibfnamefont {L.~S.}\ \bibnamefont {Bishop}}, \bibinfo {author} {\bibfnamefont {A.~W.}\ \bibnamefont {Cross}}, \emph {et~al.},\ }\href@noop {} {\bibinfo {title} {Quantum computing with {Qiskit}}} (\bibinfo {year} {2024}),\ \Eprint {https://arxiv.org/abs/arXiv:2405.08810} {arXiv:2405.08810} \BibitemShut {NoStop}%
\bibitem [{\citenamefont {Viola}\ \emph {et~al.}(1999)\citenamefont {Viola}, \citenamefont {Knill},\ and\ \citenamefont {Lloyd}}]{Viola1999PRL}%
  \BibitemOpen
  \bibfield  {author} {\bibinfo {author} {\bibfnamefont {L.}~\bibnamefont {Viola}}, \bibinfo {author} {\bibfnamefont {E.}~\bibnamefont {Knill}},\ and\ \bibinfo {author} {\bibfnamefont {S.}~\bibnamefont {Lloyd}},\ }\bibfield  {title} {\bibinfo {title} {Dynamical decoupling of open quantum systems},\ }\href {https://doi.org/10.1103/PhysRevLett.82.2417} {\bibfield  {journal} {\bibinfo  {journal} {Phys. Rev. Lett.}\ }\textbf {\bibinfo {volume} {82}},\ \bibinfo {pages} {2417} (\bibinfo {year} {1999})}\BibitemShut {NoStop}%
\bibitem [{\citenamefont {Efron}(1979)}]{Efron1979AnnStat}%
  \BibitemOpen
  \bibfield  {author} {\bibinfo {author} {\bibfnamefont {B.}~\bibnamefont {Efron}},\ }\bibfield  {title} {\bibinfo {title} {Bootstrap methods: another look at the jackknife},\ }\href {https://doi.org/10.1214/aos/1176344552} {\bibfield  {journal} {\bibinfo  {journal} {Ann. Stat.}\ }\textbf {\bibinfo {volume} {7}},\ \bibinfo {pages} {1} (\bibinfo {year} {1979})}\BibitemShut {NoStop}%
\bibitem [{\citenamefont {Dalzell}\ \emph {et~al.}(2024)\citenamefont {Dalzell}, \citenamefont {Hunter-Jones},\ and\ \citenamefont {Brand{\~a}o}}]{Dalzell2024CMP}%
  \BibitemOpen
  \bibfield  {author} {\bibinfo {author} {\bibfnamefont {A.~M.}\ \bibnamefont {Dalzell}}, \bibinfo {author} {\bibfnamefont {N.}~\bibnamefont {Hunter-Jones}},\ and\ \bibinfo {author} {\bibfnamefont {F.~G. S.~L.}\ \bibnamefont {Brand{\~a}o}},\ }\bibfield  {title} {\bibinfo {title} {Random quantum circuits transform local noise into global white noise},\ }\href {https://doi.org/10.1007/s00220-024-04958-z} {\bibfield  {journal} {\bibinfo  {journal} {Commun. Math. Phys.}\ }\textbf {\bibinfo {volume} {405}},\ \bibinfo {pages} {78} (\bibinfo {year} {2024})}\BibitemShut {NoStop}%
\bibitem [{\citenamefont {Ilse}\ \emph {et~al.}(2018)\citenamefont {Ilse}, \citenamefont {Tomczak},\ and\ \citenamefont {Welling}}]{Ilse2018ICML}%
  \BibitemOpen
  \bibfield  {author} {\bibinfo {author} {\bibfnamefont {M.}~\bibnamefont {Ilse}}, \bibinfo {author} {\bibfnamefont {J.}~\bibnamefont {Tomczak}},\ and\ \bibinfo {author} {\bibfnamefont {M.}~\bibnamefont {Welling}},\ }\bibfield  {title} {\bibinfo {title} {Attention-based deep multiple instance learning},\ }in\ \href@noop {} {\emph {\bibinfo {booktitle} {Proceedings of the 35th International Conference on Machine Learning}}},\ \bibinfo {series} {Proceedings of Machine Learning Research}, Vol.~\bibinfo {volume} {80}\ (\bibinfo {year} {2018})\ pp.\ \bibinfo {pages} {2127--2136}\BibitemShut {NoStop}%
\bibitem [{\citenamefont {Loshchilov}\ and\ \citenamefont {Hutter}(2019)}]{Loshchilov2019ICLR}%
  \BibitemOpen
  \bibfield  {author} {\bibinfo {author} {\bibfnamefont {I.}~\bibnamefont {Loshchilov}}\ and\ \bibinfo {author} {\bibfnamefont {F.}~\bibnamefont {Hutter}},\ }\bibfield  {title} {\bibinfo {title} {Decoupled weight decay regularization},\ }in\ \href {https://openreview.net/forum?id=Bkg6RiCqY7} {\emph {\bibinfo {booktitle} {International Conference on Learning Representations (ICLR)}}}\ (\bibinfo {year} {2019})\BibitemShut {NoStop}%
\bibitem [{\citenamefont {Holm}(1979)}]{Holm1979ScandJStat}%
  \BibitemOpen
  \bibfield  {author} {\bibinfo {author} {\bibfnamefont {S.}~\bibnamefont {Holm}},\ }\bibfield  {title} {\bibinfo {title} {A simple sequentially rejective multiple test procedure},\ }\href@noop {} {\bibfield  {journal} {\bibinfo  {journal} {Scand. J. Stat.}\ }\textbf {\bibinfo {volume} {6}},\ \bibinfo {pages} {65} (\bibinfo {year} {1979})}\BibitemShut {NoStop}%
\end{thebibliography}%

\newpage

\appendix

\section{Training Protocol, Hyperparameter Optimization, and Ablation Studies}
\label{app:ml_training}

This appendix provides additional details of the
dataset construction, optimization procedure,
hyperparameter selection, and ablation studies
underlying the graph neural network introduced
in Sec.~\ref{sec:ml_method}. We distinguish the
all-depth model used in the circuit-cutting experiments
from an earlier model trained exclusively on
depth-15 circuits. The latter serves as a reference
for examining the effects of training procedures
and architectural choices.

\subsection{Dataset construction and evaluation protocol}
\label{app:dataset}

\subsubsection{Dataset generation and partitioning}

The dataset consists of four-qubit random circuits
with depths $D=1,\ldots,15$, generated using the
procedure described in
Sec.~\ref{subsec:training_data}.
For each depth, we generate $400\,000$ unique
dense circuits and $200\,000$ unique sparse
circuits, giving approximately nine million
circuits across all depths.

The dense circuits at $D=15$ were retained
from the earlier depth-specific study. At
depths $1$--$14$, existing circuits generated
using the same procedure were retained, and
additional circuits were generated to complete
the dataset. Sparse circuits were deduplicated
against the dense set and against other sparse
circuits using a canonical representation of
the gate sequence. This is particularly important
at small depths, where the finite number of
possible gate combinations can produce
duplicate circuits.

Each circuit is assigned a unique identity
through a SHA-256 hash of its canonical gate
sequence, including gate types, wire labels,
and rotation angles. The dataset is partitioned
deterministically into training, validation,
and test subsets with nominal proportions
$80\%$, $10\%$, and $10\%$, respectively.

Specifically, a hash value $u(C)\in[0,1)$
is constructed from the circuit identity
and a fixed salt. A circuit is assigned
according to
\begin{equation}
\begin{aligned}
    \text{training:}&\quad u(C)<0.8,\\
    \text{validation:}&\quad 0.8\leq u(C)<0.9,\\
    \text{test:}&\quad u(C)\geq0.9.
\end{aligned}
\label{eq:app_data_split}
\end{equation}

An additional precaution is required because
qubit relabeling is used during training.
For rotation-free circuits at depths
$D=1,\ldots,14$, the split identity is
defined using the entire orbit under the
permutation group $S_4$:
\begin{equation}
    k(C)
    =
    \min_{\pi\in S_4}h(\pi C),
    \label{eq:app_orbit_hash}
\end{equation}
where $h$ is the canonical circuit hash.
This ensures that permutation-equivalent
circuits cannot appear in different data
partitions.

The need for this precaution was identified
by examining a smaller copy of the dataset.
Without orbit-based splitting, $4.1\%$ of
the depth-2 and $1.2\%$ of the depth-3
validation and test circuits had a
permutation-equivalent counterpart in the
training set. The corresponding fraction
was approximately $0.01\%$ at depth 4
and zero at depths $D\geq5$ in that
sample.

The resulting dataset contains
\begin{equation}
\begin{aligned}
N_{\mathrm{train}} &= 7\,198\,448,\\
N_{\mathrm{val}}   &= 900\,913,\\
N_{\mathrm{test}}  &= 900\,639.
\end{aligned}
\label{eq:app_dataset_sizes}
\end{equation}
Of the training examples, $4\,798\,602$
are dense circuits and $2\,399\,846$
are sparse circuits.

\subsubsection{Validation and test sets}
\begin{table}[t]
\caption{
Hyperparameter search space and selected
configuration of the all-depth GNN.
The selected model contains $1\,710\,687$
trainable parameters. The search used
$2700$ seconds per trial; final training
used a 20-hour budget. Qubit-relabeling augmentation
(\cref{subsubsec:symmetry}) is used in every trial.
}
\label{tab:app_hyperparameters}
\centering
\footnotesize
\setlength{\tabcolsep}{2.5pt}
\begin{tabular}{lll}
\hline\hline
Hyperparameter & Search values & Selected value\\
\hline
GINE layers
    & $4,6,8$ & $4$\\
Hidden dimension
    & $192,256,384$ & $256$\\
Output-head blocks
    & $2,3$ & $3$\\
Embedding dimension
    & $32,64$ & $64$\\
Edge dilation
    & $1,2,3$ & $2$\\
Batch size
    & $128,256$ & $256$\\
Peak learning rate
    & $10^{-4},3\times10^{-4},10^{-3}$
    & $10^{-3}$\\
Learning-rate schedule
    & Constant, cosine & Cosine\\
Warm-up fraction
    & $0,0.05,0.1$ & $0.1$\\
Min.\ learning-rate factor
    & $0.01,0.03,0.1$ & $0.1$\\
Weight decay
    & $10^{-5},10^{-4},10^{-3}$
    & $10^{-3}$\\
Dropout
    & $0,0.05,0.1,0.2$ & $0$\\
EMA decay
    & $0,0.995,0.999,0.9995$
    & $0.999$\\
Normalized loss weight
    & $0.3,1,3$ & $1$\\
MSE loss weight
    & $0.03,0.1,0.3$ & $0.1$\\
$\ell_1$ loss weight
    & $0,0.03$ & $0$\\
Depth-sampling exponent
    & $0,1,2$ & $1$\\
\hline\hline
\end{tabular}
\end{table}

Two validation sets are used for different
purposes. A fixed subset containing
$2000$ circuits at each depth, or
$30\,000$ circuits in total, is used
during hyperparameter optimization and
checkpoint selection. This subset was
sampled once using seed 424242 and was
shared by all search trials and training
runs.

The full validation set of $900\,913$
circuits is used to compare the final
checkpoints obtained from four independent
training seeds.

The test set of $900\,639$ circuits is
reserved for the final evaluation after
model selection. One qualification is that
its dense depth-15 component, containing
$40\,303$ circuits, had also been evaluated
during development of the earlier
depth-specific model.

A separate frozen validation set of
$19\,049$ dense depth-15 circuits is
used for the ablation studies described
below. 

\subsection{Optimization procedure}
\label{app:optimization}

\subsubsection{Training objective}

The neural network predicts the correction
$\widehat{\Delta}_C$ to the noisy output
distribution. For a circuit $C_i$, the
prediction error is
\begin{equation}
    \boldsymbol{\xi}_i
    =
    \widehat{\Delta}_{C_i}
    -
    \Delta_{C_i}.
\end{equation}

During hyperparameter optimization, we
consider the general loss
\begin{equation}
\begin{split}
\mathcal{L}
={}&
\frac{w_{\mathrm{norm}}}{B}
\sum_{i=1}^{B}
\frac{\|\boldsymbol{\xi}_i\|_2^2}
{\max(\|\Delta_{C_i}\|_2^2,10^{-8})}
\\
&+
\frac{w_{\mathrm{mse}}}{16B}
\sum_{i=1}^{B}
\|\boldsymbol{\xi}_i\|_2^2
\\
&+
\frac{w_{\ell_1}}{16B}
\sum_{i=1}^{B}
\|\boldsymbol{\xi}_i\|_1,
\end{split}
\label{eq:app_loss}
\end{equation}
where $B$ is the batch size.

The normalized squared-error term gives
greater relative importance to circuits
whose ideal-to-noisy correction is small.
The additional mean-squared-error and
$\ell_1$ terms allow the optimization
to balance relative and absolute
prediction errors.

The selected all-depth configuration uses
\begin{equation}
    (w_{\mathrm{norm}},
     w_{\mathrm{mse}},
     w_{\ell_1})
    =
    (1,0.1,0),
    \label{eq:app_loss_weights}
\end{equation}
reducing Eq.~\eqref{eq:app_loss} to the
objective used in the main text.

\subsubsection{Learning-rate schedule and weight averaging}

The network is optimized using AdamW
with momentum parameters
$(\beta_1,\beta_2)=(0.9,0.999)$,
$\epsilon=10^{-8}$, and gradient-norm
clipping at unity.

We use a warm-up and cosine-annealing
learning-rate schedule whose progress
is determined by the elapsed training
time relative to a fixed computational
budget. Let
\begin{equation}
    s=\frac{t}{T},
\end{equation}
where $t$ is the elapsed wall-clock
time and $T$ is the allotted budget.
The learning rate is
\begin{equation}
\frac{\eta(s)}{\eta_0}=
\begin{cases}
\max\left(\dfrac{s}{w},10^{-3}\right),
    & 0\leq s<w,\\[6pt]
m+\tfrac{1-m}{2}
\left[
1+\cos\left(
\pi\tfrac{s-w}{1-w}
\right)
\right],
    & w\leq s\leq1.
\end{cases}
\label{eq:app_lr_schedule}
\end{equation}
Here, $\eta_0$ is the peak learning
rate, $w$ is the warm-up fraction,
and $m$ is the minimum learning-rate
factor.

\begin{table*}[t]
\caption{Structured variational circuits using wire cutting. Values are
EV-RMSE over all $2^n-1$ Pauli-$Z$ strings. GNN, CDR, and ZNE (linear
and Richardson extrapolation) use exact fragment distributions; PEC is
statistically emulated at $10^6$ shots per fragment configuration.}
\label{tab:app_structured_variational}
\centering\footnotesize
\begin{tabular}{lrrrrrrrrrr}
\toprule
 & & & & & & & \multicolumn{2}{c}{ZNE} & & \\
\cmidrule(lr){8-9}
Family & Qubits & Depth & Wire cuts & $N_{\mathrm{var}}$ & Cut noisy & GNN & Linear & Richardson & CDR & PEC \\
\midrule
MPS staircase & 6 & 11 & 1 & 9 & 0.186 & 0.024 & 0.103 & 0.086 & 0.030 & 0.005 \\
MPS staircase & 8 & 15 & 2 & 27 & 0.118 & 0.014 & 0.064 & 0.094 & 0.013 & 0.003 \\
MPS staircase & 12 & 23 & 3 & 45 & 0.056 & 0.009 & 0.035 & 0.030 & 0.008 & 0.002 \\
QCNN & 6 & 16 & 2 & 42 & 0.196 & 0.030 & 0.144 & 0.116 & 0.046 & 0.018 \\
QCNN & 8 & 16 & 2 & 42 & 0.165 & 0.026 & 0.130 & 0.111 & 0.027 & 0.015 \\
QCNN & 12 & 21 & 3 & 225 & 0.099 & 0.020 & 0.082 & 0.068 & 0.028 & 0.010 \\
Givens staircase & 6 & 31 & 2 & 27 & 0.692 & 0.045 & 0.549 & 0.329 & 0.041 & 0.003 \\
Givens staircase & 8 & 43 & 4 & 63 & 0.858 & 0.073 & 0.681 & 0.389 & 0.003 & 0.004 \\
Givens staircase & 12 & 67 & 8 & 135 & 0.637 & 0.104 & 0.551 & 0.337 & 0.014 & 0.008 \\
\bottomrule
\end{tabular}
\end{table*}

For the selected model,
\begin{equation}
    \eta_0=10^{-3},\qquad
    w=0.1,\qquad m=0.1,
\end{equation}
with $T=72\,000$ seconds for each
final training run.

This schedule ensures that the learning
rate is annealed over the actual
computational budget. In contrast, a
schedule defined over a much larger
number of epochs may terminate before
substantial annealing occurs.

We also maintain an exponential moving
average (EMA) of the model parameters,
\begin{equation}
    \overline{\boldsymbol{\theta}}_t
    =
    \omega_t
    \overline{\boldsymbol{\theta}}_{t-1}
    +(1-\omega_t)\boldsymbol{\theta}_t,
    \label{eq:app_ema}
\end{equation}
where
\begin{equation}
    \omega_t=
    \min\left\{
    0.999,\frac{1+t}{10+t}
    \right\}.
\end{equation}
The averaged parameters are used
for validation and checkpoint selection.

\subsubsection{Depth-weighted sampling}

To control the relative frequency of
circuits at different depths during
training, we consider a family of
sampling distributions
\begin{equation}
    P_q(D)
    =
    \frac{D^q}
    {\sum_{k=1}^{15}k^q},
    \qquad
    q\in\{0,1,2\}.
    \label{eq:app_depth_sampling}
\end{equation}
Within a selected depth, circuits
are sampled uniformly with replacement
from the combined dense and sparse
training data.

The final model uses $q=1$, giving
\begin{equation}
    P_1(D)=\frac{D}{120}.
\end{equation}
Thus, depths 1 and 15 account for
approximately $0.83\%$ and $12.5\%$
of the training samples, respectively.

This choice gives deeper circuits
greater representation during training.
However, as discussed below, the
hyperparameter search did not isolate
a statistically resolved benefit
from depth-weighted sampling alone.

\subsection{Hyperparameter optimization and model selection}
\label{app:hyperparameter_search}

A hyperparameter search was performed
to identify a configuration suitable
for mitigating circuits over all
15 training depths.
We ran it as an automated,
autoresearch-style
loop~\cite{Karpathy2026autoresearch}:
the driver first trained a baseline
configuration, then a short list of
hand-chosen candidates, and then
random draws from the search
space~\cite{Bergstra2012JMLR},
skipping configurations that had
already been run. Every trial had the
same time budget and was ranked by
the same validation score. No language
model was involved in proposing
or selecting configurations.

The search consisted of 64 trials,
each trained using one random seed
and a computational budget of
$45$ minutes on an NVIDIA T4 GPU.
The trials used $40\,000$ training
circuits per depth and the same fixed
validation subset.

The search included model depth,
hidden width, edge dilation, embedding
size, output-head structure, batch size,
optimization parameters, loss weights,
and depth-sampling exponent.

Table~\ref{tab:app_hyperparameters}
summarizes the search space and the
selected configuration.

\begin{figure*}[t]
\centering
\includegraphics{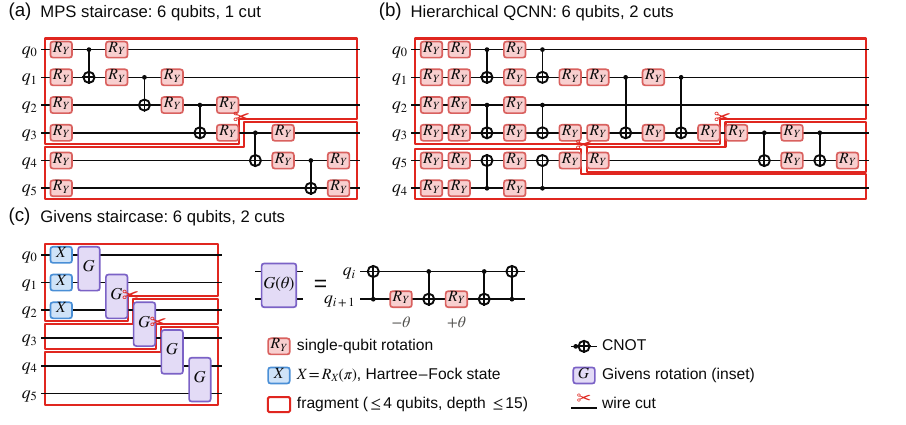}
\caption{Six-qubit instances of the three structured variational
families of \cref{app:subsec:wire_cut_variational}, drawn with the gate
symbols of \cref{fig:main}, together with their cut plans.
(a)~MPS staircase (depth 11): $R_Y$ rotations on every qubit, followed
by a CNOT staircase with an $R_Y$ rotation on both qubits after each
CNOT; one wire cut. (b)~Hierarchical QCNN (depth 16): after an $R_Y$
encoding layer, convolution ($R_Y$ rotations and CNOTs) and pooling
layers reduce the groups $\{q_0,\dots,q_3\}$ and $\{q_4,q_5\}$ to $q_3$
and $q_5$, which are then merged into the readout qubit $q_5$; two wire
cuts. Wires $q_4$ and $q_5$ are drawn in swapped order so that each
fragment occupies adjacent rows. (c)~Particle-conserving
Givens-rotation staircase (depth 31): $X$ gates prepare the
three-particle Hartree--Fock state, and Givens rotations $G(\theta)$ act
on neighboring qubit pairs; two wire cuts. Each $G(\theta)$ is drawn as
a single two-qubit block, and the inset gives its decomposition into
four CNOTs and two $R_Y$ rotations. Red outlines mark the fragments of
the cut plan (at most four qubits and depth 15 each), and scissors mark
the wire cuts. The randomly drawn rotation angles are omitted.}
\label{fig:app_ansatz_circuits}
\end{figure*}

\subsubsection{Model-selection criterion}

Candidate configurations are ranked
using a validation score that combines
expectation-value errors, distribution
errors, and the fraction of circuits
improved by mitigation.

Let $R_D$ denote the normalized
expectation-value MSE ratio at depth
$D$, defined using the pooled errors
over circuits at that depth.
We further define
\begin{equation}
    h_D=
    \frac{1}{N_D}
    \sum_{i=1}^{N_D}
    \frac{1}{2}
    \left\|
        p_{i}^{\mathrm{mit}}
        -
        p_{i}^{\mathrm{id}}
    \right\|_1,
    \label{eq:app_half_l1}
\end{equation}
and
\begin{equation}
    P_D=
    \frac{N_D^{\mathrm{improved}}}{N_D},
    \label{eq:app_improvement_fraction}
\end{equation}
where $N_D^{\mathrm{improved}}$ is the
number of circuits whose expectation-value
MSE is reduced by mitigation.

Writing
\begin{equation}
\begin{aligned}
    \overline{R}
        &=\frac{1}{15}\sum_{D=1}^{15}R_D,\\
    \overline{h}
        &=\frac{1}{15}\sum_{D=1}^{15}h_D,\\
    R_{\max}
        &=\max_D R_D,
\end{aligned}
\end{equation}
the selection score is
\begin{widetext}
\begin{equation}
\mathcal{S}
=
\ln(1+\overline{R})
+
\frac{1}{2}\ln(1+R_{\max})
+
0.2\ln(1+\overline{h})
+
2\max\left(0,0.95-\min_D P_D\right).
\label{eq:app_selection_score}
\end{equation}
\end{widetext}
Lower scores indicate better
validation performance.

This criterion penalizes both the
mean error across depths and the
worst-depth error. However, because
$R_D$ increases substantially with
depth, the score remains dominated
by the deeper circuits. For the
selected all-depth model, the
$R_{\max}$ term accounts for
approximately $59\%$ of the
full-validation score, whereas
depths 1--5 together contribute
approximately $1\%$.

Thus, although the criterion includes
an average across all depths, it
effectively places greater importance
on mitigation performance for deeper
circuits.

\subsubsection{Search results}

\begin{figure}[t]
\centering
\includegraphics{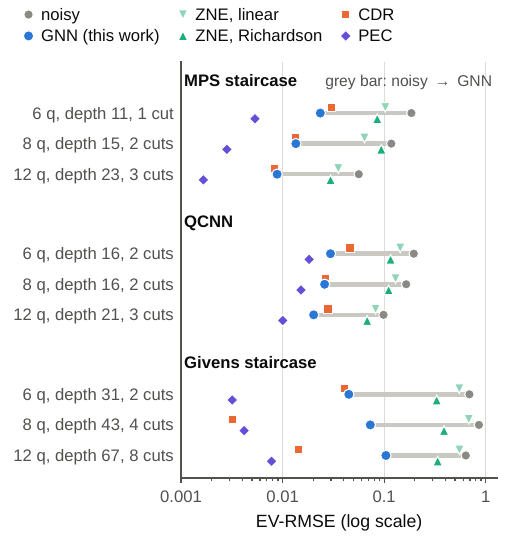}
\caption{EV-RMSE (all Pauli-$Z$ strings, logarithmic scale) of the nine
structured variational circuits per mitigation method; row labels give
the number of qubits, the circuit depth, and the number of wire cuts.
The gray bar joins the cut-noisy and GNN values, and small vertical
offsets only separate overlapping markers. ZNE is shown with linear and
Richardson extrapolation. PEC is emulated with $10^6$ shots per fragment
configuration; the GNN, ZNE, and CDR use exact fragment distributions.
\Cref{tab:app_structured_variational} lists the values.}
\label{fig:app_variational_methods}
\end{figure}
The 64-trial search combined previously
selected configurations with random
hyperparameter combinations. Trials
0--8 included reference configurations
from the earlier depth-15 search and
different depth-sampling choices.
Trials 9--63 were random draws from
the specified hyperparameter space.

The validation scores ranged from
$0.0758$ to $0.3288$, with a median
of approximately $0.1052$.
The best configuration, trial 58,
achieved a score of $0.0758$.

For the best short-budget configuration,
the mean normalized error ratio
across depths was $0.0270$, compared
with $0.0400$ for trial 0. At
depth 15, the corresponding values
were $0.0873$ and $0.1130$.

The chosen configuration also processed
approximately $9770$ samples per second,
compared with $5565$ samples per second
for the reference configuration.
Consequently, the search selects
configurations based on their accuracy
within a fixed computational budget,
rather than at a fixed number of
training samples.

The three best configurations differed
in score by less than $0.0007$ and
were evaluated with single seeds.
Their relative ordering was therefore
not statistically resolved. Trial 58
was selected according to the
predefined minimum-score criterion.

The search also did not establish a
clear advantage for a particular
depth-sampling exponent. Among the
55 random configurations, the median
scores for $q=0,1,2$ were
approximately $0.1123$, $0.1112$,
and $0.1132$, respectively.
The choice $q=1$ was retained
because it was used by the selected
configuration.

\subsection{Ablation studies}
\label{app:ablations}

We performed several ablation studies
to examine the effects of training
choices, dataset size, and circuit
representation. Unless otherwise
specified, these experiments used
dense depth-15 circuits and the
frozen validation set containing
$19\,049$ circuits.

The ablations were carried out
primarily during the development
of the earlier depth-specific model.
Their results motivate the final
training procedure but do not
constitute independent ablations
of every component of the all-depth
model.

\subsubsection{Training-procedure ablations}

We first consider a baseline
configuration consisting of a
four-layer GINE, without
qubit-relabeling augmentation or
EMA, and with a learning-rate
schedule that does not anneal
substantially within the available
training budget.

Using $40\,000$ training circuits,
the baseline achieves
\begin{equation}
    R_{\mathrm{EV}}
    =
    0.190558\pm0.002772,
    \label{eq:app_baseline_ablation}
\end{equation}
where the uncertainty is the
standard deviation across four
training seeds.

Qubit-relabeling augmentation produces
the largest individual improvement.
Its relative reduction in
$R_{\mathrm{EV}}$ is $25.5\%$,
and the 95\% confidence interval
for the seed-paired absolute change
is $[-0.0530,-0.0443]$.

EMA weight averaging produces a
smaller but resolved improvement
of $5.9\%$, with a corresponding
interval of $[-0.0143,-0.0082]$.
Both improvements are observed
for all four training seeds.

By contrast, the individual
improvements from the annealed
learning-rate schedule and stronger
regularization are not statistically
resolved in these experiments.

Combining all four modifications
reduces the error ratio by
$27.3\%$ and shifts the best
training epoch from approximately
40--60 to 970--990 within the
1000-epoch ablation budget.

These observations support the
use of symmetry augmentation and
weight averaging in the final
model. They do not establish
that each component contributes
independently or by the same
amount in the all-depth setting.

\subsubsection{Dependence on training dataset size}

To examine the effect of training
data volume, we train models on
different numbers $N$ of dense
depth-15 circuits.

\begin{figure}[t]
\centering
\includegraphics{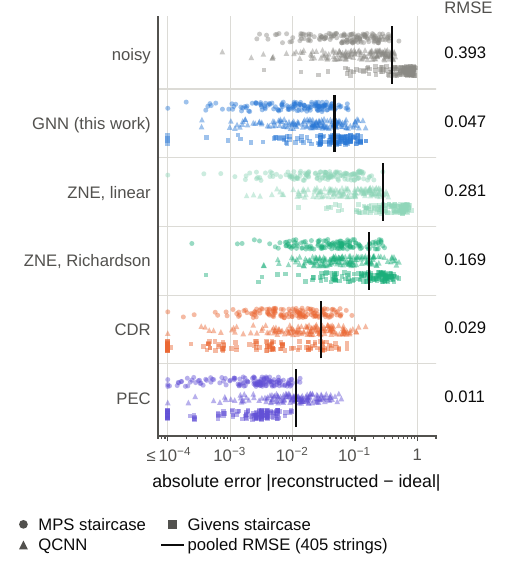}
\caption{Absolute error $|\text{reconstructed}-\text{ideal}|$ of every
one- and two-body Pauli-$Z$ expectation value (405 strings) of the nine
MPS, QCNN, and Givens circuits, per method; marker shape and vertical
offset give the circuit family. The black bar and the number on the
right give the RMSE pooled over all 405 strings, which differs from the
per-circuit EV-RMSE over all $Z$ strings in
\cref{tab:app_structured_variational}. Errors below $10^{-4}$ (CDR 35,
PEC 21, GNN 5, and linear ZNE 1 string) are drawn at $10^{-4}$. PEC is
emulated with $10^6$ shots per fragment configuration.}
\label{fig:app_variational_scatter}
\end{figure}
With the improved training recipe,
the normalized error ratio decreases
from
\begin{equation}
    R_{\mathrm{EV}}
    =
    0.183455\pm0.008352
    \qquad (N=8000)
\end{equation}
to
\begin{equation}
    R_{\mathrm{EV}}
    =
    0.100881\pm0.000993
    \qquad (N=160\,000).
\end{equation}
The quoted uncertainties are
standard deviations over
training seeds.

Over the investigated range,
the mean error ratios are
approximately described by
the empirical power law
\begin{equation}
    R_{\mathrm{EV}}(N)
    \simeq
    1.1048\,N^{-0.2002}.
    \label{eq:app_data_scaling}
\end{equation}

The fitted curve decreases
without a visibly established
plateau over the sampled
training sizes. Nevertheless,
the available data do not
exclude a nonzero asymptotic
error floor.

At the same training sizes,
the improved recipe gives
error ratios approximately
$30.5$--$36.1\%$ below those
of the baseline recipe.

The two recipes also differ
in architecture, so their
performance difference
cannot be attributed solely
to the training modifications.
A separate architecture
comparison with $16\,000$
training circuits found that
the best tested architecture
improved the error ratio by
only $2.4\%$ relative to
a four-layer GINE with
hidden dimension 256 under
the improved recipe.

The data-scaling results
therefore suggest that the
training procedure has a
substantial influence on
accuracy, although its
individual contributions
cannot be completely
separated in this comparison.

\subsubsection{Message-passing depth and receptive field}
\label{app:reach}

The message-passing depth
determines how much information
can propagate through the
circuit graph before global
pooling.

For a measured wire $w$,
let $\mathcal{C}(w)$ denote
its backward causal light cone,
defined as the set of gates
that can influence its final
measurement. Let $V_w$
denote the gates acting on
wire $w$.

We quantify the fraction
of the causal light cone
reachable through $L$
message-passing layers by
\begin{equation}
\rho_L(w)
=
\frac{
\left|
\left\{
j\in\mathcal{C}(w):
\min_{i\in V_w}
\operatorname{dist}(i,j)\leq L
\right\}
\right|
}{
|\mathcal{C}(w)|
}.
\label{eq:app_lightcone_reach}
\end{equation}
Here, $\operatorname{dist}(i,j)$
is the shortest-path distance
between two nodes in the
encoded circuit graph.

For the selected architecture,
which uses four GINE layers
and edge dilation two,
the reachable fraction at
depth 15 is approximately
$99.7\%$ for dense circuits
and $99.1\%$ for sparse
circuits, when reach is
measured from all gates
on the observed wire.

The complete light cone
is reached for approximately
$95.2\%$ of dense-circuit
wires and $92.6\%$ of
sparse-circuit wires.

These values were obtained
from $1000$ newly generated
circuits per depth in each
circuit family.

The reachable fraction is
smaller when evaluated from
the final gate on a wire
alone. Moreover, gates
outside the message-passing
receptive field may still
contribute to the final
graph-level representation
through attention pooling.

We also investigated models
with different numbers of
GINE layers and edge-dilation
distances. In a separate depth-15
ablation, increasing the
number of GINE layers from
four to six reduced
$R_{\mathrm{EV}}$ by
approximately $0.0039$,
with a 95\% interval
$[-0.0063,-0.0015]$
for the difference.

However, a four-layer model
with an additional output-head
block and comparable parameter
count achieved a similar
improvement, while eight
layers did not show a
resolved advantage over
four.

The experiments therefore
do not establish that
increasing message-passing
depth alone improves
mitigation performance.
The four-layer architecture
was retained as a favorable
balance of accuracy and
training throughput.

\begin{table*}[t]
\caption{Dependence of GNN EV-RMSE on shots per executed fragment circuit
(mean $\pm$ standard deviation across five multinomial seeds). The last
two rows give the GNN and the unmitigated EV-RMSE obtained from exact
(infinite-shot) fragment probabilities; for the uncut four-qubit circuit,
the unmitigated value is that of the uncut noisy circuit.}
\label{tab:app_shot_scan}
\centering\footnotesize
\begin{tabular}{rrrrrr}
\toprule
Shots & MPS 8q & QCNN 8q & Givens 8q & Random 8q & Uncut random 4q \\
\midrule
256 & $0.0561\pm0.0174$ & $0.0911\pm0.0061$ & $0.1205\pm0.0348$ & $0.1036\pm0.0208$ & $0.2192\pm0.0658$ \\
1024 & $0.0315\pm0.0060$ & $0.0468\pm0.0060$ & $0.0996\pm0.0448$ & $0.0606\pm0.0082$ & $0.1238\pm0.0211$ \\
4096 & $0.0189\pm0.0020$ & $0.0348\pm0.0032$ & $0.0812\pm0.0076$ & $0.0423\pm0.0097$ & $0.0755\pm0.0092$ \\
16384 & $0.0150\pm0.0013$ & $0.0284\pm0.0015$ & $0.0742\pm0.0150$ & $0.0271\pm0.0022$ & $0.0550\pm0.0077$ \\
65536 & $0.0146\pm0.0011$ & $0.0270\pm0.0011$ & $0.0762\pm0.0038$ & $0.0245\pm0.0012$ & $0.0430\pm0.0041$ \\
262144 & $0.0135\pm0.0003$ & $0.0257\pm0.0005$ & $0.0716\pm0.0020$ & $0.0218\pm0.0007$ & $0.0423\pm0.0050$ \\
1048576 & $0.0134\pm0.0001$ & $0.0262\pm0.0002$ & $0.0733\pm0.0008$ & $0.0225\pm0.0001$ & $0.0408\pm0.0007$ \\
GNN, exact probabilities & $0.0135$ & $0.0259$ & $0.0731$ & $0.0221$ & $0.0398$ \\
Noisy, exact probabilities & $0.1176$ & $0.1648$ & $0.8581$ & $0.0606$ & $0.1396$ \\
\bottomrule
\end{tabular}
\end{table*}

\begin{table*}[t]
\caption{Complete equal-per-configuration-shot comparison grid. Each
entry is the mean EV-RMSE [Eq.~\eqref{eq:ev_rmse}, all $2^n-1$ Pauli-$Z$
strings] over the two random target circuits of a cell; the lowest
mitigated value in each row is set in bold. Every executed fragment
configuration receives $10^5$ shots. ZNE uses Richardson extrapolation
over three noise scales, and CDR requires 20 additional near-Clifford
configurations per target fragment configuration. Scheme W uses
wire-only chains; G combines spatial gate cuts with temporal wire cuts.
$\gamma_{\mathrm{PEC}}$ is the PEC one-norm of the largest fragment of
each of the two circuits (one value where they coincide).}
\label{tab:app_grid_full}
\centering\footnotesize
\setlength{\tabcolsep}{5pt}
\begin{tabular}{lrrrrrrrrrr}
\toprule
Scheme & Qubits & Depth & Wire+gate cuts & $N_{\mathrm{var}}$ & $\gamma_{\mathrm{PEC}}$ & Cut noisy & GNN & ZNE & CDR & PEC \\
\midrule
W & 6 & 15 & $1+0$ & 9 & 278/594 & 0.2904 & 0.0669 & 0.2002 & \textbf{0.0336} & 0.3465 \\
W & 8 & 15 & $2+0$ & 27 & 224/478 & 0.1396 & 0.0416 & 0.0946 & \textbf{0.0316} & 0.1104 \\
W & 12 & 15 & $3+0$ & 45 & 429/533 & 0.0790 & \textbf{0.0200} & 0.0651 & 0.0252 & 0.0997 \\
W & 6 & 30 & $2+0$ & 27 & 661 & 0.1953 & \textbf{0.0432} & 0.1455 & 0.0561 & 0.4139 \\
W & 8 & 30 & $2+0$ & 27 & 661 & 0.0774 & \textbf{0.0260} & 0.0607 & 0.0411 & 0.2241 \\
W & 12 & 30 & $3+0$ & 45 & 429/594 & 0.0452 & \textbf{0.0089} & 0.0377 & 0.0161 & 0.0972 \\
W & 6 & 45 & $3+0$ & 45 & 130 & 0.1095 & 0.0574 & 0.0730 & \textbf{0.0463} & 0.0864 \\
W & 8 & 45 & $3+0$ & 45 & 661 & 0.1251 & \textbf{0.0367} & 0.0954 & 0.0426 & 0.2261 \\
W & 12 & 45 & $3+0$ & 45 & 661 & 0.0267 & \textbf{0.0089} & 0.0221 & 0.0135 & 0.1350 \\
G & 6 & 15 & $0+1$ & 12 & 180 & 0.1557 & 0.0344 & 0.1088 & \textbf{0.0270} & 0.1737 \\
G & 8 & 15 & $0+2$ & 48 & 180 & 0.1103 & 0.0412 & 0.0745 & \textbf{0.0355} & 0.1260 \\
G & 12 & 15 & $0+4$ & 120 & 180 & 0.0551 & 0.0133 & 0.0410 & \textbf{0.0101} & 0.0393 \\
G & 8 & 30 & $4+3$ & 324 & 32 & 0.1129 & \textbf{0.0179} & 0.0711 & 0.0181 & 0.0242 \\
G & 12 & 30 & $6+6$ & 1296 & 36/49 & 0.0368 & \textbf{0.0074} & 0.0275 & 0.0094 & 0.0107 \\
G & 10 & 45 & $6+4$ & 540 & 36 & 0.0454 & 0.0115 & 0.0359 & \textbf{0.0100} & 0.0139 \\
\bottomrule
\end{tabular}
\end{table*}
\subsection{Final training and reproducibility}
\label{app:final_training}

Following hyperparameter
selection, the chosen
configuration was trained
independently with four
random seeds using the
complete training dataset.

Each training run used
a computational budget
of $72\,000$ seconds
(20 hours) on an NVIDIA
T4 GPU. Training proceeded
without early stopping,
with the best checkpoint
selected from validation
evaluations performed
every 25 epochs.

The training procedure used
400 optimization steps per
epoch. A final run completed
$2\,783\,209$ optimization
steps, corresponding to
approximately $7.13\times10^8$
sample presentations with
batch size 256.

For seed 2, the selected
checkpoint occurred at
epoch 6950, after
$71\,928$ seconds of
training. Its validation-subset
macro error ratio was
approximately $0.0146$,
compared with $0.0270$
for the selected
short-budget configuration
before final training.

The improvement between
the search trial and the
final model reflects the
combined use of a larger
dataset and a longer
training budget. Their
individual effects were
not isolated.

The four final checkpoints
were compared using the
selection score in
Eq.~\eqref{eq:app_selection_score}
on the complete validation
dataset. The scores for
seeds 0--3 were
\begin{equation}
\begin{aligned}
    \mathcal{S}_0 &=0.0489480,\\
    \mathcal{S}_1 &=0.0481606,\\
    \mathcal{S}_2 &=0.0478981,\\
    \mathcal{S}_3 &=0.0479769.
\end{aligned}
\label{eq:app_seed_scores}
\end{equation}

Seed 2 achieved the lowest
score and was selected as
the final model. The
difference between seeds
2 and 3 was approximately
$0.16\%$ of their score
and should be regarded as
an effective tie rather
than a substantial
performance difference.

The selected checkpoint,
denoted
\texttt{gate11\_alldepth\_seed\_2},
contains $1\,710\,687$
parameters. Of these,
$1\,114\,961$ belong to
the active four-qubit
inference path; the
remaining parameters
include output heads
for other qubit counts
that were not trained
on the four-qubit dataset.

\begin{figure}[t]
\centering
\includegraphics{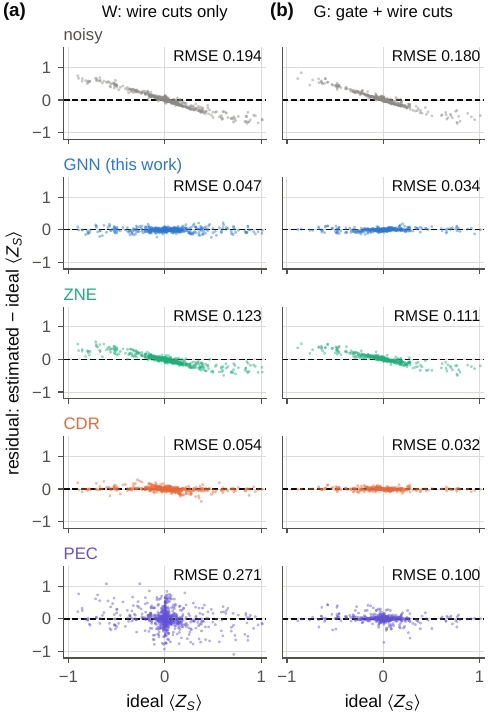}
\caption{Residual (estimated $-$ ideal) of every one- and two-body
$Z$-string expectation value against its ideal value for the comparison
grid of \cref{fig:results_methods} ($10^5$ shots per executed fragment
configuration): (a)~scheme W, 18 circuits, 810 strings; (b)~scheme G,
12 circuits, 608 strings. Rows show the cut-noisy, GNN, ZNE
(Richardson), CDR, and PEC estimates, and the dashed line marks zero
residual. The number in each panel is the RMSE over the plotted
strings; it differs from the EV-RMSE of \cref{fig:results_methods} and
\cref{tab:app_grid_full}, which runs over all $2^n-1$ $Z$ strings and is
averaged per circuit.}
\label{fig:app_grid_scatter}
\end{figure}

The overall training and
hyperparameter search used
approximately 48.5 GPU-hours
for the search and
80 GPU-hours for the
four final training runs,
all on NVIDIA T4 GPUs.

The wall-clock-based
learning-rate schedule
depends on hardware
throughput. For a
step-based reproduction
of the selected training
run, the schedule can
instead be parameterized
using the normalized step
count
\begin{equation}
    s=
    \frac{t_{\mathrm{step}}}
    {2\,783\,209},
\end{equation}
where $t_{\mathrm{step}}$
is the number
of completed optimization steps.

The numerical scalability
experiments in the main
text use the selected
all-depth model without
further parameter updates.
Their purpose is to
investigate transfer
across circuit sizes
and depths, rather than
to optimize the
neural network anew
for each target circuit.

\section{Local Mitigation, Reconstruction of Global Observables, and Residual Mitigation Error}
\label{app:reconstruction}

Once a valid cut plan is
specified, each fragment
is executed under the
preparation, measurement,
and gate-decomposition
configurations required
by its incident cuts, and
the resulting probabilities
are combined into the
fragment tensor $R_j$ of
\cref{eq:fragment_tensor}.

Each tensor entry is obtained
from a signed combination
of the probabilities of
the corresponding fragment
variants. Schematically,
\begin{equation}
\begin{split}
    R_j(\boldsymbol{\lambda}_j;\mathbf{y}_j)
    =
    \sum_{\alpha\in\mathcal{A}_j
    (\boldsymbol{\lambda}_j)}
    \sum_{\mathbf{b}_j}
    \omega_{j,\alpha}(\mathbf{b}_j)\,
    p_{j,\alpha}
    (\mathbf{y}_j,\mathbf{b}_j),
\end{split}
\label{eq:fragment_tensor_construction}
\end{equation}
where $\boldsymbol{\lambda}_j$
collects the cut indices,
$\mathcal{A}_j$ specifies
the relevant local configurations,
$\mathbf{b}_j$ denotes internal
measurement outcomes, and
$\omega_{j,\alpha}$ contains
the eigenvalue signs and
local coefficients associated
with the decomposition.

Let $K_{\mathrm{w}}$ and
$K_{\mathrm{g}}$ denote the
numbers of wire and gate
cuts, respectively. The
global output distribution
is reconstructed as
\begin{widetext}
\begin{equation}
\begin{split}
    p_C(\mathbf{y})
    =
    \frac{1}{2^{K_{\mathrm{w}}}}
    \sum_{\mathbf{P}\in
    \{I,X,Y,Z\}^{K_{\mathrm{w}}}}
    \sum_{\mathbf{k}\in
    \{1,\ldots,6\}^{K_{\mathrm{g}}}}
    \left(
        \prod_{g=1}^{K_{\mathrm{g}}}
        c_{g,k_g}
    \right)
    \prod_{j=1}^{N_F}
    R_j
    \left(
        \mathbf{P}_j^{\mathrm{in}},
        \mathbf{P}_j^{\mathrm{out}},
        \mathbf{k}_j;
        \mathbf{y}_j
    \right),
\end{split}
\label{eq:cut_global_contraction}
\end{equation}
\end{widetext}
where $c_{g,k_g}$ are the
coefficients of the gate-cut
decomposition. The factor
$2^{-K_{\mathrm{w}}}$
arises from the Pauli
decomposition of the identity
channel at each wire cut.

Eq.~\eqref{eq:cut_global_contraction}
is a tensor-network contraction
with one four-valued index
per wire cut and one
six-valued index per gate cut.
When the fragment probabilities
are ideal, it reconstructs
the ideal statistics of
the uncut circuit.

\begin{table*}[t]
\caption{Fixed-total-budget comparison, $B=10^5$ shots per target
circuit, with three target circuits per cell. $N_{\mathrm{var}}$ is the
number of fragment configurations per target and $S_{\mathrm{GNN}}$ the
shots per GNN (and PEC) configuration; CDR receives about eleven times
fewer shots per configuration. Values are means over the three circuits
of the EV-RMSE [all $2^n-1$ Pauli-$Z$ strings, given to three
significant figures] and of the half-$\ell_1$ discrepancy (TV) from the
ideal distribution.}
\label{tab:app_budget_full}
\centering\footnotesize
\setlength{\tabcolsep}{4pt}
\begin{tabular}{lrrrrrrrrrrrrr}
\toprule
 & & & & & & \multicolumn{4}{c}{EV-RMSE} & \multicolumn{4}{c}{TV} \\
\cmidrule(lr){7-10}\cmidrule(lr){11-14}
Target & Qubits & Depth & Wire+gate cuts & $N_{\mathrm{var}}$ & $S_{\mathrm{GNN}}$ & Cut noisy & GNN & CDR & PEC & Cut noisy & GNN & CDR & PEC \\
\midrule
Chain & 4 & 15 & $0+0$ & 1 & 100000 & 0.218 & 0.0634 & 0.102 & 0.615 & 0.34 & 0.10 & 0.15 & 0.70 \\
Chain & 6 & 15 & $1+0$ & 9 & 11111 & 0.120 & 0.0461 & 0.0948 & 0.436 & 0.32 & 0.13 & 0.27 & 0.81 \\
Chain & 8 & 15 & $2+0$ & 27 & 3703 & 0.101 & 0.0424 & 0.0773 & 0.308 & 0.45 & 0.17 & 0.33 & 0.85 \\
Chain & 12 & 15 & $3+0$ & 45 & 2222 & 0.0302 & 0.0192 & 0.0287 & 0.373 & 0.49 & 0.29 & 0.45 & 0.99 \\
Chain & 16 & 15 & $4+0$ & 63 & 1587 & 0.0110 & 0.00816 & 0.0105 & 0.368 & 0.57 & 0.39 & 0.52 & 1.00 \\
Chain & 20 & 15 & $6+0$ & 99 & 1010 & 0.00760 & 0.00559 & 0.00728 & 0.281 & 0.70 & 0.48 & 0.64 & 1.00 \\
Chain & 24 & 15 & $7+0$ & 117 & 854 & 0.00193 & 0.00187 & 0.00181 & 0.225 & 0.71 & 0.56 & 0.60 & 1.00 \\
\addlinespace
Chain & 8 & 30 & $2+0$ & 27 & 3703 & 0.0999 & 0.0568 & 0.0841 & 0.611 & 0.50 & 0.25 & 0.37 & 0.93 \\
Chain & 8 & 45 & $3+0$ & 45 & 2222 & 0.0884 & 0.0530 & 0.0804 & 0.380 & 0.42 & 0.25 & 0.38 & 0.92 \\
Chain & 8 & 60 & $5+0$ & 81 & 1234 & 0.0833 & 0.0463 & 0.0745 & 0.217 & 0.43 & 0.21 & 0.36 & 0.74 \\
\addlinespace
Gate ladder & 6 & 15 & $0+1$ & 12 & 8333 & 0.165 & 0.0525 & 0.0747 & 0.347 & 0.46 & 0.13 & 0.18 & 0.66 \\
Gate ladder & 8 & 15 & $0+2$ & 48 & 2083 & 0.0983 & 0.0411 & 0.0625 & 0.275 & 0.47 & 0.18 & 0.29 & 0.88 \\
Gate ladder & 12 & 15 & $0+4$ & 120 & 833 & 0.0541 & 0.0288 & 0.0359 & 0.129 & 0.55 & 0.32 & 0.35 & 0.89 \\
Gate ladder & 8 & 30 & $4+3$ & 324 & 308 & 0.0960 & 0.0619 & 0.0712 & 0.226 & 0.46 & 0.29 & 0.31 & 0.81 \\
Gate ladder & 12 & 30 & $6+6$ & 1296 & 77 & 0.0405 & 0.0490 & 0.0333 & 0.192 & 0.58 & 0.62 & 0.43 & 0.98 \\
\bottomrule
\end{tabular}
\end{table*}

In practice, every variant
is executed under the
specified noise model
and sampled using a finite
number of shots. The
pretrained model then
predicts the corrected
probability distribution
of each variant,
\begin{equation}
    \widehat{p}_{j,\alpha}^{\mathrm{mit}}
    =
    \Pi_{\triangle_{16}}
    \left[
        \widetilde{p}_{j,\alpha}
        -
        \widehat{\Delta}_{j,\alpha}
    \right],
    \label{eq:cut_mitigation_variant}
\end{equation}
where $\Pi_{\triangle_{16}}$
denotes projection onto
the four-qubit probability
simplex.

These corrected distributions
replace the unmitigated
distributions in
Eq.~\eqref{eq:fragment_tensor_construction}.
The resulting tensors are
then contracted using
Eq.~\eqref{eq:cut_global_contraction}
to obtain the mitigated
global statistics.

Importantly, the projection
onto the probability simplex
is performed separately
for each fragment variant.
Because the global
reconstruction contains
signed coefficients, the
reconstructed estimate
need not itself be
non-negative in every
component. Nevertheless,
it provides an estimator
for the desired observables
through the same linear
functionals used for
ordinary measurement
distributions.
For the wire-cut transfer study
of \cref{tab:results_transfer},
the reconstructed full-circuit
distribution is additionally
projected onto the probability
simplex before $\REV$ and TV
are evaluated. Without this final
projection, the mean $\REV$ of
each set in \cref{tab:results_transfer}
changes by at most $0.017$
(12 qubits, depth 60: $0.163$
instead of $0.146$).

\paragraph{Residual mitigation error.}
The residual mitigation
error is also propagated
through the contraction.
Writing the corrected
fragment tensor as
\begin{equation}
    \widehat{R}_j
    =
    R_j^{\mathrm{id}}
    +
    \delta R_j,
\end{equation}
the difference between
the reconstructed
estimate and the
ideal result admits
a first-order expansion
\begin{equation}
\begin{split}
    \delta p_C
    =
    \sum_{j=1}^{N_F}
    \mathcal{R}_{\mathcal{P}}
    \left(
        R_1^{\mathrm{id}},
        \ldots,
        \delta R_j,
        \ldots,
        R_{N_F}^{\mathrm{id}}
    \right)
    +
    O(\delta R^2),
\end{split}
\label{eq:cut_error_expansion}
\end{equation}
where the higher-order
terms contain products
of errors from
different fragments.

\subsection{Direct reconstruction of observables}
\label{app:direct_observables}

Constructing the full
$2^n$-dimensional output
distribution is not always
necessary. When the goal
is to estimate a particular
observable, the terminal
output indices of each
fragment can be contracted
with the corresponding
observable coefficients
before the global
reconstruction.

For example, consider
a Pauli-$Z$ string
$Z_S=\prod_{i\in S}Z_i$.
For a fragment $j$, let
$S_j$ denote the subset
of its terminal output
wires belonging to $S$.
We define the reduced
fragment tensor
\begin{equation}
\begin{split}
    R_j^{(S)}
    (\boldsymbol{\lambda}_j)
    =
    \sum_{\mathbf{y}_j}
    (-1)^{
    \sum_{i\in S_j}y_i
    }
    R_j
    (\boldsymbol{\lambda}_j;
    \mathbf{y}_j).
\end{split}
\label{eq:cut_reduced_tensor}
\end{equation}

The expectation value
$\langle Z_S\rangle$
is then obtained by
performing the same
contraction as in
Eq.~\eqref{eq:cut_global_contraction},
with $R_j$ replaced by
$R_j^{(S)}$.
No full output distribution
is required.

\begin{table}[t]
\caption{Energies per site, $E/(NJ)$, of the 24-site holographic Ising
VQE (\cref{fig:results_holo}): exact ground state, noise-free
transferred ansatz, cut-noisy estimate, and mitigated estimates, with
$2\times10^5$ shots per measurement basis.}
\label{tab:results_holographic}
\centering\footnotesize
\setlength{\tabcolsep}{2.8pt}
\begin{tabular}{lrrrrrr}
\toprule
$h/J$ & Exact & Ansatz & Cut noisy & GNN & CDR & PEC \\
\midrule
0.5 & $-1.0273$ & $-1.0253$ & $-0.3237$ & $-1.0003$ & $-0.9804$ & $-0.3498$ \\
1.0 & $-1.2583$ & $-1.2351$ & $-0.4020$ & $-1.1318$ & $-1.0401$ & $-0.7542$ \\
1.5 & $-1.6643$ & $-1.6096$ & $-0.5127$ & $-1.5830$ & $-1.4812$ & $-0.5230$ \\
\bottomrule
\end{tabular}
\end{table}

More general Pauli
observables can be
evaluated using appropriate
measurement-basis rotations
on the terminal wires.
Consequently, for a
Hamiltonian expressed as
\begin{equation}
    H=\sum_{\mu}h_\mu P_\mu,
    \label{eq:cut_hamiltonian}
\end{equation}
where $P_\mu$ are Pauli
strings, the energy can
be estimated from the
reconstructed expectations
of its constituent terms,
\begin{equation}
    \widehat{E}
    =
    \sum_{\mu}
    h_\mu\,
    \widehat{\langle P_\mu\rangle}.
    \label{eq:cut_energy}
\end{equation}

\begin{table}[t]
\caption{Selected observables in the 20-qubit dimerized Ising quench at
$10^5$ shots per fragment configuration and, in the indented rows, at
$2\times10^4$ shots. CDR executes eleven circuits per fragment
configuration at the same number of shots per circuit. The EV-RMSE and
the half-$\ell_1$ discrepancy at both budgets are given in
\cref{fig:results_quench}(b,c).}
\label{tab:results_quench}
\centering\footnotesize
\setlength{\tabcolsep}{2pt}
\begin{tabular}{lrrrrr}
\toprule
Observable & Exact & Cut noisy & GNN & CDR & PEC \\
\midrule
Magnetization $m_z$ & $-0.142$ & 0.209 & $-0.292$ & $-0.364$ & $-0.194$ \\
\quad $2{\times}10^4$ shots & & 0.208 & $-0.285$ & $-0.431$ & $-0.188$ \\
Intradimer $\langle ZZ\rangle$ & 0.920 & 0.122 & 0.895 & 1.000 & 0.960 \\
\quad $2{\times}10^4$ shots & & 0.122 & 0.893 & 0.985 & 0.989 \\
Interdimer $\langle ZZ\rangle$ & 0.057 & 0.046 & 0.133 & 0.129 & 0.058 \\
\quad $2{\times}10^4$ shots & & 0.045 & 0.127 & 0.205 & 0.062 \\
Domain-wall density & 0.244 & 0.457 & 0.233 & 0.206 & 0.234 \\
\quad $2{\times}10^4$ shots & & 0.457 & 0.235 & 0.192 & 0.225 \\
\bottomrule
\end{tabular}
\end{table}
This observable-level
contraction avoids
explicitly constructing
the exponentially large
output distribution.
It is the strategy
employed for the
28-qubit energy
calculations.

For the full-distribution
benchmarks, the fragment
tensors are contracted
using a greedy ordering
that limits the size of
intermediate tensors.
Cut indices are summed
as soon as they no
longer connect to any
uncontracted fragment.
This reduces the memory
cost relative to
contracting fragments
strictly in circuit order,
particularly for circuits
containing many temporal
cuts.

Every cut plan is
validated before noisy
simulation by reconstructing
the exact ideal fragment
statistics. For circuits
of up to 24 qubits,
the reconstructed
probabilities agree
with direct statevector
simulation to approximately
$10^{-12}$. For the
28-qubit calculations,
the corresponding
one- and two-body
expectation values are
checked to approximately
$10^{-10}$.

These checks verify
the algebraic correctness
of the chosen cut
decompositions and
reconstruction procedure.
They do not, by themselves,
guarantee the accuracy
of the reconstructed
observables when the
fragment distributions
are noisy or imperfectly
mitigated.

\section{Additional Numerical Results}
\label{app:additional_numerics}

This appendix records additional numerical data underlying the
comparisons in Sec.~\ref{sec:results}. Unless otherwise stated,
fragment circuits are simulated using the fixed depolarizing plus
amplitude-damping noise model, and no parameters of the pretrained
four-qubit GNN are updated. The cut-noisy reference denotes
unmitigated reconstruction from noisy fragment outputs. The values
reported as $\REV$ for circuit ensembles are means of per-circuit
normalized MSE ratios, whereas EV-RMSE values are reported as in the
underlying simulation campaigns.

\subsection{Wire-cut controls, structured variational circuits, and method comparison}\label{app:subsec:wire_cut_variational}

Here we test whether the generalization to circuits larger than those used in training extends to variational circuits, and compare our method with established mitigation techniques.
The circuits are decomposed using wire cuts only.
Three variational families are considered: a matrix-product-state (MPS)
staircase, a hierarchical quantum convolutional neural network (QCNN),
and a particle-conserving Givens-rotation staircase; their six-qubit
instances are shown in \cref{fig:app_ansatz_circuits}. These contain
6--12 qubits and depths from 11 to 67. Each is decomposed through wire
hand-offs, with exact noisy fragment probabilities used for the GNN,
ZNE, and CDR, and with PEC statistically emulated with $10^6$ shots per
configuration. The EV-RMSEs are listed in \cref{tab:app_structured_variational} and shown in
Fig.~\ref{fig:app_variational_methods}, and \cref{fig:app_variational_scatter} shows the absolute
error of every reconstructed one- and two-body expectation value. Pooled over these 405 strings, the RMSE falls from $0.393$ (cut noisy) to $0.047$ with the GNN, compared with $0.029$ for CDR and $0.011$ for PEC.
The GNN substantially reduces the EV-RMSE for all nine targets.
For instance, the Givens staircase at 12 qubits and depth 67
has EV-RMSE $0.637$ before mitigation and $0.104$ afterward, whereas
CDR and PEC achieve $0.014$ and $0.008$, respectively. The near-Clifford Givens fragments are especially well suited to CDR, since its near-Clifford training circuits closely approximate the target
fragments. PEC is also effective in these cases because the fragments
contain comparatively few noise locations and its total quasiprobability overhead is modest. PEC gives the lowest EV-RMSE
for eight of the nine circuits and CDR for the remaining one (the eight-qubit Givens staircase); the GNN is the most
accurate method for none of them, although it reduces the error substantially in every case.

\begin{figure*}[t]
\centering
\includegraphics{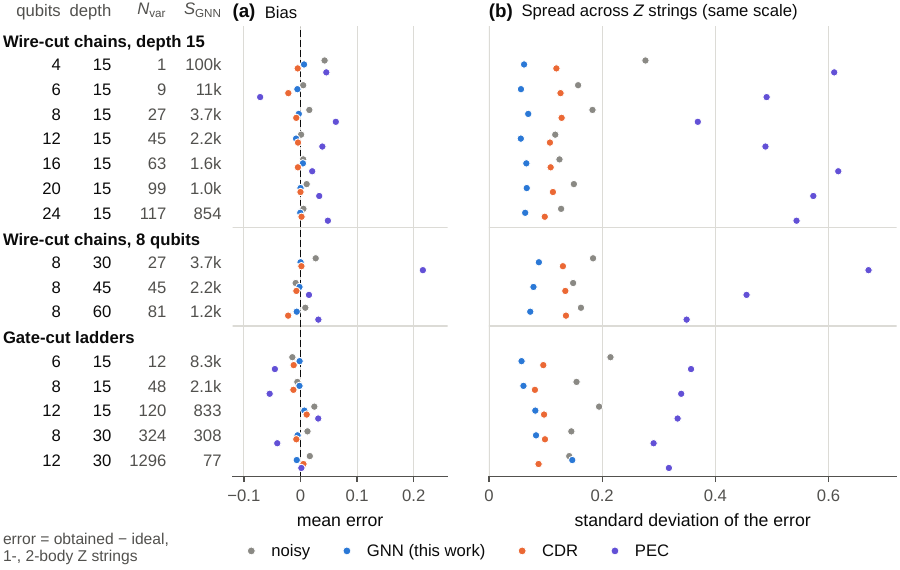}
\caption{Error decomposition under the fixed-total-budget protocol
($B=10^5$ shots per target circuit). Each row is one of the 15 cells
of the fixed-budget study (three target circuits each), labeled by its
numbers of qubits, depth, fragment configurations per target
$N_{\mathrm{var}}$, and shots per GNN configuration $S_{\mathrm{GNN}}$.
The errors (obtained $-$ ideal) of all one- and two-body Pauli-$Z$
expectation values of the cell's three circuits are pooled. (a)~Mean
of the error (bias); the dashed line marks zero bias. (b)~Standard
deviation of the error across these strings. Both panels use the same
scale. The RMS error over the same strings is within $5\%$ of (b) for
every method, so a small bias does not imply a small mean-squared
error. These statistics use only one- and two-body strings, whereas the
EV-RMSE of \cref{fig:results_fixed_budget} uses all $2^n-1$ strings.}
\label{fig:app_budget_bias}
\end{figure*}

For the equal-per-configuration-shot comparison of \cref{subsec:results_comparison}, the
complete EV-RMSE grid is listed in \cref{tab:app_grid_full}, and \cref{fig:app_grid_scatter}
shows the residual of every mitigated one- and two-body $Z$ expectation value against its
ideal value. Pooled over these strings, the GNN lowers the RMSE from $0.194$ to $0.047$ for scheme W and from $0.180$ to $0.034$ for scheme G, close to CDR ($0.054$ and $0.032$), whereas the PEC residuals of scheme W are widely spread (RMSE $0.271$, above the cut-noisy value).

\subsection{Finite-shot study and fixed-budget errors beyond EV-RMSE}\label{app:subsec:fixed_budget_errors}
The dependence of the GNN EV-RMSE on the number of shots per executed fragment circuit,
shown in Fig.~\ref{fig:results_shots}, is listed in \cref{tab:app_shot_scan}. The
fixed-total-budget study of \cref{subsec:results_budget} assigns $B=10^5$ shots to each
target circuit; the complete results, including PEC and the half-$\ell_1$ discrepancy,
are listed in \cref{tab:app_budget_full}.

In the main text, Fig.~\ref{fig:results_fixed_budget} shows that the GNN
gives the lowest EV-RMSE for most cells until the number of fragment
configurations becomes too large. At 20 qubits, the model reduces the
EV-RMSE from $0.00760$ to $0.00559$ with approximately 1010 shots per
configuration; the TV discrepancy decreases from $0.70$ to $0.48$. At 24
qubits, with only 854 shots per configuration, the GNN reduces the
EV-RMSE only from $0.00193$ to $0.00187$, and CDR is marginally lower
($0.00181$), although the GNN TV discrepancy decreases from $0.71$ to
$0.56$. At the other extreme, a 12-qubit depth-30 ladder requires 1296
fragment configurations, leaving only 77 shots per GNN input: the
EV-RMSE increases from $0.0405$ to $0.0490$. CDR achieves $0.0333$ in
that cell, whereas PEC ($0.192$) remains less accurate at its allocated
budget.

In addition to the RMSE, \cref{fig:app_budget_bias} separates the bias of the one- and two-body $Z$ expectation values from their spread. For the cases studied above, the bias of the unmitigated estimates is already small, so the RMSE is dominated by the variance.

\subsection{Holographic Ising VQE}
\label{app:subsec:holo}

\begin{figure*}[t]
\centering
\includegraphics{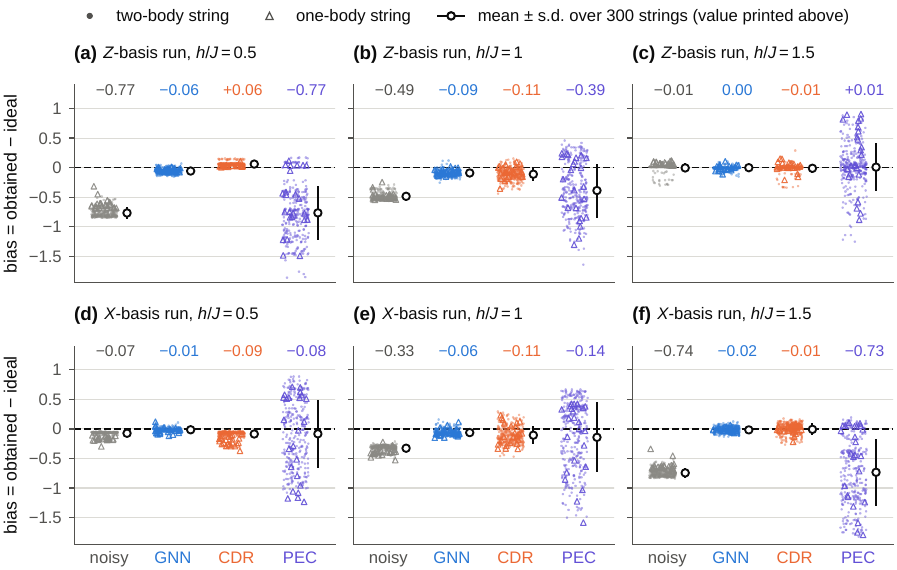}
\caption{Per-string errors of the holographic Ising VQE at $h/J=0.5$,
$1$, and $1.5$ (columns) in the $Z$-basis runs ($Z$ and $ZZ$ strings,
top) and $X$-basis runs ($X$ and $XX$ strings, bottom). Each marker is the
error (obtained $-$ ideal) of one of the 300 Pauli strings of a run
(triangles: 24 one-body strings; dots: all 276 two-body strings, not
only the 23 nearest-neighbor pairs that enter the energy). Black
markers and whiskers give the mean $\pm$ one standard deviation over the
300 strings, with the mean printed above each strip. ZNE was not run for
this experiment.}
\label{fig:app_holo_bias}
\end{figure*}

The energies underlying \cref{fig:results_holo} are listed in
\cref{tab:results_holographic}. \Cref{fig:app_holo_bias} shows the
errors of the individual one- and two-body Pauli expectation values from
which they are assembled. The cut-noisy errors form a narrow, coherent
offset that a learned correction can remove, whereas the PEC errors are
spread widely around their mean.

\begin{figure*}[t]
\centering
\includegraphics{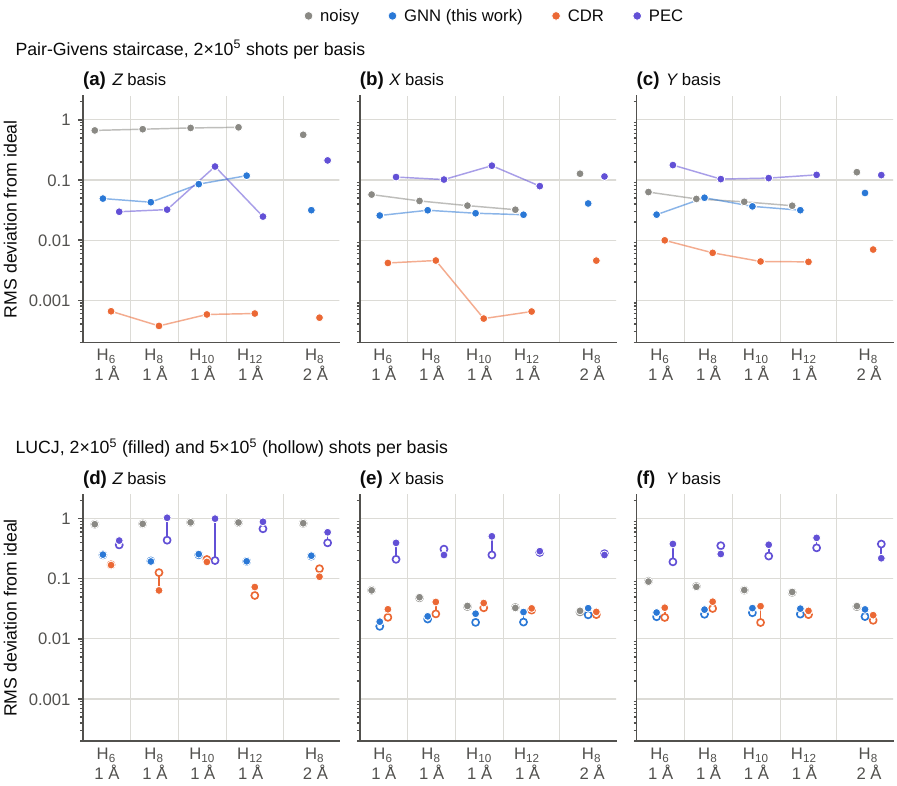}
\caption{Size of the Pauli-expectation errors for the number-conserving
chemistry benchmarks: RMS deviation from the ideal value over all one-
and two-body Pauli strings of each measurement-basis run ($Z$, $X$,
$Y$), per chain and method, on a logarithmic scale. (a)--(c)~Pair-Givens
staircase with $2\times10^5$ shots per basis; lines join the
$R=1$\,\AA{} chains. (d)--(f)~LUCJ with $2\times10^5$
(filled) and $5\times10^5$ (hollow) shots per basis. ZNE was not run for
these benchmarks.}
\label{fig:app_pairs_rms}
\end{figure*}

\subsection{Many-body dynamics with temporal and gate cuts}
\label{app:subsec:results_quench}

A complementary test concerns real-time dynamics of a 20-qubit
dimerized Ising chain, for which the qubits remain active throughout
the evolution. The Hamiltonian is
\begin{equation}
H=-J\sum_k Z_{2k}Z_{2k+1}
  -J'\sum_k Z_{2k+1}Z_{2k+2}
  -h\sum_i X_i,
\label{eq:results_dimer_ham}
\end{equation}
with ten dimers, $J=1$, $J'=0.25$, and $h=1$. The initial state
$\ket{0}^{\otimes20}$ is evolved to $t=1.4/J$ using nested Trotter
steps. The weak inter-dimer $ZZ$ interactions are gate-cut, and each
outer Trotter step defines a temporal wire-cut boundary. The
implementation contains seven time slices, 120 wire cuts, 32 $ZZ$
gate cuts, 70 fragments of two or three wires, and 40\,380 fragment
configurations, using the four-state wire-cut decomposition. The full
circuit depth is 77.
The four-state decomposition is carried out using
\begin{equation}
    \mathcal{S}_4
    =
    \left\{
    \ket{0},\ket{1},
    \ket{+},\ket{+i}
    \right\}.
    \label{eq:wire_four_states}
\end{equation}
This follows from the identities
\begin{equation}
\begin{aligned}
    X &= 2\ket{+}\bra{+}-I,\\
    Y &= 2\ket{+i}\bra{+i}-I.
\end{aligned}
\label{eq:wire_four_state_identity}
\end{equation}
The number of local configurations
is then reduced to
\begin{equation}
    N_{\mathrm{wire}}^{(4)}
    =
    4^a3^b.
\end{equation}

The reduction in the number of
preparation settings comes at the
cost of larger quasiprobability
coefficients. For the decompositions
used here, the associated one-norm
weights are
\begin{equation}
    \gamma_{\mathrm{w}}
    =
    \begin{cases}
        4, & \text{six-state scheme},\\
        6, & \text{four-state scheme}.
    \end{cases}
    \label{eq:wire_gamma}
\end{equation}
These weights characterize the
potential amplification of
statistical errors arising from
the signed reconstruction.

\begin{figure}[t]
\centering
\includegraphics{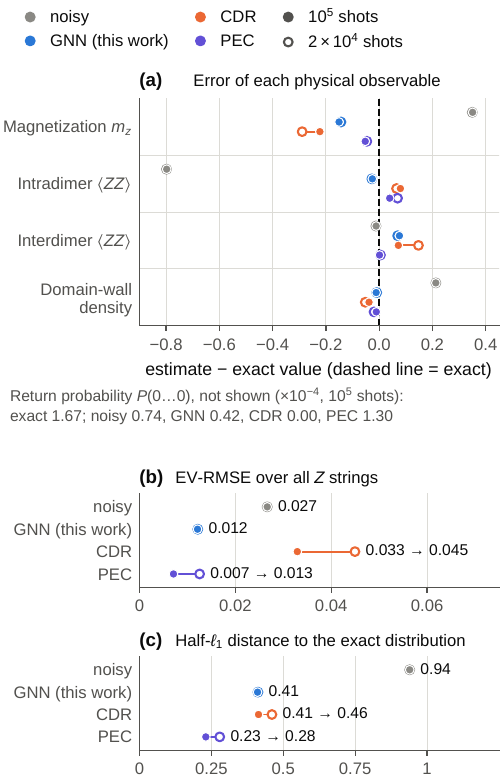}
\caption{Errors of physical observables (estimate minus exact noise-free
value) for the 20-qubit dimerized Ising-chain quench at depth 77.
(a)~Magnetization, intra- and interdimer $\langle ZZ\rangle$, and
domain-wall density; the dashed line marks zero error. Filled markers
use $10^5$ and open markers $2\times10^4$ shots per fragment
configuration. The return-probability values are listed below (a).
(b)~EV-RMSE over all $2^n-1$ $Z$ strings and (c)~half-$\ell_1$
distance to the exact 20-qubit distribution at the same two budgets;
labels give the $10^5\rightarrow2\times10^4$ values where they differ.
The same four-qubit mitigator is applied to all fragment configurations.}
\label{fig:results_quench}
\end{figure}

The results are summarized in \cref{tab:results_quench} and \cref{fig:results_quench}.
The cut-noisy data reduce the intradimer
correlation from the ideal value $0.920$ to $0.122$ and even give the wrong
sign for the magnetization. The learned mitigator restores the
intradimer correlation to $0.895$ and brings the domain-wall density from
$0.457$ to $0.233$, close to the ideal value $0.244$. However, the
magnetization is overcorrected to $-0.292$ rather than $-0.142$, and
the interdimer correlation is also less accurate than its unmitigated
value. PEC gives the smallest EV-RMSE in this example
[\cref{fig:results_quench}(b)], consistent with the relatively modest
inverse-noise overhead of the sparse fragments. CDR executes eleven
circuits per fragment configuration (444\,180 circuit executions in
total, compared with 40\,380 for the other methods), so at equal shots
per circuit it uses about eleven times the measurement budget.
Reducing the per-variant shots from $10^5$ to $2\times10^4$ changes
the GNN intradimer correlation only from $0.895$ to $0.893$ and its
EV-RMSE remains $0.012$, whereas CDR and PEC exhibit larger
shot-dependent changes. The example therefore demonstrates both the
usefulness of mixed temporal/gate cutting for many-body dynamics and
the observable dependence of residual learned-mitigation bias.

\subsection{Pauli-expectation errors in the number-conserving chemistry benchmarks}\label{app:subsec:bias_plots}

\begin{figure}[t]
\centering
\includegraphics{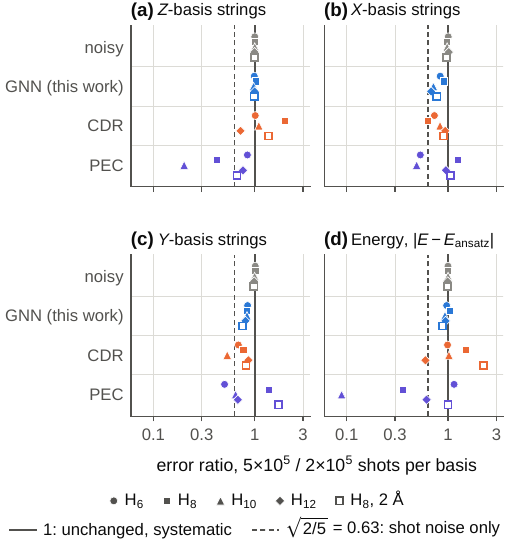}
\caption{Effect of increasing the measurement budget of the
LUCJ ansatz from $2\times10^5$ to $5\times10^5$ shots per basis:
ratio of the RMS deviation from the ideal value over the one- and
two-body Pauli strings of the (a)~$Z$-, (b)~$X$-, and (c)~$Y$-basis
runs, and (d)~of the absolute energy error
$|E-E_{\mathrm{ansatz}}^{\mathrm{id}}|$. The solid line at 1 marks an
unchanged (systematic) error, and the dashed line at
$\sqrt{2/5}\approx0.63$ marks pure shot-noise scaling (all per-circuit
shot counts scale by $2.5$). Marker shape identifies the chain. Each
ratio compares single runs. ZNE was not run.}
\label{fig:app_lucj_bias_500k}
\end{figure}

Complementing the discussion in \cref{subsec:results_pair_chemistry}, \cref{fig:app_pairs_rms} shows the
RMS deviation of the one- and two-body Pauli expectation values in all three measurement bases for both
number-conserving ans\"atze. In the $Z$ basis of the pair-Givens staircase, CDR stays below $10^{-3}$,
whereas the GNN leaves $0.03$--$0.12$; on the LUCJ ansatz, the CDR $Z$-basis error rises to
$0.06$--$0.2$ and that of the GNN to $0.2$--$0.26$. Fig.\ref{fig:app_lucj_bias_500k} tests whether these
errors are limited by sampling by increasing the budget of the LUCJ ansatz from $2\times10^5$
to $5\times10^5$ shots per basis. The GNN error ratios stay close to 1 in the $Z$ basis ($0.97$--$1.02$)
and for the energy ($0.88$--$1.04$), as for the cut-noisy control, so these errors are systematic. The GNN
$X$- and $Y$-basis ratios ($0.68$--$0.90$) move toward shot-noise scaling, whereas the PEC ratios scatter
between $0.09$ and $1.7$, as expected for sampling-limited estimates.

\begin{figure}[t]
\centering
\includegraphics{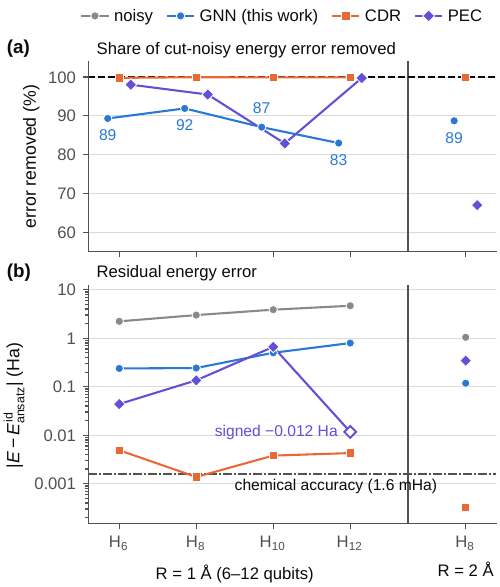}
\caption{Pair-Givens staircase versus chain length (one qubit per H
atom). (a)~Percentage of the cut-noisy energy error removed,
$100\,(1-|E_{\mathrm{m}}-E_{\mathrm{ansatz}}^{\mathrm{id}}|/|E_{\mathrm{noisy}}-E_{\mathrm{ansatz}}^{\mathrm{id}}|)$,
with the GNN values printed. The axis starts at $55\%$, the dashed line
marks complete removal, and markers are offset slightly horizontally for
legibility. (b)~Absolute residual energy error relative to the noise-free
ansatz energy on a logarithmic scale, with the cut-noisy estimate and
chemical accuracy ($1.6$\,mHa, dashed). Lines join the $R=1$\,\AA{}
chains; $\mathrm{H}_8$ at $R=2$\,\AA{} is shown separately. The hollow
diamond marks a negative signed error (PEC, $\mathrm{H}_{12}$:
$-0.012$\,Ha). Single runs with $2\times10^5$ shots per basis.}
\label{fig:app_givens_energy}
\end{figure}

\subsection{Extension to 28 qubits: nitrogen in the electron-pair space}
\label{subsec:results_nitrogen}

Finally, we test the same fixed neural mitigator on a 28-qubit,
depth-103 chemistry circuit. The target is $\mathrm{N}_2$ in a cc-pVDZ
basis with 28 spatial orbitals and 14 electrons (seven pairs), at
bond lengths $R=1.098$ and $1.6\,\text{\AA}$. We again work within the
seniority-zero subspace; the complete \emph{pair-space projected}
Hamiltonian is measured, and DOCI is the exact reference within that
subspace. The full-space coupled-cluster reference is not the target
energy of this restricted ansatz.

The LUCJ ansatz contains nine overlapping four-orbital
windows, eight wire cuts, nine fragments, and 135 fragment
configurations per measurement basis. Its full $Z$-measurement circuit
has 361 elementary gates, including 90 CNOTs, and depth 103. Because
the energy does not require explicit reconstruction of the
$2^{28}$-component distribution, the one- and two-body Pauli expectations
entering the pair Hamiltonian are contracted directly through the
fragment tensors, as described in \cref{app:direct_observables}.
The ideal reference and the variational optimization are obtained in
the seven-pair configuration space, which has
$\binom{28}{7}=1\,184\,040$ basis states. The ideal cut contraction
agrees with the direct ansatz expectations to within approximately
$10^{-10}$.

The measurement budget is $10^6$ shots for each of the $Z$-, $X$-, and
$Y$-basis runs, corresponding to approximately 7407 shots per
fragment variant for the GNN. CDR, with ten near-Clifford copies,
receives about 673 shots per configuration. The same pretrained
four-qubit network is used at both geometries, without further
training or changes to its input dimension.

\begin{table*}[t]
\caption{Energies in the electron-pair (seniority-zero) subspace, in
hartree. $E_{\mathrm{ansatz}}^{\mathrm{id}}$ is the noise-free energy of
the respective ansatz and $E_{\mathrm{DOCI}}$ the exact minimum in the
pair subspace. The last four columns give the signed energy errors
$E_{\mathrm{method}}-E_{\mathrm{ansatz}}^{\mathrm{id}}$. Hydrogen chains
(STO-3G): $2\times10^5$ shots per measurement basis; $\mathrm{N}_2$
(cc-pVDZ, 28 qubits, depth 103): $10^6$ shots per basis. All methods
draw on the same budget, so the CDR training copies reduce the shots per
CDR circuit. Each entry is a single run. The full-space FCI energies are
$-3.23607$ ($\mathrm{H}_6$), $-4.30757$ ($\mathrm{H}_8$, $1$\,\AA), and
$-3.79669$\,Ha ($\mathrm{H}_8$, $2$\,\AA); FCI was not computed for
$\mathrm{H}_{10}$ and $\mathrm{H}_{12}$.}
\label{tab:app_pair_energies}
\centering\footnotesize
\begin{tabular}{llrrrrrr}
\toprule
 & & & & \multicolumn{4}{c}{$E_{\mathrm{method}}-E_{\mathrm{ansatz}}^{\mathrm{id}}$} \\
\cmidrule(lr){5-8}
Ansatz & System & $E_{\mathrm{ansatz}}^{\mathrm{id}}$ & $E_{\mathrm{DOCI}}$ & Cut noisy & GNN & CDR & PEC \\
\midrule
Pair-Givens & $\mathrm{H}_6$, $1$\,\AA & $-3.15836$ & $-3.17086$ & $+2.2307$ & $+0.2382$ & $+0.0049$ & $+0.0440$ \\
Pair-Givens & $\mathrm{H}_8$, $1$\,\AA & $-4.19421$ & $-4.20911$ & $+3.0037$ & $+0.2430$ & $+0.0014$ & $+0.1356$ \\
Pair-Givens & $\mathrm{H}_{10}$, $1$\,\AA & $-5.23158$ & $-5.24833$ & $+3.8819$ & $+0.5015$ & $+0.0038$ & $+0.6644$ \\
Pair-Givens & $\mathrm{H}_{12}$, $1$\,\AA & $-6.26983$ & $-6.28798$ & $+4.6750$ & $+0.7956$ & $+0.0043$ & $-0.0117$ \\
Pair-Givens & $\mathrm{H}_8$, $2$\,\AA & $-1.98203$ & $-3.51198$ & $+1.0467$ & $+0.1181$ & $+0.0003$ & $+0.3454$ \\
\midrule
LUCJ & $\mathrm{H}_6$, $1$\,\AA & $-3.15968$ & $-3.17086$ & $+2.5836$ & $+0.6633$ & $+0.4067$ & $+0.5738$ \\
LUCJ & $\mathrm{H}_8$, $1$\,\AA & $-4.19512$ & $-4.20911$ & $+3.3264$ & $+0.6411$ & $+0.2289$ & $+3.6721$ \\
LUCJ & $\mathrm{H}_{10}$, $1$\,\AA & $-5.23306$ & $-5.24833$ & $+4.3504$ & $+0.8861$ & $+0.3112$ & $+4.2565$ \\
LUCJ & $\mathrm{H}_{12}$, $1$\,\AA & $-6.27073$ & $-6.28798$ & $+5.0965$ & $+0.8795$ & $+0.2056$ & $+4.0501$ \\
LUCJ & $\mathrm{H}_8$, $2$\,\AA & $-2.44412$ & $-3.51198$ & $+0.9764$ & $+0.2184$ & $+0.0453$ & $+0.5544$ \\
\midrule
LUCJ & $\mathrm{N}_2$, $1.098$\,\AA & $-108.9993$ & $-109.0364$ & $+62.113$ & $+6.888$ & $+2.778$ & $+18.700$ \\
LUCJ & $\mathrm{N}_2$, $1.6$\,\AA & $-108.7618$ & $-108.8016$ & $+59.607$ & $+7.050$ & $+3.173$ & $+69.547$ \\
\bottomrule
\end{tabular}
\end{table*}

\begin{figure}[t]
\centering
\includegraphics{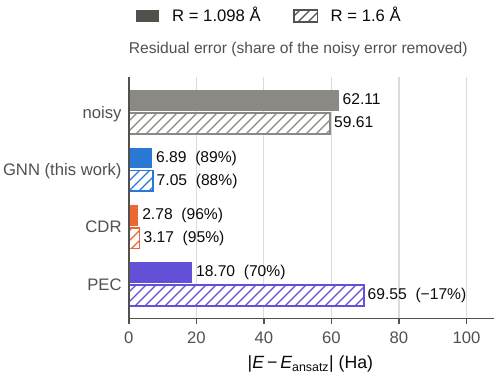}
\caption{Residual energy error $|E-E_{\mathrm{ansatz}}^{\mathrm{id}}|$
(Ha) of the 28-qubit, depth-103 $\mathrm{N}_2$ LUCJ ansatz in
the seniority-zero subspace, relative to the noise-free ansatz energy,
for $R=1.098$\,\AA{} (solid) and $1.6$\,\AA{} (hatched). Numbers in
brackets give the share of the cut-noisy error removed,
$100\,(1-|E_{\mathrm{m}}-E_{\mathrm{ansatz}}^{\mathrm{id}}|/|E_{\mathrm{noisy}}-E_{\mathrm{ansatz}}^{\mathrm{id}}|)$;
a negative share means the mitigated energy is further from the ansatz
energy than the cut-noisy one. All residuals exceed chemical accuracy
($1.6$\,mHa) by more than three orders of magnitude.}
\label{fig:results_nitrogen}
\end{figure}

The results are shown in \cref{fig:results_nitrogen} and listed in \cref{tab:app_pair_energies}.
At the equilibrium geometry, the unmitigated energy is $62.113$ Ha
above the ideal ansatz value. The GNN reduces this discrepancy to
$6.888$ Ha, recovering approximately $89\%$ of the initial error;
CDR leaves $2.778$ Ha and PEC $18.700$ Ha. At the stretched geometry,
the corresponding errors are $59.607$, $7.050$, $3.173$, and $69.547$
Ha for cut noisy, GNN, CDR, and PEC, respectively. Thus the proposed
method remains useful on a circuit much larger and deeper than those
used in training, but it is not the most accurate of the tested
mitigators for this chemical observable. The residual errors are also
much larger than the chemical-accuracy scale and should not be
interpreted as quantitatively accurate molecular energies.

\begin{figure}[t]
\centering
\includegraphics{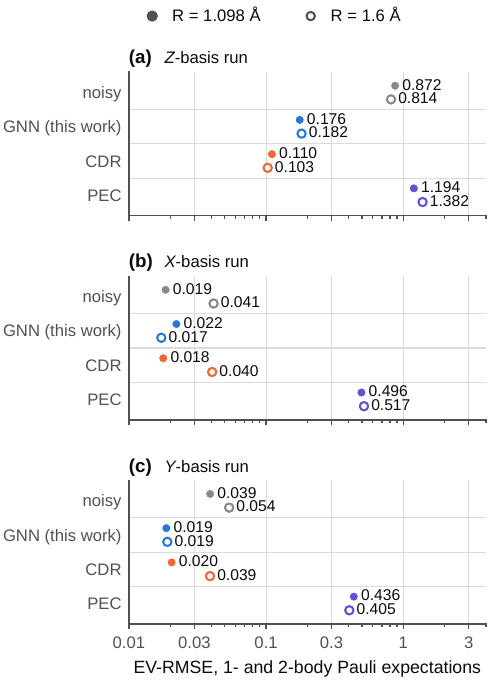}
\caption{RMS error of the recombined one- and two-body Pauli expectation
values (28 one-body and 378 two-body strings per basis run) for the two
28-qubit $\mathrm{N}_2$ geometries (filled: $R=1.098$\,\AA; open:
$R=1.6$\,\AA) in the (a)~$Z$-, (b)~$X$-, and (c)~$Y$-basis runs, on a
logarithmic scale.}
\label{fig:app_n2_bias}
\end{figure}

The decomposition of the energy error explains this outcome. The
noisy execution overpopulates nominally empty pair orbitals, producing
average pair numbers of $10.23$ and $10.28$ at the two geometries
instead of seven. The GNN reduces these to $6.80$ and $6.75$, whereas
CDR gives $7.23$ and $7.21$. PEC gives $7.24$ at equilibrium but
$10.29$ at the stretched geometry, where its energy error exceeds the
cut-noisy one. However, the
learned correction still under-occupies all seven occupied pair
orbitals, by $0.05$--$0.16$ in occupation. The deficit matters most for
the two N\,1s core orbitals, which are under-occupied by $0.05$ and
$0.11$. Their orbital energies are about $-53$\,Ha, so they alone
contribute about $+8$\,Ha to the one-body energy error, of which the
two-body term offsets about $4$\,Ha. In the
$Z$-basis runs, the RMS error of the reconstructed one- and two-body expectation values decreases
from $0.872$ to $0.176$ at $R=1.098\,\text{\AA}$ and from $0.814$
to $0.182$ at $R=1.6\,\text{\AA}$; CDR reaches $0.110$ and $0.103$,
respectively (see also \cref{fig:app_n2_bias}). The smaller errors in the $X$ and $Y$ measurement runs
contribute much less to the total energy discrepancy. These data are
consistent with the state-dependent bias observed in the hydrogen
chains and show why accurate local distribution reconstruction alone
is not sufficient to guarantee chemical accuracy for a weighted
many-body observable.

The computational distinction is nevertheless important. The
28-qubit target is never provided directly to the neural network;
it is processed as 135 four-qubit fragment configurations per basis.
The reported classical mitigation and observable-contraction costs are
approximately 10--13 seconds and two seconds, respectively, per
measurement basis, after the fragment statistics have been generated.
These timings do not include the quantum shot cost or the one-time
neural-network training cost. The example illustrates a scalable
\emph{interface} between a small learned mitigator and a larger
structured circuit, while simultaneously exposing the residual
observable bias that limits its accuracy.

\section{Hardware Test on an 11-Qubit Grover Circuit}\label{sec:grover}

\begin{figure*}[t]
  \centering
  \includegraphics{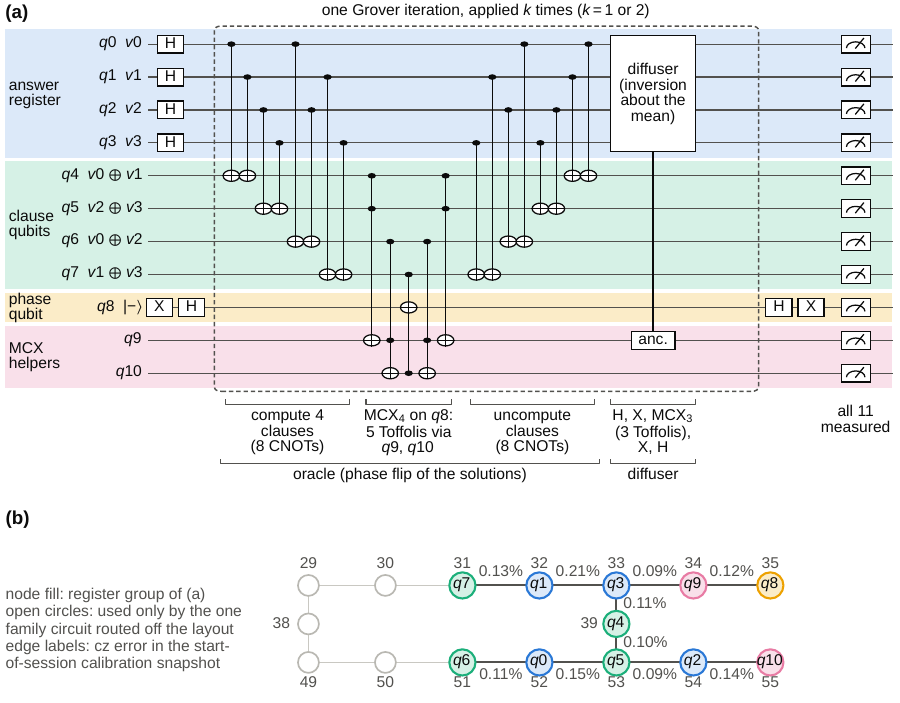}
  \caption{The hardware target. (a) Logical circuit of the 11-qubit $2\times2$-Sudoku Grover target; one iteration is
  shown, applied $k=1$ or 2 times. $q_0$--$q_3$: answer register (cells $v_0v_1/v_2v_3$). $q_4$--$q_7$: one rule each
  ($v_0\neq v_1$, $v_2\neq v_3$, $v_0\neq v_2$, $v_1\neq v_3$), computed as an XOR with two CNOTs and uncomputed
  afterwards. $q_8$: phase qubit in $|{-}\rangle$. $q_9$, $q_{10}$: helpers for the multi-controlled X gates (5
  Toffolis in the oracle, 3 in the diffuser). Post-selection discards shots with any of $q_4$--$q_{10}$ equal to 1.
  (b) Physical qubits of \texttt{ibm\_quebec}: the shared layout maps $q_0,\dots,q_{10}$ to 52, 32, 54, 33, 39, 53,
  51, 31, 35, 34, 55; the open circles were used only by the one family circuit that routing kept off the layout.
  Edge labels give \texttt{cz} errors in the start-of-session calibration snapshot (median 0.117\%). Node positions
  are schematic.}
  \label{fig:grover_circuit}
\end{figure*}

In this appendix, we test learned mitigation on an 11-qubit Grover-search circuit run on IBM's 156-qubit superconducting
processor \device{}. The networks of this appendix are separate models with their own device-level graph encoding,
trained only on hardware data from one session. No application of the \alldepth{} to hardware
data is recorded, and neither applies directly: both are built for at most four qubits, 15 layers, and the
gates H, S, T, $R_X$, $R_Y$, $R_Z$, and CNOT, whereas the compiled target consists of native
\cz{}, \texttt{rz}, \texttt{sx}, and \texttt{x} operations at depths 284 and 579, and qubit-dependent hardware noise
breaks the relabeling symmetry (\cref{subsubsec:symmetry}). We ask whether \emph{a network trained on same-session circuits that share a target's structure can
mitigate the held-out target from its compiled circuit and noisy counts}. The criterion fixed beforehand, a result closer to
the exact ideal than the raw one, is already met by post-selection, so we compare against strong baselines.

\subsection{Target, training family, and session}\label{sec:gr-setup}

The target (\cref{fig:grover_circuit}(a)) solves a $2\times2$ binary Sudoku~\cite{QiskitTextbook2023} with
four answer qubits, four clause qubits (one per rule), a phase-kickback qubit, and two helpers; with six-CNOT
Toffolis~\cite{Nielsen2010,Barenco1995PRA}, one Grover iteration~\cite{Grover1996STOC} contains 64 logical two-qubit
gates. The solutions are 0110 and 1001. For $|\mathcal S|$ marked outcomes among 16,
the ideal probability of a marked outcome after $k$ iterations is~\cite{Boyer1998FdP,Nielsen2010}
\begin{equation}
  P_k(|\mathcal S|)=\sin^2\!\big((2k+1)\theta\big),\qquad \sin\theta=\sqrt{|\mathcal S|/16},
  \label{eq:grover-ideal}
\end{equation}
so $\Psol=0.78125$ at $k=1$ and $0.9453125$ at $k=2$. Every qubit outside the answer register ($q_4$--$q_{10}$)
ideally returns to 0, so a shot with any of them equal to 1 is a detected error, and the exact ideal of every compiled circuit is computable. Family circuits keep the
target's registers, oracle skeleton, and diffuser and change only the rules (drawn from a vocabulary of 68 rules). The family has 1,761
circuits on 416 answer sets other than the Sudoku's; 315 circuits share the target's structure of four two-cell
linear rules, and 7 answer sets (339 circuits) share the Sudoku's complement-pair structure $\{s, s\oplus1111\}$,
the near-twins of the target.

All circuits share one initial layout (\cref{fig:grover_circuit}(b)), chosen by the mean estimated success
probability~\cite{Nishio2020JETC}, and were compiled with Qiskit~\cite{JavadiAbhari2024arXiv}; the target has 137
\cz{} gates at depth 284 ($k=1$) and 287 at depth 579 ($k=2$), in the family's shallow tail (83\% and 72\% of family circuits
have more). For zero-noise extrapolation (ZNE) the target was also run with every \cz{} tripled. The 23
reorderings, which assign the Sudoku's rules to the clause qubits in the other 23 orders, are physically distinct
circuits for the same problem, never used for training. All 1,870 circuits ran on 23 September 2026 in one
38-minute batch of 9 jobs, with 4,096 shots per circuit, one target copy per $k$ in every job, and XY4 dynamical
decoupling~\cite{Viola1999PRL}.

\begin{figure*}[t]
  \centering
  \includegraphics{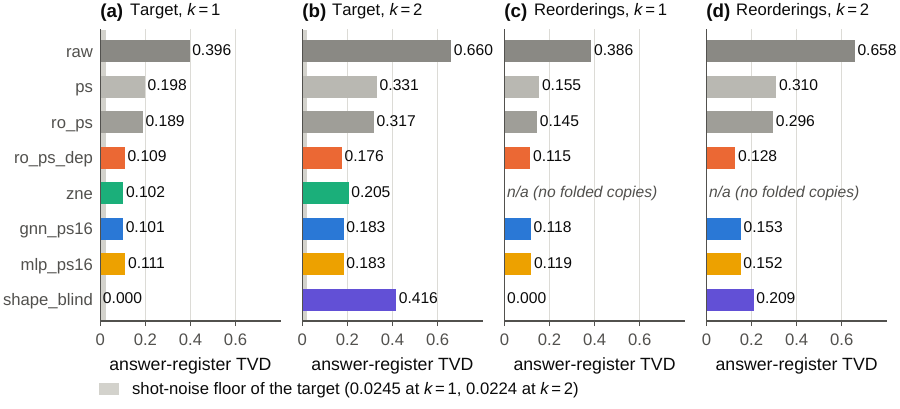}
  \caption{Answer-register TVD (\cref{eq:tvd}) to the exact ideal on \texttt{ibm\_quebec} (one session, 4,096 shots per
  circuit; lower is better) for the methods of \cref{sec:gr-methods}: (a, b) the held-out target at $k=1$ and 2,
  mean over its 9 job copies; (c, d) its 23 reorderings, for which \texttt{zne} is not available. The neural arms are
  separate models trained only on hardware data from this session, not the all-depth model; their values are
  3-seed ensembles averaged over 3 answer-set splits. Gray band in (a, b): shot-noise floor of the target, i.e., the
  expected TVD from sampling the ideal with the shots left after post-selection (0.0245 at $k=1$, 0.0224 at
  $k=2$). Resampling one target copy's counts moves its \texttt{ro\_ps\_dep} TVD by a standard deviation of 0.013
  ($k=1$) and 0.019 ($k=2$); differences between methods of this size are not resolved.}
  \label{fig:grover_methods}
\end{figure*}

\subsection{Methods compared}\label{sec:gr-methods}

The input of each circuit is its measured distribution over $2^{11}=\num{2048}$ outcomes, and its label is the
exact ideal of its compiled noiseless circuit. The family is split by answer set (292/62/62 training/validation/test
answer sets, about 1,265/250/245 circuits); three split seeds are crossed with three initialization seeds, and the
target and reorderings are always held out. Every estimate is projected onto the probability simplex and scored by
the answer-register total variation distance (TVD)
\begin{equation}
  \TVDfour(\hat p)=\frac12\sum_{x\in\{0,1\}^4}\big|\hat p_{[4]}(x)-p_{\mathrm{ideal},[4]}(x)\big|,
  \label{eq:tvd}
\end{equation}
where $[4]$ denotes the marginal on $q_0$--$q_3$; \REV{} is the 11-qubit analog of \cref{eq:relative_ev_error}, and \Psol{}
is the mitigated mass on $\{0110, 1001\}$. Arms are compared with the reference baseline \ropsdep{} by the
per-circuit change in \TVDfour{}, averaged over the three splits; each bracketed interval is the envelope (lowest
lower and highest upper bound) of the three per-split 95\% cluster-bootstrap intervals~\cite{Efron1979AnnStat}
(clusters are answer sets on the family test set and jobs on the target and reorderings). With at most 9 job
clusters, the intervals for the target and reorderings are approximate, and none includes training variance. This reference and this
comparison are not documented as fixed in advance.

\paragraph{Baselines.} The baseline \rawarm{} is the measured distribution. Post-selection, \psarm{}, keeps the 16
outcomes with $q_4$--$q_{10}$ all equal to 0 and renormalizes (a check designed into the circuit~\cite{BonetMonroig2018PRA}). The arm \rops{} first applies
readout correction~\cite{Bravyi2021PRA} with per-qubit assignment matrices measured in the same job, giving a
16-outcome distribution $r$. The reference \ropsdep{} then inverts a global depolarizing
channel~\cite{Vovrosh2021PRE},
\begin{equation}
\begin{aligned}
  \hat p_{\mathrm{dep}}&=\Psimp\!\left[\frac{r-\lambda\one/16}{1-\lambda}\right],\\
  \lambda&=\mathrm{clip}\!\left(a+b\,\frac{n_{2q}}{300}+c\,\frac{D}{600},\;0,\;0.95\right),
\end{aligned}
  \label{eq:depinv}
\end{equation}
with $n_{2q}$ the compiled \cz{} count, $D$ the compiled depth, and $(a,b,c)$ fitted by least squares to the
depolarizing strengths of the training circuits, a strong first model because random circuits turn local noise
into approximately global white noise~\cite{Dalzell2024CMP}. Zero-noise extrapolation, \znearm{}, is linear
extrapolation~\cite{Temme2017PRL,Li2017PRX,GiurgicaTiron2020QCE,Kandala2019Nature,Kim2023Nature} from the target
and its folded copy in the same job, $\hat p_{\mathrm{zne}}=\Psimp\big(\tfrac32\,r^{(1)}-\tfrac12\,r^{(3)}\big)$;
it needs no training data but exists only for the target. Finally, \shapeblind{} fits the two-level distribution of
\cref{eq:grover-ideal} to \rops{}, with the number of solutions chosen by least squares, so it uses only $k$.

\begin{figure*}[tbp]
  \centering
  \includegraphics{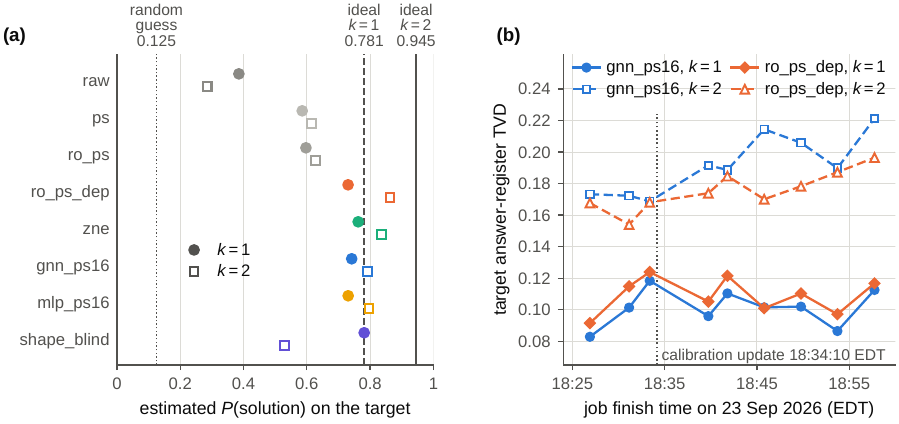}
  \caption{(a) Estimated probability of a Sudoku solution for the held-out target on \texttt{ibm\_quebec}, mean over its 9
  job copies (filled circles $k=1$, open squares $k=2$; methods as in \cref{fig:grover_methods}). The neural arms
  are models trained on hardware data, not the all-depth model; their values are means of 3 per-split 3-seed ensembles,
  with bars spanning the per-split values where these are wider than the marker. Vertical lines: exact ideal
  (0.78125 at $k=1$, dashed; 0.9453125 at $k=2$, solid) and random guessing (0.125, dotted). (b) Per-job target TVD
  across the session at each job's finish time (EDT), for \texttt{gnn\_ps16} (split-0 ensemble; circles $k=1$,
  open squares $k=2$) and \texttt{ro\_ps\_dep} (diamonds $k=1$, open triangles $k=2$); solid lines $k=1$, dashed
  lines $k=2$. Dotted line: a calibration update stamped 18:34:10 EDT, between the third and fourth jobs (finished
  18:33:23 and 18:39:44 EDT), whose start and end snapshots hold identical per-qubit and per-edge values. Because (b)
  uses split 0 only, its \texttt{gnn\_ps16} mean at $k=2$ (0.192) differs from the 3-split value (0.183).}
  \label{fig:grover_psolution}
\end{figure*}

\paragraph{Networks.} All networks receive the raw \num{2048}-outcome distribution and its post-selected
16-outcome version. Graph networks add the compiled circuit as a graph with one node per native operation, 72 node
features (one-hot over \cz{}, \texttt{rz}, \texttt{sx}, \texttt{x}, and measurement, 5; sine and cosine of the \texttt{rz}
angle, 2; layout-slot one-hots of both wires, 16+16; classical-bit one-hot, 11; start-of-session $T_1$, $T_2$,
readout, and \texttt{sx} errors of both wires, 8; \cz{} error and duration, 2; position, 5; circuit context, 7) and,
as in \cref{subsec:circuit_graph}, edges along each wire in both directions plus dilated edges three steps apart, read by a
6-layer GINE~\cite{Hu2020ICLR} of width 256 with an attention readout~\cite{Ilse2018ICML} that adds a learned
vector (a virtual node that passes no messages) and a three-layer MLP head. The
primary model \gnnps{} (1,546,769 parameters) predicts a zero-sum correction to the post-selected 16-outcome
distribution, so post-selection is built into it; \mlpps{} (598,544) is the same without the circuit, and \gnncirc{}
replaces the counts by a uniform distribution (a leakage test). One untuned default recipe was fixed before any
results (mean-squared-error loss on the zero-sum correction; AdamW~\cite{Loshchilov2019ICLR}, learning rate $10^{-3}$
with 3\% warm-up, cosine schedule over 1,500 epochs to 0.03 of the peak rate, weight decay $10^{-4}$, batch 16,
gradient clipping at 1.0, EMA 0.999, dropout 0.1; validation every 5 epochs, early stopping on validation \REV{}
after 40 evaluations without improvement). Of the encoding and training recipe of
\cref{sec:ml_method,app:ml_training}, only the GINE code, the bidirectional and dilated edge types, and the EMA carry over
(not the relabeling augmentation, depth-weighted sampling, budget schedule, or search), so this appendix does not
test that training recipe.

\subsection{Results}\label{sec:gr-results}

The raw target \Psol{} was $0.385\pm0.013$ ($k=1$) and $0.285\pm0.021$ ($k=2$) over the 9 job copies, yet the
solutions remained the two most likely outcomes on the target and all reorderings. All baselines and all networks
that use the counts are much closer to the ideal than raw hardware (\cref{fig:grover_methods}). The primary network
\gnnps{} lowers the target TVD from 0.396 to 0.101 at $k=1$ and from 0.660 to 0.183 at $k=2$. The baselines come as close: \ropsdep{}
reaches 0.109 and 0.176, and \znearm{}, without training data, 0.102 and 0.205, differences comparable to the
shot-noise spread of one target copy (about 1,950 and 1,160 shots survive post-selection; \cref{fig:grover_methods}). On the family test set
\gnnps{} and \ropsdep{} reach 0.155 and 0.151 (raw 0.463), and, on the reorderings at $k=2$, 0.153 and 0.128. Apart from \shapeblind{}, \Psol{}
(\cref{fig:grover_psolution}(a)) is best estimated by \znearm{} at $k=1$ (0.762, against 0.742 for \gnnps{} and
0.730 for \ropsdep{}) and by \ropsdep{} at $k=2$ (0.863), where the networks under-correct (\gnnps{} 0.791,
\mlpps{} 0.796, ideal 0.945). The two-level fit \shapeblind{} is exact on the target at $k=1$, because the device still ranks the
true answers first, but picks three solutions in 7 of 9 jobs at $k=2$ (mean TVD 0.416).

\paragraph{Paired comparison and drift.} No learned arm beats \ropsdep{} robustly. The TVD change of \gnnps{} is
$+0.004$ [$-0.004$, $+0.010$] on the family test set and $+0.025$ [$+0.008$, $+0.042$] on the reorderings at $k=2$.
The one favorable interval is on the target at $k=1$, $-0.008$ [$-0.014$, $-0.002$], and it is not robust: it is no
larger than the gap between the graph and counts-only models on that circuit (0.101 against 0.111), and it includes
zero when the near-twins are removed from training ($-0.005$ [$-0.012$, $+0.001$]).
An exact sign test gives $p=0.039$ for \gnnps{} (split 0) being better in 8 of 9 jobs at $k=1$, above
the Holm threshold~\cite{Holm1979ScandJStat} of 0.0011 for the 45 paired comparisons made (9 learned
arms, some not described here, $\times$ 5 cells: target and reorderings at $k=1$ and 2, family test set) at level 0.05. Across the session
(\cref{fig:grover_psolution}(b); split-0 ensembles) the mitigated TVD shows no trend at $k=1$, whereas at $k=2$ both
arms worsen and the split-0 \gnnps{} ensemble is worse than \ropsdep{} in all 9 jobs, by up to 0.044 (sign test
$p=0.0039$, also above the Holm threshold; the split-1 ensemble is better in 8 of 9).

\paragraph{The network ignores the circuit.} Shuffling graphs among held-out circuits of the same $k$ leaves the
per-run \REV{} of \gnnps{} essentially unchanged (family test set 0.1181 to 0.1189). The counts-only \mlpps{} matches
\gnnps{} on the family test set and the reorderings with about 40\% of the parameters and a seventh of the training
time. The circuit leaks no answers: \gnncirc{} is worse than raw (family test TVD 0.680).

\paragraph{Interpretation.}\label{sec:gr-interp} A network trained on a same-session, structurally matched family mitigates the held-out target well, but no
better than standard mitigation. The largest step, post-selection on $q_4$--$q_{10}$ (TVD 0.396 to 0.198 at $k=1$), is built into the
\texttt{ps16} head; the learned correction on top (to 0.101) matches what readout correction with a
three-coefficient depolarizing inversion (0.109) or ZNE without training data (0.102) achieves. The graph is not
uninformative: \cz{} count and depth explain little of the per-circuit depolarizing strengths ($R^2=0.265$), and at
$k=2$ the target reads 0110 ahead of 1001 in all 9 jobs (mean difference $+0.135$ after
\rops{}), whereas over the reorderings this difference ranges from $-0.163$ to
$+0.273$. A global depolarizing model cannot remove this circuit-specific bias, and the networks remove it only
partly. That they ignore the graph may partly reflect a structural limitation: with six layers and dilation 3,
message passing spans at most 18 operations along a wire (\cref{app:reach}), a small fraction of the target's 284
and 579 compiled layers, and the virtual-node readout passes no messages; whether an encoder with more reach would
use the graph was not tested.

\end{document}